\documentclass[pra,reprint,twocolumn,superscriptaddress,showpacs,floatfix]{revtex4-1}
\usepackage{amsmath}
\usepackage{mathrsfs}
\usepackage{txfonts}
\usepackage{amssymb}
\usepackage{graphicx}
\usepackage{hyperref}
\usepackage{ulem}
\usepackage{xcolor}
\usepackage{color}
\usepackage{overpic}
\usepackage{psfrag}
\usepackage{tabularx}
\usepackage{multirow}
\usepackage{array}
\usepackage{placeins}
\newcommand{\PreserveBackslash}[1]{\let\temp=\\#1\let\\=\temp}

\usepackage{bm}
\usepackage{dsfont}
\usepackage{mathtools}

\newcommand{\ket} [1] {| #1 \rangle}

\makeatletter

\begin{document}

\title{Green parafermions as emergent flat-band excitations in condensed matter}

\author{Huan-Qiang Zhou}
\affiliation{Centre for Modern Physics, Chongqing University, Chongqing 400044, The People's Republic of China}

\author{Ian P. McCulloch}
\affiliation{Department of Physics, National Tsing Hua University, Hsinchu 30013, Taiwan}
\affiliation{Frontier Center for Theory and Computation, National Tsing Hua University, Hsinchu 30013, Taiwan}

\author{Murray T. Batchelor}
\affiliation{Mathematical Sciences Institute, The Australian National University, Canberra ACT 2601, Australia}
\affiliation{Centre for Modern Physics, Chongqing University, Chongqing 400044, The People's Republic of China}

\begin{abstract}
	Green parafermions, originally introduced by Green and extended by Greenberg and Messiah through trilinear and relative trilinear commutation relations beyond Bose-Fermi statistics, are generally regarded as mathematical curiosities without physical realization. We show that these paraparticles can in fact emerge as composite excitations in a broad class of condensed-matter systems undergoing spontaneous symmetry breaking with type-B Goldstone modes. The key ingredient is the introduction of auxiliary Majorana fermions defined on emergent unit cells produced by partial translational-symmetry breaking. When the auxiliary Majoranas are treated as physical degrees of freedom, the resulting Green parafermion states (up to a projection operator) correspond to flat-band excitations, whose creation and annihilation operators satisfy the trilinear algebra. When they are regarded as fictitious, the same construction explains the appearance of exponentially many degenerate ground states and reveals a surprising correspondence between Green parafermions and self-similar geometric objects, such as the golden spiral. Explicit realizations are demonstrated for the ferromagnetic spin-1  biquadratic model and the ferromagnetic $\rm {SU}(2)$ flat-band Tasaki model, showing that condensed-matter systems with type-B Goldstone modes provide a natural setting for Green parafermions as emergent, possibly observable quasiparticles.
\end{abstract}
\maketitle

\section{Introduction}~\label{intro}
One of the fundamental principles in quantum mechanics is the symmetrization postulate stating that a quantum state of many identical particles must be either symmetric or antisymmetric under permutations, leading to either bosons or fermions~\cite{messiahbook}. For both of them, there exist two distinct but equivalent formalisms to describe a quantum theory of identical particles, namely the first and second quantization formalisms. In the first quantization formalism a quantum state of identical particles is characterized by a wave function, whereas in the second quantization formalism a multiparticle state is identified by a vector in the Fock space generated from the action of creation operators on a unique
Fock vacuum. However, there is no  {\it a priori} reason to exclude the existence of particles other than bosons and fermions, although there has long been an argument against their existence in nature, under some physical assumptions that are seemingly reasonable~\cite{Haag}.

Mathematically, there are two ways pointing towards a consistent quantum theory of identical particles other than bosons and fermions. One way is to consider representations of the braid group instead of the symmetric group, which leads to anyons in two spatial dimensions~\cite{leinaas,wilczek1}, with a remarkable realization in the fractional quantum Hall effect as emergent excitations~\cite{wilczek2, nayak}.  The other way is to consider a representation of the symmetric group beyond one dimension. Along this line, there are two schools of thought that lead to two distinct types of paraparticles, depending on the mathematical nature of representation spaces of the symmetric group, given
both yield representations of the symmetric group that are generically not one-dimensional. The first is Green paraparticles~\cite{green}, featuring the trilinear commutation relations for paraparticle creation and annihilation operators. Fruitful outcomes have been born out of an extensive investigation into  Green paraparticles in the past decades~\cite{messiah,greenberg,kamefuchi0,hartle,stolt,bialynicki,Yamada,landshoff,kamefuchi,tony,mark,mark1}, mainly focused on the inherent  mathematical structure underlying the defining trilinear commutation relations. The second is a novel type of paraparticles associated with an $R$ matrix as a solution to the constant Yang-Baxter equation~\cite{cyabe1,cyabe2,cyabe3}, with the bilinear commutation relations for paraparticle creation and annihilation operators as a notable feature~\cite{hazzard}.  Although both types of paraparticles share formally identical form in the first quantization formalism,  a distinction between them is embodied in the second quantization formalism, when they are reduced to ordinary bosons and fermions. In fact, Green paraparticles become ordinary particles when their order is equal to one~\cite{green, greenberg}. Note that the equivalence between the first and second quantization formalisms has been established for Green paraparticles of finite order, but it remains unclear for Green paraparticles of infinite order~\cite{stolt}.  In contrast, for $R$ matrix parastatistics, the defining bilinear commutation relations contain a built-in $R$ matrix, with its special choices corresponding to ordinary bosons and fermions.

Although the existence of both types of paraparticles in nature as elementary particles remains speculative, it was demonstrated that paraparticles associated with an $R$ matrix can be realized in condensed matter systems, which in turn are experimentally feasible in three-level Rydberg atom or molecule systems~\cite{hazzard,hazzard1,hazzard2}. Instead, as stressed in Ref.~\cite{hazzard}, a realization of Green paraparticles in a condensed matter system as low-lying excitations remains elusive.
As a consequence, a natural question arises as to whether or not Green paraparticles  emerge as low-lying excitations in condensed matter.

In this article, we aim to address this intriguing question, at least partially. As we shall argue,  Green parafermions  emerge as composite particles, with auxiliary Majorana fermions as a key ingredient, in a broad class of condensed matter systems 
undergoing spontaneous symmetry breaking  (SSB)  with type-B  Goldstone modes (GMs), as long as the ground state degeneracies are exponential with system size. Such exponential degeneracies have been investigated recently in the context of SSB with type-B GMs~\cite{goldensu3,spinorbitalsu4,dimertrimer,finitesize,TypeBtasaki,jesse}.  As it turns out,  Green parafermions manifest themselves in Green parafermion states (up to a projection operator) as flat-band excitations. If auxiliary Majorana fermions are physical, only one flat-band excitation is realized for a particular realization of Green parafermions.  In contrast, if auxiliary Majorana fermions are fictitious, then exponentially many flat-band excitations are exposed for the original model Hamiltonian, thus demystifying many puzzles in exponentially many degenerate ground states and revealing an unexpected connection between Green parafermions and self-similar geometric objects, with the celebrated golden spiral as the simplest example.
For auxiliary Majorana fermions, the dichotomy between  physical versus fictitious stems from an observation that the model Hamiltonian alone is not {\it sufficient} to fully specify such a condensed matter system. Instead, a definition of the Hilbert space is mandatory to make it clear what particles are involved in a specific model. Note that a similar observation has been made by Wen~\cite{wen} in a different context.

A conceptual framework developed here heavily relies on the possibility for a representation of Green parafermions in terms of ordinary spin or fermion degrees of freedom and auxiliary Majorana fermions, which in turn is deeply rooted in the fact that, for condensed matter systems undergoing SSB  with type-B  GMs,   if the ground state degeneracies are exponential with system size, then a lattice may be partitioned into emergent unit cells, regardless of being periodic or non-periodic, as a result of partial SSB of the translation symmetry under one lattice unit cell. Here by partial SSB we mean that for a condensed matter system with exponentially many degenerate ground states arising from SSB with type-B GMs, some of them are translation-invariant under a certain number of lattice unit cells, whereas others are not translation-invariant~\cite{goldensu3,spinorbitalsu4,dimertrimer,finitesize,TypeBtasaki,jesse}. Indeed, it is the presence of emergent unit cells that makes it possible to form Green parafermions by introducing auxiliary Majorana fermions as some extra degrees of freedom in a condensed matter system undergoing SSB  with type-B  GMs. In principle, our conceptual framework works for condensed matter systems   undergoing SSB  with type-B  GMs on any lattices in any spatial dimension, although our discussion is restricted to one spatial dimension for simplicity. 

Specifically, for  a condensed matter system  undergoing SSB  with type-B  GMs, it is possible to realize a set of Green parafermion fields on an emergent unit cell in a logically consistent way, if the ground state degeneracies are exponential with system size under both periodic boundary conditions (PBCs) and open boundary conditions (OBCs). Their creation and annihilation operators are realized by introducing auxiliary Majorana fermions on emergent unit cells, together with spin or fermion degrees of freedom, for a specific partition of a lattice into emergent unit cells.  Generically, such a partition is typically non-periodic.  However, if we restrict to an atypical partition with emergent unit cells arranged periodically, then the situation is drastically simplified, with the number of Green parafermion fields depending on the dimension of the local Hilbert space and the symmetry group of the model Hamiltonian under investigation. Note that the creation and annihilation operators for the same field satisfy the trilinear commutation relations, originally introduced by Green~\cite{green}, and  the  relative trilinear commutation relations for two different fields, developed by Greenberg and Messiah~\cite{greenberg}.  Here we note that not all realizations are Hermitian in condensed matter, in contrast to what has been implicitly assumed in the literature. Consequently, it is necessary to further develop the theory of parafield quantization, in order to accommodate a non-Hermitian realization of the trilinear and relative trilinear commutation relations.

Notably, a hierarchical structure is revealed at a Hamiltonian level in a well-known scenario that SSB is regarded as a limit of explicit symmetry breaking. 
As usual, an extra term is introduced into a specific model Hamiltonian, with a constant coefficient as an external control parameter.  Here both the model Hamiltonian and the extra term are defined in the original unconstrained Hilbert space $V_0$ and are translation-invariant under one lattice unit cell -- a feature that is tied with the ferromagnetic order. However, if some proper constraints, enforced by a projection operator, are imposed on the Hilbert space, then the projected Hamiltonian, with the extra term being included, becomes essentially the total Green parafermion number operator, namely the sum of all the Green parafermion number operators (up to a projection operator). As a result, one may focus on a Green parafermion system, with its Hamiltonian being the total Green parafermion number operator.

This Green parafermion system is invariant under subsystem gauge transformations -- a notion that interpolates between local gauge transformations and global ones, in contrast to the model Hamiltonian and the extra term. Here subsystem gauge symmetries originate from an internal symmetry group acting on local (internal) degrees of freedom located at different lattice unit cells inside an emergent unit cell, so they are not {\it inherent} in nature. Instead, an internal symmetry group is {\it induced} from the presence of auxiliary Majorana fermions that are arranged to make it possible for local spin or fermion degrees of freedom located at lattice unit cells inside different emergent unit cells to communicate with each other for a specific partition. As a result, for an atypical partition,  we have to introduce a set of auxiliary (physical or fictitious) Majorana fermions (arranged periodically), each of which acts as a medium for spin or fermion degrees of freedom located at different emergent unit cells to communicate. Note that communication between spin or fermion degrees of freedom, with auxiliary Majorana fermions acting as a medium, represents a novel type of interaction that is well beyond any traditional characterization of couplings between different types of particles, in the sense that auxiliary Majorana fermions do not manifest themselves in the Hamiltonian explicitly.  In addition, there is a peculiar role of a Green parafermion system in a hierarchical structure at a Hamiltonian level: all the sectors labeled by the total number of Green parafermions have to be taken into account such that
they yield exponentially many degenerate ground states, when an external control parameter vanishes in the scenario that SSB is a limit of explicit symmetry breaking. A connection between a condensed matter system and a Green parafermion system is thus established via the constraints imposed on the Hilbert space, though the former is not invariant under subsystem gauge transformations. Mathematically, these constraints may be characterized in terms of two projection operators that commute with each other. As such, 
flat-band excitations in condensed matter systems undergoing SSB with type-B GMs may be constructed as Green parafermion states in the momentum space representation, up to a projection operator, as a result of the translation symmetry of the model Hamiltonian under one lattice unit cell, when PBCs are adopted. Here Green parafermion states appear as eigenstates of the total Green parafermion number operator for an Hermitian realization or left and right eigenstates of the total Green parafermion number operator for a non-Hermitian  realization.

As a salient feature, auxiliary Majorana fermions do not appear in the model Hamiltonian and the extra term as well as in the total Green parafermion number operator. Auxiliary Majorana fermions thus simply behave as if they are {\it spectators} instead of {\it participants}.  As already mentioned above, auxiliary Majorana fermions fall into two categories: one is physical and the other fictitious. For auxiliary but physical Majorana fermions,
the Hilbert space must be enlarged to accommodate both the subspace for spin or fermion degrees of freedom and the subspace for auxiliary Majorana fermions. In contrast, for auxiliary but fictitious Majorana fermions, the Hilbert space only involves spin or fermion degrees of freedom, depending on what model is under investigation. 
Accordingly,  if auxiliary Majorana fermions are physical, then they may be exploited to form {\it real}
Green parafermions on emergent unit cells for a specific partition. On the other hand, if auxiliary Majorana fermions are fictitious, then they may be exploited to form {\it imaginary} Green parafermions on emergent unit cells for all possible partitions. 

For auxiliary (physical) Majonara fermions, if they are introduced on  emergent unit cells arranged periodically for an atypical partition, then the extra term introduced in this scenario selects Green parafermion states or their linear combinations in the momentum space representation (up to a projection operator) as flat-band excitation states in condensed matter systems undergoing SSB with type-B GMs,  as a result of the translation symmetry of the model Hamiltonian under one emergent unit cell. Indeed,  for a specific partition, there are different ways to organize local spin or fermion degrees of freedom inside an emergent unit cell; one way corresponds to one choice of gauge for a Green parafermion system and different ways are connected via a gauge transformation, under the condition that {\it only} observables invariant under subsystem gauge transformations are allowed. Otherwise, inaccessibility to internal degrees of freedom inside an emergent unit cell is violated so that one has to work in another partition. In other words, real Green parafermions are defined in emergent unit cells for one single partition, so any local spin or fermion degrees of freedom belong to {\it only} one emergent unit cell, in contrast to imaginary Green parafermions.

For auxiliary (fictitious) Majorana fermions,  the absence of auxiliary (physical) Majorana fermions defined on an emergent unit cell of {\it fixed} size for a specific partition is equivalent to the presence of auxiliary (fictitious) Majorana fermions defined on emergent unit cells of any different sizes  for all possible partitions.  It follows that no local spin or fermion degrees of freedom could be regarded as internal degrees of freedom inaccessible to local measurements, since all possible partitions are on the same footing. Indeed, local degrees of freedom inside an emergent unit cell are inaccessible {\it only} in {\it one} particular partition, but not if all possible partitions are allowed to be present. As such,  for imaginary Green parafermions, any given local degrees of freedom belong to different  emergent unit cells, depending on what partition is adopted. The projected Green parafermion states may thus be expanded in terms of exponentially many monomials constructed from auxiliary (fictitious) Majorana fermions, if we resort to a specific realization of Green parafermions. As a consequence, the coefficients in front of these monomials must yield exponentially many degenerate ground states of the model Hamiltonian. This construction offers a natural explanation for many puzzling features in exponentially many degenerate ground states arising from SSB with type-B GMs  in a condensed matter system. In particular, type-B GMs as a result of SSB themselves are not {\it fundamental}, in the sense that they may be reduced to exponentially many flat bands, which emerge from the identification of the projected Green parafermion states with the primary family consisting of the highest weight state(s) and primary generalized highest weight states.  All other secondary families stem from  the primary family, thus revealing a hierarchical structure behind exponentially many degenerate ground states. This statement in turn reflects the aforementioned hierarchical structure at a Hamiltonian level in the scenario that SSB is regarded as a limit of explicit symmetry breaking. As a result, this identification stems from conceptual cross-fertilization between disparate research areas, ranging from parastatistics to quantum states of matter and quantum phase transitions.

The layout of this article is as follows.
In Section~\ref{ssbtype-b}, we review a few general properties of condensed matter systems undergoing SSB with type-B GMs, emphasizing the structure of their ground state manifolds, which are exactly solvable as a result of the frustration-free nature of the model Hamiltonians. We present the spin-1 ferromagnetic biquadratic model~\cite{barber,barber1,barber2}  (also cf. Refs.~\onlinecite{saleur,aufgebauer,moudgalya} for other relevant approaches to this model) and the  $\rm {SU}(2)$ flat-band ferromagnetic Tasaki model~\cite{tasaki} as two typical examples for condensed matter systems undergoing SSB  with exponentially many degenerate ground states, one for quantum many-body spin systems and the other for strongly correlated itinerant electron models. 
Section~\ref{projection} introduces the projection-operator framework,  with a focus on the two illustrative examples.
In Section~\ref{gscheme},  we turn to specific realizations of Green parafermions in condensed matter,
with auxiliary Majorana fermions as a key ingredient, establishing the hierarchical structure that connects the original microscopic Hamiltonian to the emergent Green-parafermion description. The distinction between auxiliary physical and fictitious Majorana fermions and its implications for the structure of the Hilbert space are discussed in detail. 
In Section~\ref{ssbscenario}, Green parafermions are realized in a scenario that SSB is a limit of explicit symmetry breaking. This construction leads to two equations that are necessary to identify an extra term breaking the symmetry group explicitly, which is fundamental to realize Green parafermions (up to a projection operator) in condensed matter. 
In Section~\ref{real} and Section~\ref{fictitious}, real and imaginary Green parafermion states are constructed from auxiliary physical and fictitious Majorana fermions, respectively.   In addition to the two illustrative models,  we mention the applicability of our framework to a few other condensed matter systems undergoing SSB with type-B GMs~\cite{dtmodel,So4}. Consequently,
condensed matter systems undergoing SSB with type-B GMs provide a playground for realizing Green parafermion states (up to a projection operator), as long as the ground state degeneracies are exponential with system size.
Section~\ref{conclusion} summarizes the results and discusses their broader implications. Technical derivations and supplementary material are collected in the Appendices.

\section{Condensed matter systems undergoing SSB with type-B GMs}~\label{ssbtype-b} 
 
Consider a condensed matter system described by a specific (frustration-free) model Hamiltonian $\mathscr{H}$, which undergoes continuous SSB from the symmetry group $G$ to the residual symmetry group $H$. Here $\mathscr{H}$ itself is always Hermitian and translation-invariant under one lattice unit cell. By frustration-free we mean that there is a ground state of the Hamiltonian $\mathscr{H}$, which is a simultaneous ground state of each and every Hamiltonian density~\cite{tasakibook}. Actually, it is possible to state that a model Hamiltonian undergoing SSB with type-B GMs is frustration-free,  under a reasonable assumption that there is a unique highest weight state $\vert \psi_0 \rangle$, which is translation-invariant under one lattice unit cell, for a semi-simple symmetry group $G$ or the semi-simple subgroup of the non-semi-simple symmetry group $G$. Indeed, this assumption is valid for all known condensed matter systems undergoing SSB with type-B GMs, regardless of the ground state degeneracies being polynomial or exponential with system size~\cite{FMGM,hqzhou,2dtypeb,goldensu3,spinorbitalsu4,dimertrimer,finitesize,TypeBtasaki,jesse}. 
Hence the model Hamiltonian $\mathscr{H}$ may be decomposed into a sum of projection operators (up to a positive multiplicative constant and an additive constant). Indeed, it is positive semi-definite. The frustration-free nature makes it possible to investigate the underlying physics of condensed matter systems undergoing SSB with type-B GMs from their exactly solvable degenerate ground states.
Indeed, there are quite a few condensed matter systems that appear as a representation of the Temperley-Lieb algebra in their one-dimensional guises~\cite{tla,baxterbook,martin}, so they are completely integrable in the Yang-Baxter sense~\cite{baxterbook,faddeev}. However, we stress that the occurrence of SSB with type-B GMs does not rely on complete integrability.  Instead it is strongly tied with the frustration-free nature of the model Hamiltonians. 

A proper classification of GMs has been achieved after a long term pursuit~\cite{goldstone,goldstone1,goldstone2,nielsen, schafer, miransky, nicolis,nambu,   brauner-watanabe, brauner-watanabe1,watanabe,watanabe1,NG,NG1,NG2}. This pursuit culminated in the introduction of type-A GMs and type-B GMs, thus leading to the establishment of the GM counting rule~\cite{watanabe,watanabe1,NG}: $N_A+2N_B=N_{BG}$,
where $N_A$ and $N_B$ are, respectively, the numbers of type-A and type-B GMs, and $N_{BG}$ is equal to the dimension of the coset space $G/H$. 
For our purpose, we restrict ourselves to one spatial dimension, with system size $L$ (measured in terms of the lattice unit cells), though it is possible to make an extension to two and higher spatial dimensions. We remark that occasionally it is convenient to measure system size in terms of the number of lattice sites, it thus should be multiplied by the number of lattice sites contained in one lattice unit cell.  As a convention,  we use $j$ or $q$  to label lattice unit cells for quantum many-body spin systems or strongly correlated itinerant electron systems. One of the advantages for choosing to work in one spatial dimension is that
SSB with type-A GMs is forbidden as a result of the Mermin-Wagner-Coleman theorem~\cite{mwc,mwc1}. This observation drastically simplifies the GM counting rule, given no type-A GMs emerges, namely $N_A=0$.

As a consequence, condensed matter systems undergoing SSB with type-B GMs may be classified into two categories~\cite{goldensu3,spinorbitalsu4,dimertrimer,finitesize,TypeBtasaki,jesse}: for one category,  the ground state degeneracies under PBCs and OBCs are the same, and are polynomial with system size, with the ferromagnetic spin-$1/2$ ${\rm SU}(2)$  Heisenberg model as a paradigmatic example; for the other category, the ground state degeneracies under PBCs and OBCs are different, and are exponential with system size, with  the ferromagnetic spin-1 biquadratic model and the  $\rm {SU}(2)$ flat-band ferromagnetic Tasaki model as two typical examples. 

The  spin-1 ferromagnetic biquadratic model~\cite{barber,barber2} is described by the Hamiltonian
\begin{equation}
	\mathscr{H}=\sum_{j}{\left(\textbf{S}_j \cdot \textbf{S}_{j+1}\right)^2}. \label{hambq}
\end{equation}
Here $\textbf{S}_j=(S^x_j,S^y_j,S^z_j)$ is the vector of the spin-1 operators at lattice site $j$, acting on the local Hilbert space spanned by $\vert + \rangle_j$, $\vert 0 \rangle_j$ and $\vert - \rangle_j$, which are eigenvectors of $S^z_j$ with respective eigenvalues 1,0 and $-1$.
The sum over $j$ is taken from 1 to $L$ for PBCs.
This model constitutes a representation of the Temperley-Lieb algebra, so it is exactly solvable by means of the Bethe Ansatz and related methods~\cite{barber,barber2}.
It possesses staggered ${\rm SU(3)}$ symmetry~\cite{afflecksun} if $L$ is even, which has a total of 8
generators, and  uniform ${\rm SU(2)}$ symmetry if $L$ is odd,  which has only a total of 3
generators. The eight generators for the staggered ${\rm SU(3)}$ group and the three generators for the uniform ${\rm SU(2)}$ group can be expressed in terms of the spin-1 operators $S^x_j,S^y_j$, and $S^z_j$ at lattice site $j$~\cite{goldensu3,jesse}. The SSB pattern is from ${\rm SU(3)} $ to ${\rm U(1)}\times {\rm U(1)}$, with $N_B=2$, if $L$ is even and from ${\rm SU(2)} $ to ${\rm U(1)}$, with $N_B=1$, if $L$ is odd. 
Hence this model exhibits the odd-even parity effect. Note that the lattice unit cell contains only one lattice site, so the model Hamiltonian (\ref{hambq}) is translation-invariant under one lattice site.

The $\rm {SU}(2)$ flat-band ferromagnetic Tasaki model may in general be defined on a $d$-dimensional decorated hypercubic lattice  $\Lambda$. Here $\Lambda$ is the union of two sublattices $\mathscr{E}$ and  $\mathscr{I}$, namely $\Lambda=\mathscr{I} \cup \mathscr{E}$, where $\mathscr{E}$ denotes the set of sites in the lattice with unit lattice spacing, and $\mathscr{I}$ consists of all the sites located at the middle of each bond of the sublattice $\mathscr{E}$. As a convention, 
$ \mathscr{E}$ is called the external sublattice and  $\mathscr{I}$ is called the internal sublattice. In fact, this model is a variant of the Hubbard model and is described by the Hamiltonian~\cite{tasaki}
\begin{equation}
	\mathscr{H}= t\sum_{u\in \mathscr{I},\sigma=\uparrow, \downarrow}{\hat b}_{u,\sigma}^\dagger{\hat b}_{u,\sigma}+
	U\sum_{x\in \Lambda}{\hat n}_{x,\uparrow}{\hat n}_{x,\downarrow}, \label{hamtasaki}
\end{equation}
where  ${\hat b}_{u,\sigma}={\hat c}_{u,\sigma}+\nu\sum_{q \in \mathscr{E},(|q-u|=1/2)}{\hat c}_{q,\sigma}$ and  ${\hat b}^\dagger_{u,\sigma}$ is the adjoint for $u\in \mathscr{I}$ and ${\hat n}_{x,\sigma} ={\hat c}^\dagger_{x,\sigma} {\hat c}_{x,\sigma}$ for $x\in \Lambda$, with ${\hat c}^\dagger_{x,\sigma}$ and  ${\hat c}_{x,\sigma}$ the creation and annihilation operators of electrons for spin $\sigma$ ($\sigma =\uparrow$ and $\downarrow $) at lattice site $x$, respectively. In addition, $t$  is a hopping  parameter, $U$ describes the on-site repulsive Coulomb interaction, and $\nu$ is a real number. From now on, we assume that $t>0, U> 0$ and  $\nu>0$. We emphasize that in the above definition of ${\hat b}_{u,\sigma}$, the sum over $q$, subject to the constraints $|q-u|=1/2$,  depends on the type of boundary condition adopted~\cite{TypeBtasaki}.  Our construction here is restricted to PBCs.   Following Tasaki~\cite{tasaki,tasakibook},  we define $\hat{a}_{q,\sigma}^\dagger={\hat c}_{q,\sigma}^\dagger-\nu \sum_{u\in \mathscr{I},(|q-u|=1/2)}{\hat c}_{u,\sigma}^\dagger$ ($\sigma = \uparrow$ and $\downarrow$). As we shall see later on, these operators are suitable for a realization of Green parafermions in this model.

The Hamiltonian (\ref{hamtasaki}) possesses the symmetry group  $\rm{U(1)} \times\rm{SU(2)}$, where $\rm{U(1)}$ is generated by the electron number operator ${\hat N}= {\hat N}_\uparrow + {\hat N}_\downarrow$ in the charge sector, with $ {\hat N}_\sigma = \sum _{x\in \Lambda} {\hat n}_{x,\sigma}$,
and  $\rm{SU(2)}$  is generated by the three generators  ${\hat S}^\pm=\sum_{x\in \Lambda} {\hat S}^\pm_x$ and ${\hat S}^z=\sum_{x\in \Lambda}{\hat S}^z_x$ in the spin sector, satisfying $[{\hat S}^z,{\hat S}^+]={\hat S}^+$, $[{\hat S}^z,{\hat S}^-]=-{\hat S}^-$ and $[{\hat S}^+,{\hat S}^-]=2{\hat S}^z$. Here ${\hat S}^z_x$, ${\hat S}^+_x$ and ${\hat S}^-_x$ are defined as ${\hat S}^z_x=({\hat n}_{x \uparrow}-{\hat n}_{x \downarrow})/2$, $ {\hat S}^+_x={\hat c}_{x\uparrow}^\dagger {\hat c}_{x\downarrow}$ and $ {\hat S}^-_x={\hat c}_{x\downarrow}^\dagger {\hat c}_{x\uparrow}$. Note that ${\hat S}^z= ({\hat N}_\uparrow - {\hat N}_\downarrow)/2$. From now on, we restrict ourselves to the one-dimensional version of the Tasaki model on a decorated lattice $\Lambda$, where the lattice unit cells contain two nearest-neighbor lattice sites,
with one from the external sublattice  $\mathscr{E}$ and  the other from the internal sublattice $\mathscr{I}$. For brevity,  the system size $|\Lambda|$ is identified as $2L$ (measured in terms of the number of lattice sites), so the total number of the lattice unit cells is $L$. Here the lattice sites in $\mathscr{E}$ are labeled as $q \in \{ 1,2,\ldots,L\}$ and the lattice sites in $\mathscr{I}$ are labeled as $ u \in \{1/2,3/2,\ldots,L+1/2\}$.
As a convention, the unit cell labeled by $q$ consists of two lattice sites, one labeled by $q$ from  the external sublattice  $\mathscr{E}$ and the other by $u=q+1/2$ from the internal sublattice $\mathscr{I}$.
As shown recently~\cite{TypeBtasaki}, SSB from $\rm{SU(2)}$ to  ${\rm U(1)}$ with type-B GMs
occurs, where $N_B=1$, and the ground state degeneracies are exponential in system size, but different under PBCs and OBCs.

As a common feature,  both of the models (\ref{hambq}) and (\ref{hamtasaki}) exhibit a hierarchical structure among exponentially many degenerate ground states. As argued in Refs.~\cite{goldensu3,jesse,TypeBtasaki}, this hierarchical structure has already appeared in exponentially many generalized highest weight states,  including the highest weight state as a special case. Generically,  atypical generalized highest weight states are periodic and typical generalized highest weight states are non-periodic. Physically, the occurrence of atypical degenerate ground  states with period $p$ follows from partial SSB from the cyclic group $\mathscr{Z}_L$  generated by the translation symmetry operation under one lattice unit cell to the
cyclic group $\mathscr{Z}_{L/p}$  generated by the translation symmetry operation under $p$ lattice unit cells for all possible $p$'s, as long as $p$ divides $L$. This implies that the ground state subspace $\mathscr{V}$ is split into a direct sum $\bigoplus_p \mathscr{V}_p$ of  distinct sectors $\mathscr{V}_p$ that are translation invariant under $p$ lattice unit cells. 
This partial (discrete) SSB always accompanies continuous SSB from $G$ to $H$ with type-B GMs. As a consequence, we are led to a notion - emergent unit cells consisting of $p$ adjacent lattice unit cells.
In addition, the time-reversal symmetry group $\mathscr{Z}_2$ as a discrete symmetry group is also (partially)  broken spontaneously,  accompanying continuous SSB from $G$ to $H$, in the sense that some degenerate ground states are invariant (up to a minus sign) and others are not invariant under the time-reversal symmetry operation.

Here we emphasize that there are many other condensed matter systems undergoing SSB with type-B GMs, with the ground state degeneracies being exponential with system size~\cite{goldensu3,spinorbitalsu4,dimertrimer,finitesize,TypeBtasaki,jesse}.  In fact, they fall into two subclasses, depending on whether or not a continuous symmetry group $G$ (modulo a discrete symmetry subgroup) is semi-simple, as far as the origin of the exponenial ground state degeneracies with system size is concerned. Notably, the origin may be traced back to the presence of emergent subsystem invertible symmetries~\cite{dimertrimer,jesse}, as follows from an equivalent restatement of the Elitzur theorem. 
For quantum many-body spin models investigated in Refs.~\cite{FMGM,hqzhou,goldensu3,spinorbitalsu4,2dtypeb},  the symmetry groups are simple or semi-simple. If the symmetry group is semi-simple,  there is only one highest weight state, up to a unitary transformation induced from the symmetry group $G$.  Note that
only polynomially many degenerate ground states are generated from the action of the generator(s) of the symmetry group $G$ on the unique highest weight state. The number of generalized highest weight states is thus exponential.
As shown in Ref.~\cite{jesse}, it is the presence of exponentially many generalized highest weight states in these systems that accounts for the exponential ground state degeneracies, but they are different under PBCs and OBCs. For strongly correlated itinerant electron models, such as the  $\rm {SU}(2)$ flat-band Tasaki model and the Mielke model~\cite{mielke}, the symmetry groups (modulo a discrete symmetry subgroup) are non-semi-simple, given that they contain a $\rm {U}(1)$ subgroup generated by  the electron number operator. As such, in each sector labeled by the eigenvalues of the electron number operator and the Cartan generator(s) of the semi-simple subgroup of the non-semi-simple symmetry group $G$, there is at least one highest weight state, but generalized highest weight states are also possibly present. In fact, the total number of highest weight states is exponential, though the total number of sectors  is polynomial (depending on spatial dimensionality)~\cite{TypeBtasaki}.

Recent developments in condensed matter systems undergoing SSB with type-B GMs may boil down to a scaling behavior of the entanglement entropy, when a condensed matter system is bipartitioned into a block and its environment. In this regard,  the entanglement entropy for the ferromagnetic quantum spin-$1/2$ ${\rm SU}(2)$ Heisenberg model has been systematically investigated in Refs.~\cite{popkov,popkov1,doyon,doyon1,FMGM,hqzhou,2dtypeb}. 
It was found that  the entanglement entropy scales logarithmically with the block size  in the thermodynamic limit, with prefactor half the number of type-B GMs, as far as the orthonormal basis states for an irreducible representation space of the symmetry group $G$ as degenerate ground states are concerned. Consequently, we were led to the identification of the fractal dimension $d_f$ with the number of type-B GMs for the orthonormal basis states~\cite{FMGM,hqzhou,2dtypeb}. 
However, for a condensed matter system undergoing SSB with type-B GMs, if the ground state degeneracies are exponential with system size, an atypical generalized highest weight state defined on an emergent unit cell with period $p$ is invariant under the symmetric group $S_{L/p}$, if $p$ divides $L$. In particular, the highest weight state is invariant under the symmetric group $S_{L}$, if it is chosen properly to ensure that $p=1$. Note that the period $p$ is nothing to do with PBCs, since it is still well-defined under OBCs, given that all degenerate ground states under PBCs only constitute a subset of those under OBCs, as a result of the frustration-free nature of the model Hamiltonians~\cite{jesse}. In particular,  
when $p=L$, it should be treated as non-periodic, regardless of what types of boundary conditions are adopted. This is particularly so in the thermodynamic limit.
In contrast, a {\it typical} degenerate ground state is not permutation-invariant, thus pointing towards a connection to Green parafermions~\cite{green,greenberg}, given they are relevant to high-dimensional representations of the symmetric group. We shall return to this point in Section~\ref{conclusion}, after a specific realization of Green parafermions is discussed in condensed matter. As a result,  the logarithmic scaling behavior of the entanglement entropy with the block size  in the thermodynamic limit, with the prefactor being half the number of type-B GMs, may be established for atypical (periodic) degenerate ground states generated from atypical generalized highest weight states~\cite{goldensu3,TypeBtasaki}, though a proper modification is needed for atypical (non-periodic) degenerate ground states.

A hierarchical structure is hidden behind exponentially many degenerate ground states, which may be traced back to partial SSB of the translation symmetry under one lattice unit cell of the model Hamiltonian $\mathscr{H}$. This leads to a projection operator formalism, which we now turn to.

\section{A projection-operator formalism}~\label{projection}

To begin, we note that the ground state degeneracies are different under PBCs and OBCs for condensed matter systems undergoing SSB with type-B GMs, as long as they are exponential with system size~\cite{goldensu3,spinorbitalsu4,dimertrimer,finitesize,TypeBtasaki,jesse}. Here we mainly focus on condensed matter systems under PBCs, unless otherwise stated.

\subsection{Projection operators $\Pi_1$ and $\Pi_2$}

As a convention,  we  may set a multiplicative constant to be one and an additive constant to be zero, unless otherwise stated, given the  Hamiltonian $\mathscr{H}$ is a sum of local projection operators. From now on, we refer to a model Hamiltonian $\mathscr{H}$ as a canonical form if this convention is satisfied.
We further assume that there is a projection operator $\Pi_1$ that maps the Hamiltonian $\mathscr{H}$ into a projected Hamiltonian $\mathscr{\bar H}$, where $\mathscr{\bar H} = \Pi _1\mathscr {H} \Pi_1$. Here the original (unconstrained) Hilbert space, denoted as $V_0$, is mapped onto the constrained Hilbert space, denoted as $V_1$. Usually, this projected Hamiltonian $\mathscr{\bar H}$ itself is simply a sum of local projection operators, acting on the constrained Hilbert space $V_1$.  It is thus possible to go one step further to introduce another projection operator $\Pi_2$ such that the projected Hamiltonian  $\mathscr{\bar H}$ becomes the zero operator in a constrained Hilbert space, denoted as $V$. Mathematically, this amounts to requiring that $\Pi_2 \mathscr{\bar H}\Pi_2 = 0$. Here we remark that $\Pi_2$ commutes with $\Pi_1$: $\Pi_1 \Pi_2 = \Pi_2 \Pi_1 $.  

Mathematically, $\Pi_1$ is, by definition, the projection operator  that decomposes the original (unconstrained) Hilbert space $V_0$  into the direct sum of the constraint Hilbert space $V_1$ and its (orthogonal) complement $V_1^c$ in $V_0$. Similarly,  $\Pi_2$ is the projection operator that decomposes $V_1$ into the direct sum of $V$ and its (orthogonal) complement $V^c$ in $V_1$. 
Hence the two projection operators $\Pi_1$ and $\Pi_2$ are, by definition, {\it factorized} into a sequence of  projection operators that project out all the orthonormal basis states in $V_1$ and $V$, respectively. In other words,  explicit mathematical expressions for the projection operators $\Pi_1$ and $\Pi_2$ follow from  enumerating all the orthonormal basis states in $V_0$, $V_1$ and $V$.
The subtlety here lies in the fact that naturally occurring basis states in a condensed matter system are not always orthogonal to each other. This stems from the fact that, for a specific system, a set of basis states may be generated from the action of a sequence of operators on one chosen highest weight state, some of which in turn contain operators defined on a region $R$ consisting of a few adjacent lattice sites. As it turns out,  a region $R$ is not necessarily commensurate with the lattice unit cell, as far as their sizes are concerned.
As a result, a set of basis states are not orthogonal to each other, as happens for the ferromagnetic $\rm {SU}(2)$ flat-band Tasaki model.  For simplicity, we restrict the size of the lattice unit cell to be one for a quantum many-body spin system and to be two for a strongly correlated itinerant electron system. This assumption is sufficient for all condensed matter systems under investigation. A region $R$ is thus always commensurate with the lattice unit cell  for a quantum many-body spin system, as happens for the spin-1 ferromagnetic biquadratic model. 

Meanwhile, we stress that both $\Pi_1$ and $\Pi_2$  depend on what types of boundary conditions are adopted, since the orthonormal basis states to be projected out are generated by a sequence of operators that are usually defined on a region consisting of more than  one lattice site.  In addition,  $\Pi_2$ might be trivial, in the sense that $\Pi_2$ is the identity operator when $\mathscr{\bar H}$ is already the zero operator.

\subsection{The combined projection operator $\Pi = \Pi_1 \Pi_2$: a mathematical lemma}

Once the two projection operators $\Pi_1$ and $\Pi_2$ are introduced, we may define the combined projection operator $\Pi \equiv \Pi_2\; \Pi_1$. By definition, we have $\Pi \mathscr{H}\Pi = 0$. A remarkable observation is that $\Pi \mathscr{H}\Pi = 0$ is equivalent to the condition that  $\Pi \mathscr{H}=0$ or  $\mathscr{H} \Pi =0$, according to the mathematical lemma in Appendix~\ref{lemma} for a condensed matter system described by the model Hamiltonian $\mathscr{H}$ (in a canonical form). In other words,  the model Hamiltonian $\mathscr{H}$ simply becomes the zero operator after the projection operation $\Pi$ acts from either the left-hand side or the right-hand side.  As a result, the model Hamiltonian $\mathscr{H}$ commutes with this combined projection operator $\Pi$.  Any state in the form $\vert v \rangle = \Pi \vert \eta \rangle$ is thus a degenerate ground state of  the model Hamiltonian $\mathscr{H}$, with $ \vert \eta \rangle$ being any vector in the original unconstrained Hilbert space $V_0$, if and only if  $ \vert \eta \rangle$ is not orthogonal to the ground state subspace. 

Obviously, there are many different choices for $ \vert \eta \rangle$, if one is only content with finding a (degenerate) ground state. Generically, if we randomly choose  $\vert \eta \rangle$, then $\Pi \vert \eta \rangle$ is a linear combination of many degenerate ground states that fall into different (orthogonal) sectors labeled by the eigenvalues of the Cartan generator(s) of the symmetry group $G$, given  the dimension of the ground state subspace for such a condensed matter system is exponential in system size. However,  our aim is to enumerate all of exponentially many degenerate ground states arising from SSB with type-B GMs. In fact,  even if we take the symmetry group $G$ into account, there are still exponentially many  degenerate factorized ground states.

\subsection{Two illustrative examples}

We now turn to construct the projection operators $\Pi_1$ and $\Pi_2$ for the two illustrative models. Here we restrict ourselves to define the projection operators $\Pi_1$ and $\Pi_2$ in a rather informal way. Although such an  informal discussion  might be sufficient for understanding how the projection operators $\Pi_1$ and $\Pi_2$ emerge in a specific condensed matter system, a convincing mathematical proof for their existence is still needed, which is relegated to Appendix~\ref{pipi}.  This is particularly so for the $\rm {SU}(2)$ flat-band ferromagnetic Tasaki model, 
because the basis states generated from the action of ${\hat a}_{q,\sigma}^\dagger$  and ${\hat b}_{u,\sigma}^\dagger$ on the fermionic Fock vacuum $\vert \otimes_{x \in \Lambda} 0_x\rangle$ are only linearly independent, but not orthogonal.  Here the fermionic Fock vacuum $\vert \otimes_{x \in \Lambda} 0_x\rangle$ is defined by ${\hat c}_{x,\sigma}  \vert \otimes_{y \in \Lambda} 0_y\rangle = 0$ for any $x\in \Lambda$.

\subsubsection{The spin-1 ferromagnetic biquadratic model}

For  the spin-1 ferromagnetic biquadratic model, the Hamiltonian  (\ref{hambq}) may be reshaped into the form $\mathscr {H} = \sum_j \mathscr {H}_{j j+1}$ (up to a multiplicative constant and an additive constant), where 
\begin{align*}
	\mathscr {H}_{j j+1}= &\vert - + \rangle_{j j+1}  \langle -+ \vert  +
	\vert 00 \rangle_{j j+1} \langle 00 \vert+ \vert +- \rangle_{j j+1} \langle +- \vert \cr 
	&- [\vert 00 \rangle_{j j+1} \langle -+ \vert - \vert +- \rangle_{j j+1} \langle -+ \vert + \cr
	&\vert +- \rangle_{j j+1} \langle 00 \vert +{\rm h.c.}].
\end{align*}
Here we have dropped the subscript $j j+1$ from all the bra states for brevity and ${\rm h.c.}$ is the shorthand for the Hermitian conjugation.
If we choose $\Pi_1$ to be the projection operator that projects out all the  states containing $\vert - \rangle_j$, which is defined on a region $R$ consisting of one lattice site labeled by $j$, then we have 
\begin{equation*}
	\mathscr{\bar H}_{j j+1} = 
	\Pi_1 \vert 0 \rangle_j  \vert 0 \rangle_{j+1}  {_j}\langle 0~\vert {_{j+1}} \langle 0 \vert \Pi_1. 
\end{equation*} 
It is readily seen that the projected Hamiltonian $\mathscr{\bar H}$ is simply a sum of local projection operators, subject to the projection operator $\Pi_1$. Now we define $\Pi_2$ to be
the projection operator that projects out all the  states containing $\vert 0 \rangle_j \vert 0 \rangle_{j+1}$, which is defined on a region consisting of two adjacent lattice sites labeled by $j$ and $j+1$, then we have $\Pi_1 \Pi_2 = \Pi_2 \Pi_1$. 
As such, we are led to the combined projection operator $\Pi$ that commutes with the Hamiltonian (\ref{hambq}):
$\Pi \mathscr {H} = \mathscr {H} \Pi$. The constrained Hilbert space $V$ thus consists of all vectors $\vert v \rangle$ that do not contain any local states $\vert - \rangle_j$ and $\vert 0 \rangle_j \vert 0 \rangle_{j+1}$. In other words, any state $\vert v \rangle$, generated from the action of a sequence of operators that do not contain $S^-_j S^-_{j+1}$ on the highest weight state $ \otimes_j\vert +\rangle_j$,   is a degenerate ground state of the Hamiltonian  (\ref{hambq}).

\subsubsection{The $\rm {SU}(2)$ flat-band ferromagnetic Tasaki model}

For the $\rm {SU}(2)$ flat-band ferromagnetic Tasaki model (\ref{hamtasaki}), one chooses $\Pi_1$ to be the projection operator that projects out all the states generated from the action of a sequence of operators that consist of ${\hat a}_{q,\sigma}^\dagger$ and  ${\hat b}_{u,\sigma}^\dagger$, as long as it contains at least ${\hat b}_{u,\sigma}^\dagger$ for $u \in \mathscr{I}$. Recall that both ${\hat a}_{q,\sigma}^\dagger$ and  ${\hat b}_{u,\sigma}^\dagger$ are
defined on a region $R$ consisting of three lattice sites,  one lattice site labeled by $q \in \mathscr{E}$  and the two nearest neighbors labeled $u\in \mathscr{I}$ for ${\hat a}_{q,\sigma}^\dagger$ and one lattice site labeled by  $u\in \mathscr{I}$ and the two nearest neighbors labeled by $q \in \mathscr{E}$ for ${\hat b}_{u,\sigma}^\dagger$, with $|q-u|=1/2$.   Note that the basis states generated from the action of ${\hat a}_{q,\sigma}^\dagger$ on the fermionic Fock vacuum $\vert \otimes_{x \in \Lambda} 0_x\rangle$ are orthogonal to those generated from the action of  ${\hat b}_{u,\sigma}^\dagger$.

It is readily seen that the hopping term in the Hamiltonian (\ref{hamtasaki}) vanishes in the constrained
Hilbert space $V_1$. The projected Hamiltonian $\mathscr{\bar H}$ now takes the form
\begin{equation*}
	\mathscr{\bar H} = U\sum_{x\in \Lambda}\Pi_1{\hat n}_{x,\uparrow}{\hat n}_{x,\downarrow}\Pi_1. \label{hamtasaki1}
\end{equation*}
It is simply a sum of local projection operators, subject to the projection operator $\Pi_1$. To go further, we need to introduce another operator 
$\hat{a}_{q,\sigma}^\dagger$ and its Hermitian conjugation $\hat{a}_{q,\sigma}$. Instead of the standard basis states generated by acting with $\hat{c}_{x,\sigma}^\dagger$ on $\vert \otimes_{y \in \Lambda} 0_y\rangle$, it is convenient to work in another set of non-orthogonal but linearly independent basis states generated by acting $\hat{a}_{q,\sigma}^\dagger$ and $\hat{b}_{u,\sigma}^\dagger$ on 
$\vert \otimes_{x \in \Lambda} 0_x\rangle$.  In fact, the non-orthogonal set generated by  the action of ${\hat a}_{u,\sigma}^\dagger$ and  ${\hat b}_{u,\sigma}^\dagger$ on the fermionic Fock vacuum $\vert \otimes_{x \in \Lambda} 0_x\rangle$ are equivalent to  the orthogonal set generated by  the action of ${\hat c}_{x,\sigma}^\dagger$. Mathematically, ${\hat c}_{x,\sigma}^\dagger$ may be expressed in terms of  ${\hat a}_{u,\sigma}^\dagger$ and  ${\hat b}_{u,\sigma}^\dagger$ linearly, given the definitions of ${\hat a}_{u,\sigma}^\dagger$ and  ${\hat b}_{u,\sigma}^\dagger$ in terms of  ${\hat c}_{x,\sigma}^\dagger$.
Note that $[\hat{b}_{u,\sigma},\hat{a}_{q,\tau}^\dagger]_+=0$ for any $u \in \mathscr{I}$, $q \in \mathscr{E}$ and $\sigma, \tau = \uparrow, \downarrow$.  As a result, we have $[\hat{b}_{u,\sigma}^\dagger\hat{b}_{u,\sigma},\hat{a}_{q,\tau}^\dagger]_-=0$ and $[\hat{b}_{u,\sigma}^\dagger\hat{b}_{u,\sigma},\hat{a}_{q,\tau}]_-=0$.
Since the subspace generated from the action of ${\hat a}_{q,\sigma}^\dagger$ on the fermionic Fock vacuum $\vert \otimes_{x \in \Lambda} 0_x\rangle$ are orthogonal to that generated from the action of  ${\hat b}_{u,\sigma}^\dagger$, the introduction of the projection operator $\Pi_1$ amounts to stating that the constrained Hilbert space $V_1$ is spanned by all the  states generated from the action of $\hat{a}_{q,\sigma}^\dagger$ ($q=1,2, \ldots, L$) on the  fermionic Fock vacuum $\vert \otimes_{x \in \Lambda} 0_x\rangle$. Here we remark that such a basis state generated by the action of a sequence of operators containing $\hat{a}_{q,\tau}^\dagger$ and ${\hat b}_{u,\sigma}^\dagger$ on $\vert \otimes_{x \in \Lambda} 0_x\rangle$ are yet to be normalized.
Note that $[\Pi_1,\hat{a}_{q,\sigma}^\dagger]_-=0$ and $[\Pi_1,\hat{a}_{q,\sigma}]_-=0$. Physically, this follows from  the so-called zero-energy condition that
$\hat{b}_{u,\sigma}$ nullifies any degenerate ground state, which in turn stems from the frustration-free nature of the Hamiltonian (\ref{hamtasaki})~\cite{tasakibook}.

We now define the projection operator $\Pi_2$ that projects out  all the  states generated by a sequence of operators that consist of ${\hat a}_{q,\sigma}^\dagger$, as long as it contains 
${\hat a}_{q,\sigma}^\dagger{\hat a}_{q',{\bar \sigma}}^\dagger$ 
defined on a region $R_1$ consisting of three lattice sites, one lattice site labeled by $q \in \mathscr{E}$  and the two nearest neighbors labeled by $u\in \mathscr{I}$, with $|u-q|=1/2$ if $|q-q'|=0$ or  on a region $R_2$ consisting of five lattice sites, two lattice sites labeled by $q$ and $q'$, where $q,q' \in \mathscr{E}$  and the three nearest neighbors labeled by $u\in \mathscr{I}$, with $|u-q|=1/2$ and $|u-q'|=1/2$  if $|q-q'|=1$.
Here ${\bar \sigma}_\alpha$ denotes the image of $\sigma_\alpha$ under the time-reversal symmetry operation. We remark that these basis states are yet to be normalized. Physically, this projection operator  $\Pi_2$ makes it possible to avoid two physical effects, which play a crucial role in the explanation for the emergence of the saturated ferromagnetism at quarter filling~\cite{tasakibook}: any state repulsion arising from repulsion in the real space representing the Coulomb interaction on the lattice sites in the external sublattice $\mathscr{E}$ and any ferromagnetic exchange interaction representing the Coulomb interaction on the lattice sites in  the internal sublattice  $\mathscr{I}$.   Mathematically, this follows from the fact that the Hamiltonian (\ref{hamtasaki}) is frustration-free and positive semi-definite. 

Note that $\Pi_1 \Pi_2 = \Pi_2 \Pi_1$, as a result of the orthogonality between the two subspaces -- one is generated by the action of ${\hat a}_{q,\sigma}^\dagger$  on $\vert \otimes_{x \in \Lambda} 0_x\rangle$  and the other is generated by the action of ${\hat b}_{u,\sigma}^\dagger$. In addition, $\Pi$ commutes with $\mathscr {H}$, namely
$\Pi \mathscr {H} = \mathscr {H} \Pi$.  The constrained Hilbert space $V$ thus consists of all vectors $\vert v \rangle$ that do not accommodate any states generated by the action of a sequence of operators containing ${\hat a}_{q,\uparrow}^\dagger {\hat a}_{q,\downarrow}^\dagger$,  ${\hat a}_{q\pm 1,\downarrow}^\dagger {\hat a}_{q,\uparrow}^\dagger$ and ${\hat b}_{u,\sigma}^\dagger$ on $|\otimes_{x \in \Lambda} 0_x\rangle$. In other words, such a state $\vert v \rangle$ is a degenerate ground state of the Hamiltonian (\ref{hamtasaki}), with the ground state energy being zero.

\vspace{10mm}
As mentioned above, we present a constructive proof for the uniqueness and existence of the projection operators $\Pi_1$ and $\Pi_2$ in the two illustrative models (cf.~Appendix~\ref{pipi}). In principle, this proof may be adapted to any condensed matter system undergoing SSB with type-B GMs, as long as the ground state degeneracies are exponential under PBCs and OBCs. 
For the two illustrative models, the explicit mathematical expressions for the projection operators $\Pi_1$ and $\Pi_2$  are constructed in Appendix~\ref{pipi}. Note that we have adopted some abstract notations to label fully factorized (degenerate) ground states there, with a connection to the notations used here being explained in detail.  Here we mention that there is a subtle difference between fully factorized ground states in these two models. For the ferromagnetic spin-1  biquadratic model, fully factorized ground states are unentangled, whereas for the ferromagnetic $\rm {SU}(2)$ flat-band Tasaki model, fully factorized ground states, expressed in terms of the creation operators $\hat{a}_{q,\sigma}^\dagger$, are entangled, due to the fact that $\hat{a}_{q,\sigma}^\dagger$ are defined on a region $R$ consisting of three lattice sites, so there is always an overlap if two of  the creation operators $\hat{a}_{q,\sigma}^\dagger$ act on the nearest-neighbor lattice unit cells labeled by $q$'s from the external sublattice $\mathscr{E}$. As we shall see later on, the importance of fully factorized ground states in both cases lies in the fact that they are either highest and generalized highest weight states or  lowest and generalized lowest weight states, which are connected to each other via the time-reversal symmetry operation.
In addition to the above two models, we have also explicitly constructed the projection operators $\Pi_1$ and $\Pi_2$ for other condensed matter systems~\cite{shi-so,shi-dtmodel,shiqq}. In particular, the projection operator formalism works for any condensed matter systems undergoing SSB with type-B GMs, even if the ground state degeneracies are polynomial.

One might wonder why we do not deal directly with the projection operator $\Pi$ as an independent mathematical entity. Instead, it is split into the two projection operators $\Pi_1$ and $\Pi_2$. As we shall show below, this may be attributed to the fact that,
in order to bridge Green parafermion states and flat-band excitations in a condensed matter system,  we have to require that the projected Hamiltonian $\mathscr{\bar H}$ possesses a commutative  set of  conserved operators that may be re-interpreted as the Green parafermion number operators.  In fact, it is this physical requirement that makes a connection with Green parafermions possible. Indeed, it is usually impossible for a nontrivial Hamiltonian to possess so many conserved operators, if no further constraints are imposed on the Hilbert space.  As it turns out, one way around this issue is to demand that the projected Hamiltonian $\mathscr{\bar H}$ becomes the zero operator after another (nontrivial) projection operator $\Pi_2$ is implemented,  if $\mathscr{H}$ is in a canonical form, unless the projected Hamiltonian $\mathscr{\bar H}$ itself is already the zero operator. In this sense, the necessity for introducing the two projection operators $\Pi_1$ and $\Pi_2$ may be justified by a strong tie with a realization of Green parafermions in condensed matter.

\section{Realizations of Green parafermions: Hermitian and non-Hermitian}~\label{gscheme}

Having outlined the necessary formalism, we now turn to a realization of Green parafermions in condensed matter systems  undergoing SSB with type-B GMs. To this end,  it is necessary to develop a conceptual framework, with auxiliary Majorana fermions defined on an emergent unit cell as a key ingredient, either periodic or non-periodic. Actually, it is their presence that makes it possible to realize Green parafermions from local spin or fermion degrees of freedom in condensed matter systems. As a result, we are capable of  building a bridge between condensed matter systems undergoing SSB with type-B GMs and a set of Green parafermion fields, as long as the ground state degeneracies are exponential with system size $L$. 

Here we stress that the model Hamiltonian $\mathscr{H}$ under investigation {\it only} involves spin or fermion degrees of freedom.  According to the conventional wisdom, one should adopt an implicit assumption that a condensed matter system is fully specified by the model Hamiltonian $\mathscr{H}$. This is certainly true if the Hilbert space only involves spin or fermion degrees of freedom already contained in the Hamiltonian. However, there is no {\it a priori} reason to justify that this is a full picture. In other words,  it is possible that the model Hamiltonian alone is not necessarily sufficient to specify a condensed matter system. 

The implications for this possibility are far-reaching.
Physically, this points towards a way to introduce some extra degrees of freedom into the Hilbert space without any change in the Hamiltonian itself.  It turns out that it is possible to turn one condensed matter system into another by introducing extra degrees of freedom, with their difference only being manifested in the Hilbert spaces. We may take one step further afterwards if extra degrees of freedom thus introduced are removed eventually from the Hilbert spaces, in order to bring all these different systems back to the {\it same} original system. In other words, even
if only spin or fermion degrees of freedom already contained in the Hamiltonian are involved, it is still possible to introduce extra degrees of freedom into the Hilbert space, as long as they are eventually removed.

Remarkably, a condensed matter system undergoing SSB with type-B GMs provides a natural platform for this picture to emerge, if the ground state degeneracies are exponential in system size. In this picture, extra degrees of freedom introduced into the Hilbert space are auxiliary Majorana fermions.
Physically, the flexibility for introducing auxiliary Majorana fermions as extra degrees of freedom into the Hilbert space heavily relies on the presence of emergent unit cells, either periodic or non-periodic.
Depending on whether or not auxiliary Majorana fermions are removed from the Hilbert space eventually, there are two types of auxiliary Majorana fermions --  physical or fictitious.

\subsection{ Auxiliary Majorana fermions: physical or fictitious}

To set a stage for the notation,  we  choose an integer $p$ that divides $L$.  Here now $p$ represents the size of an emergent unit cell for such an atypical partition.  A (one-dimensional) lattice is thus partitioned into $N$ emergent unit cells labeled by $l$, where $l = 0,1,2,\ldots,N-1$, with $N=L/p$. As a result, a lattice unit cell labeled by $j$ or $q$ for quantum many-body spin systems or  strongly correlated itinerant electron models
is now labeled as $(l,\alpha)$, if $j= pl+\alpha$ or $q= pl+\alpha$, with $\alpha =0,1, \ldots,p-1$. In other words, one subscript $j$ or $q$ labeling lattice unit cells for quantum many-body spin systems or strongly correlated itinerant electron systems is now replaced by two subscripts $(l,\alpha)$, where $\alpha$ labels $p$ lattice unit cells inside an emergent unit cell labeled by $l$. Note that we did not include a subscript labeling a lattice site inside a specific lattice unit cell for simplicity.  Physically, this amounts to dividing spin or fermion degrees of freedom labeled by $j$ or $q$ into external degrees of freedom labeled by $l$ and internal degrees of freedom labeled by $\alpha$, so the dichotomy between external and internal degrees of freedom is defined with respect to an emergent unit cell. This marks a significant deviation from the traditional definition of internal degrees of freedom that are inherent in nature.

The introduction of auxiliary Majorana fermions makes it possible to attach an alternative meaning to an emergent unit cell.
We remark that an emergent unit cell, as a result of partial SSB of the translation symmetry under one lattice unit cell, stems from cooperative phenomena for both atypical and typical partitions. In fact, it is the presence of this partial SSB that makes it possible to introduce auxiliary Majorana fermions as some extra degrees of freedom, which are arranged periodically for atypical partitions and non-periodically for typical partitions. In particular,  an emergent unit cell consists of $p$ adjacent lattice unit cells for an atypical partition. Here we mainly focus on a specific periodical arrangement of auxiliary Majorana fermions,  denoted as $\gamma_\alpha$ ($\alpha=0,1,\ldots,p-1$), by attaching $\gamma_\alpha$ to a lattice unit cell labeled by $\alpha$ for any $l$ for both quantum many-body spin systems and strongly correlated itinerant electron systems. However, an extension to a non-periodical arrangement of auxiliary Majorana fermions for a typical partition will also be covered after a detailed discussion of a periodical arrangement of auxiliary Majorana fermions for an atypical partition (cf. Subsection~\ref{non-periodic}).
In addition, 
we also introduce $n$ sets of auxiliary Majorana fermions labeled by $\mu$ ($\mu=1,\ldots,n$) at any lattice unit cell labeled by $\alpha$ inside an emergent unit cell labeled by $l$,  denoted as $\gamma_{\mu,l,\alpha}$, for quantum many-body spin systems, in order to ensure that the combined objects formed from spin degrees of freedom and auxiliary Majorana fermions $\gamma_{\mu,l,\alpha}$ at different lattice unit cells (labeled by $l,\alpha$) anti-commute with each other. To this end, one has to make sure that spin degrees of freedom must manifest themselves in a set of the Pauli matrices. Here we have assumed that the local Hilbert space accommodates the tensor product space of $n$ two-dimensional subspaces.

Now we are ready to realize Green parafermions for condensed matter systems undergoing SSB with type-B GMs. As we shall see later on,  the introduction of a set of auxiliary Majorana fermions labeled by $\alpha$ ($\alpha =1,2,\ldots, p-1$) is necessary to realize one species of  Green parafermions, which includes $n$ Green parafermion fields labeled by $\mu$. In fact,  Green parafermions as composite particles are formed from auxiliary Majorana fermions, together with spin or fermion degrees of freedom. 
For our purpose, we have to distinguish two cases: one is that auxiliary Majorana fermions are present in the Hilbert space, so they are {\it physical}; the other is that auxiliary Majorana fermions are {\it fictitious} so that they are not included in the Hilbert space. An essential difference between the two cases lies in the fact that for auxiliary (physical) Majorana fermions, the Hilbert space  accommodates both the subspace for spin or fermion degrees of freedom and the subspace for auxiliary Majorana fermions, depending on the specifics of a condensed matter system, 
in contrast to auxiliary (fictitious) Majorana fermions. Indeed, the presence of  auxiliary (fictitious) Majorana fermions does not change the Hilbert space, so they are introduced as a mathematical device and have to be removed eventually, in order to ensure that the Hilbert space only involves spin or fermion degrees of freedom, depending on what model is under investigation.

\subsection{Green parafermions  on periodic emergent unit cells\\ 
	for an atypical partition}~\label{greenpara-tcr}

We assume that there are $n$ Green parafermion fields $\phi_\mu$ labeled by $\mu$ ($\mu =1,2,\ldots,n$), which are defined on periodic emergent unit cells for an atypical partition.   The creation and annihilation operators  of Green parafermions are denoted by $a^*_{\mu, p, l}$ and $a_{\mu, p, l}$, which are defined on an emergent unit cell labeled by $l$.  Here we have introduced three subscripts $\mu$, $p$ and $l$ in total, in order to ensure that the notations are adapted to a realization of Green parafermions in condensed matter. For simplicity, we refrain from a detailed discussion of the defining trilinear and relative trilinear commutation relations for Green parafermions. Instead, we only write down the 
trilinear commutation relations for the creation and annihilation operators  $a^*_{\mu, p, l}$ and $a_{\mu, p, l}$  for the same field and the relative trilinear commutation relations, with further details relegated to Appendix~\ref{tcr}.
Following Green~\cite{green}, they satisfy the trilinear commutation relations
\begin{align}
	[[a^*_{\mu, p,l}, a_{\mu, p,l^{'}}]_-, a_{\mu, p,l^{''}}]_- &= -2 \delta_{l l^{''}} a_{\mu, p,l^{'}}, \cr
	[[a_{\mu,p, l}, a_{\mu, p,l^{'}}]_-, a_{\mu, p,l^{''}}]_- &= 0, \label{tcr2}
\end{align}
in addition to other relations that follow by swapping  $a^*_{\mu,p, l}$ and $a_{\mu, p,l}$ in both sides of the above equations. Here we have adopted the notation $*$ for the creation operators. Note that $*$ is not necessarily identical to the adjoint operation $\dagger$. This is due to the fact that {\it not only} Hermitian {\it but also} non-Hermitian realizations of the trilinear and relative trilinear commutation relations are relevant to condensed matter systems undergoing SSB with type-B GMs.  As follows from Jacobi's identity and the trilinear commutation relations (\ref{tcr2}), we have
\begin{equation}
	[[a_{\mu, p,l}, a_{\mu, p,l'}]_{-}, a^*_{\mu,p, l''}]_- = 2 \delta_{l' l^{''}} a_{\mu,p, l} - 2 \delta_{l l^{''}} a_{\mu,p, l'}. \label{jacobi}
\end{equation}
The relative trilinear commutation relations involving $\phi_\mu$ twice and $\phi_{\mu'}$  take the form
\begin{align}
	[[a^*_{\mu,p,l}, a_{\mu,p, l^{'}}]_-, a_{\mu',p, l^{''}}]_- &=0, \cr
	[[a_{\mu,p, l}, a_{\mu,p, l^{'}}]_-, a_{\mu',p,l^{''}}]_- &=0, \cr
	[[a^*_{\mu,p, l}, a^*_{\mu,p, l^{'}}]_-, a_{\mu',p, l^{''}}]_- &= 0.	\label{tcr3}
\end{align}
and
\begin{align}
	[[a_{\mu',p, l''}, a^*_{\mu,p, l}]_-, a_{\mu,p, l^{'}}]_- &= 2\delta_{l l^{'}} a_{\mu',p,l^{''}}, \cr
	[[a_{\mu, p,l'}, a_{\mu',p, l^{''}}]_-, a^*_{\mu,p, l}]_- &= - 2\delta_{l l^{'}} a_{\mu', p,l^{''}}.
	\label{tcr4}
\end{align}
We remark that  the number of Green parafermion fields $n$ involved depends on the symmetry group $G$ and the dimension of the local Hilbert space for a specific condensed matter system.

The  creation and annihilation operators $a^*_{\mu, p, l}$ and $a_{\mu, p, l}$ in turn may be constructed in terms of the Green components  $b^*_{\mu,p, l,\alpha}$ and $b_{\mu,p,l,\alpha}$, according to the Green Ansatz for one single parafermion field~\cite{green} and its variant for two and more parafermion fields~\cite{greenberg}. Notably, the Green components are neither bosons nor fermions. More precisely, we have 
\begin{align}
	a^*_{\mu,p, l} &= \sum_{\alpha =0}^{p-1} b^*_{\mu,p,l,\alpha}, \cr
	a_{\mu,p, l} &= \sum_{\alpha =0}^{p-1} b_{\mu,p,l,\alpha}. \label{greenansatz}
\end{align}
Here  $b^*_{\mu,p,l,\alpha}$ and $b_{\mu,p,l,\alpha}$ satisfy the  anti-commutation or commutation relations (cf.~Eqs.~(\ref{greencomp1}),~(\ref{greencomp2}),~(\ref{greencomp3}) and (\ref{greencomp4}) in Appendix~\ref{tcr}).
 As a result, it is necessary to introduce auxiliary Majorana fermions, in order to realize the creation and annihilation operators  $a^*_{\mu, p, l}$ and $a_{\mu, p, l}$ for Green parafermions in terms of ordinary spin or fermion operators. In other words, it would be impossible to realize the creation and annihilation operators for Green parafermions in terms of ordinary spin or fermion operators alone, without auxiliary Majorana fermions.
 Mathematically, a realization of the Green components  $b^*_{\mu,p, l,\alpha}$ and $b_{\mu,p,l,\alpha}$ in terms of ordinary spin  degrees of freedom, together with auxiliary Majorana fermions  $\gamma _{\mu,l,\alpha}$ and $\gamma_{\alpha}$ or  fermion degrees of freedom, together with auxiliary Majorana fermions $\gamma_{\alpha}$, makes it possible to introduce an internal symmetry group ${\rm U}(p)$. We remark that this internal symmetry group ${\rm U}(p)$ is induced from a unitary transformation of combined objects of spin degrees of freedom and auxiliary Majorana fermions $\gamma _{l,\alpha}$ for quantum many-body spin systems or fermion degrees of freedom for strongly correlated electron systems.  Indeed, the defining relations for the Green components  $b^*_{\mu,p, l,\alpha}$ and $b_{\mu,p,l,\alpha}$ are invariant under 
a unitary transformation, with its entries involving auxiliary Majarana fermions $ \gamma_{\alpha}$. More precisely, for any $p \times p$ unitary matrix $U$,  the  anti-commutation or commutation relations (cf.~Eqs.~(\ref{greencomp1}),~(\ref{greencomp2}),~(\ref{greencomp3}) and (\ref{greencomp4}) for the Green components $b^*_{\mu,p,l,\alpha}$ and $b_{\mu,p,l,\alpha}$ in Appendix~\ref{tcr}) are invariant under a unitary transformation:
\begin{equation}
{\tilde b}_{\mu,p,l,\alpha} = \sum _{\alpha'=0} ^{p-1} U_{\alpha \alpha'} \gamma_{\alpha} \gamma_{\alpha'} b_{\mu,p,l,\alpha'},\label{bogoliubov}
\end{equation}
meaning that ${\tilde b}^*_{\mu,p,l,\alpha}$ and ${\tilde b}_{\mu,p,l,\alpha}$ satisfy the same  anti-commutation or commutation relations as $b^*_{\mu,p,l,\alpha}$ and $b_{\mu,p,l,\alpha}$.
The involvement of auxiliary Majorana fermions $\gamma_\alpha$ in this internal symmetry group ${\rm U}(p)$ constitutes an extension of the Bogoliubov transformations for ordinary fermions~\cite{pcoleman}, thus explaining the difference between the Green components  $b^*_{\mu,p, l,\alpha}$ and $b_{\mu,p,l,\alpha}$ and ordinary fermions. 

However, this extension of  the Bogoliubov transformations for the Green components  $b^*_{\mu,p,l,\alpha}$ and $b_{\mu,p,l,\alpha}$ is induced from the Bogoliubov transformations for ordinary fermions~\cite{pcoleman} with the help of auxiliary Majorana fermions $\gamma_\alpha$, as far as a realization of Green parafermions in a specific model is concerned. This will become clear when a realization of Green parafermions in the ferromagnetic $\rm {SU}(2)$ flat-band Tasaki model follows below.
In fact, the presence of auxiliary Majorana fermions $\gamma_\alpha$ in an extension of the Bogoliubov transformations for the Green components justifies the necessity for introducing them in a realization of Green parafermions.  Indeed,  this also justifies why auxiliary Majorana fermions $\gamma_{\mu,l,\alpha}$ have to be introduced for quantum many-body spin models, since their presence turns spin degrees of freedom located at different lattice unit cells (labeled by $l,\alpha$) into combined objects, thus ensuring that these combined objects behave as if they are ordinary fermions. 
Generically, spin degrees of freedom  are represented as $n$ sets of  $\sigma^{\pm}_{\mu,p,l,\alpha}$ and  $\sigma^z_{\mu,p,l,\alpha}$, each of which is related with a set of $\sigma^{\pm}$ and  $\sigma^z$ constructed from the Pauli matrices (up to a multiplicative constant and an additive constant). The combined objects may thus be represented as  $\gamma_{\mu,l,\alpha} \sigma^{\pm}_{\mu,p,l,\alpha}$. 
Note that a concrete realization of  Green parafermions depends on the size of the lattice unit cell, in addition to the dimension of the local Hilbert space for a specific model. 
In other words, a realization of Green parafermions is model-dependent, though our construction reveals a universal mathematical structure behind condensed matter systems undergoing SSB with type-B GMs, if the ground state degeneracies are exponential.

For a specific realization, the creation and annihilation operators $a^*_{\mu, p, l}$ and $a_{\mu, p, l}$ may be defined in the constrained Hilbert space $V_1$ specified by the projection operator $\Pi_1$ or in the original unconstrained Hilbert space $V_0$, depending on the symmetry group $G$, the dimensionality of the local Hilbert space and the number of Green parafermion fields involved. In other words, Green parafermions live in the constrained Hilbert space $V_1$ or in the original unconstrained Hilbert space $V_0$, depending on what system is under investigation.  
As we have shown in Appendix~\ref{tcr},  for an atypical partition, Green parafermions share the same order, identical to the emergent unit cell size  $p$. Indeed, this is necessary to ensure that they belong to the same subset, according to Greenberg and Messiah~\cite{greenberg}.

In addition, a commutative set of the Green parafermion number operators $n_{\mu, p, l}$ may be defined that are expressed in terms of creation and annihilation operators $a^*_{\mu, p, l}$ and $a_{\mu, p, l}$.
Generically, the Green parafermion number operators $n_{\mu, p, l}$ are not necessarily Hermitian, since a non-Hermitian realization of Green parafermions is  allowed. For an Hermitian realization, we have $n^\dagger_{\mu, p, l} = n_{\mu, p, l}$. For a non-Hermitian realization, we have   $n^\dagger_{\mu, p, l} \neq n_{\mu, p, l}$. Assume that the parafermion Fock vacuum, denoted as $\vert \Omega_0 \rangle$, is unique~\cite{greenberg}.
As a result, we may construct Green parafermion states  $a^\dagger_{\mu_1,p, l_1}a^\dagger_{\mu_2,p, l_2} \ldots a^\dagger_{\mu_m,p,l_m} \vert \Omega_0 \rangle$ ($m=1,\ldots,M$) as the eigenstates of $\sum_{\mu,l} n_{\mu,p, l}$ for  an Hermitian realization and Green parafermion states  $a^\dagger_{\mu_1, p, l_1}a^\dagger_{\mu_2, p, l_2} \ldots a^\dagger_{\mu_m, p,l_m} \vert \Omega_0 \rangle$ ($m=1,\ldots,M$) as the right eigenstates of $\sum_{\mu,l} n^\dagger_{\mu,p, l}$  and 
Green parafermion states  $a^*_{\mu_1, p, l_1}a^*_{\mu_2, p, l_2} \ldots a^*_{\mu_m, p,l_m} \vert \Omega_0 \rangle$ ($m=1,\ldots,M$) as the right eigenstates of $\sum_{\mu,l} n_{\mu,p, l}$ for a non-Hermitian realization. 
Here $M$ denotes the maximum occupation number of  one species of Green parafermions of order $p$, which includes $n$  Green parafermion fields.
We remark that $*$ may be replaced by $\dagger$ to construct the eigenstates of $\sum_{\mu,l} n_{\mu,p, l}$ for a Hermitian realization, given that they are identical. In contrast, it is necessary to tackle two different sets of the right eigenstates,  one for $\sum_{\mu,l} n_{\mu,p, l}$ and the other for $\sum_{\mu,l} n^\dagger_{\mu,p, l}$, for a non-Hermitian realization. 

Like ordinary fermions, Green parafermions obey the Pauli exclusion principle for Green parafermions~\cite{greenberg}, stating that there are at most $p$ parafermions of order $p$ in the same state. This follows from the fact that $(a^\dagger_{\mu,p, l})^p \vert \Omega_0 \rangle\neq 0$, but $(a^\dagger_{\mu,p, l})^{p+1} \vert \Omega_0 \rangle=0$~\cite{greenberg}. This implies that  Green parafermions of order $p$ are drastically different from ordinary fermions,  if the order $p$ is not less than two. Mathematically, one may trace this unusual parastatistics to the fact that Green parafermions follow from high-dimensional representation of the symmetric group. Physically, Green parafermions are  realized on an emergent unit cell of size $p$ in condensed matter systems under investigation. As a result, we have $M=npN=nL$.

\subsection{Green parafermions on non-periodic emergent unit cells\\ for a typical partition}~\label{non-periodic}
Up to the present, our focus is on atypical partitions, so $p$ always divides $L$. There are thus $N$ ($N=L/p$) emergent unit cells such that they are arranged periodically on a (one-dimensional) lattice, with no emergent unit cells intersecting with each other.  In this setting, auxiliary Majorana fermions are arranged periodically on emergent unit cells, so Green parafermions thus realized  are identical, in the sense that 
the creation and annihilation operators  $a^*_{\mu, p, l}$ and $a_{\mu, p, l}$ of two Green parafermions located at two emergent unit cells are identical up to a translation symmetry operation. Note that there is no intersection between any two emergent unit cells, so internal degrees of freedom are identified as spin  degrees of freedom plus auxiliary Majorana fermions $\gamma_{\mu,l,\alpha}$ or fermion degrees of freedom inside an emergent unit cell. For a specific realization,  internal degrees of freedom are subject to an internal symmetry group ${\rm U}(p$), which may be interpreted as a subsystem gauge group such that local (internal) degrees of freedom inside an emergent unit cell is inaccessible to any local measurements, if one is restricted to physical observable invariant under subsystem gauge transformations. Note that Green parafermions defined on two emergent unit cells of different sizes are not identical, since their orders are different.

However, this does not even exhaust all possible atypical partitions.  Indeed, for a specific atypical partition, with emergent unit cells consisting of $p$ lattice unit cells arranged periodically on a (one-dimensional) lattice, there are  other $p-1$ different but equivalent atypical partitions, which are labeled by $p'$, with $p'=1,2, \ldots, p-1$. They are equivalent in the sense that they become identical to the original one, after a translation operation under $p'$ lattice unit cells is performed.
If one chooses two of these partitions and put them together to form a new partition, then there is an intersection of two emergent unit cells for this new partition, with one original emergent unit cell being split into two emergent unit sub-cells. This construction may be extended to more than two partitions. Note that emergent unit sub-cells are still periodic, with period $p$. For a typical partition,  a (one-dimensional) lattice is divided into a union of non-periodic emergent unit cells that are not intersecting with each other.

To proceed, we explain what consequence may be drawn from an intersection between two emergent unit cells coming from the two different but equivalent atypical partitions. Suppose one partition becomes another under a translation operation under $p'$ lattice unit cells, with $p'=1, \ldots, p-1$, then an emergent unit cell consisting of $p$  adjacent lattice unit cells is split into two new emergent unit sub-cells, with their sizes being  $p'$ and $p-p'$.
As a result, one species of Green parafermion of  order $p$ undergoes fission into two  species of Green parafermions of  orders $p'$ and $p-p'$, with their internal symmetry groups being ${\rm U}(p')$ and ${\rm U}(p-p')$, respectively. Physically, this may be understood from inaccessibility to internal degrees of freedom, since the presence of an intersection between two emergent unit cells that are identical after a translation operation under $p'$ lattice unit cells implies that 
only spin or fermion degrees of freedom inside two emergent unit sub-cells consisting of $p'$ and $p-p'$ are internal degrees of freedom.
Similarly, if one puts three atypical partitions among $p$ different but equivalent ones together to form a new partitions. Suppose the first partition becomes the second one under a translation operation of $p'$ lattice unit cells and  becomes the third one under a translation operation of $p''$ lattice unit cells, subject to $p'' > p'$, then one species of Green parafermion of  order $p$ undergoes fission into three species of Green parafermions of  orders $p'$, $p''-p'$  and $p-p''$, with their internal symmetry groups being ${\rm U}(p')$, ${\rm U}(p'-p'')$ and ${\rm U}(p-p'')$, respectively. This may be extended further, until   one  species of Green parafermion of  order $p$  undergoes fission into $p$ species of Green parafermions of order 1.  Again, this may be understood from inaccessibility to internal degrees of freedom.
As a result, the restriction that $p$ divides $L$ is not necessary, since one may add a certain number of Green parafermions of order 1, with the total number being less than $p$ to ensure that $L$ is allowed to be any positive integer. As such, we are led to a typical partition of a (one-dimensional) chain into a union of nonperiodic emergent unit cells that are not intersecting with each other. The price we have to pay is to deal with different species of Green parafermion fields with different orders. 

Generically,  $L$ is partitioned into a sum of a sequence of positive integers $p_\nu$ ($\nu=1,2,\ldots,Q$), with multiplicities being $N_\nu$, namely $L=\sum_\nu N_\nu p_\nu$, valid for an atypical or a typical partition. Without loss of generalities, one may demand that $p_\nu$ are ordered such that $p_1 < p_2 < \dots < p_Q$. In this way a one-dimensional lattice is  partitioned into  $Q$ types of emergent unit cells of different sizes $p_\nu$, labeled by $(l_\nu, \alpha_{l_\nu})$, with $\alpha_{l_\nu}= 0,1,\ldots,p_l-1$. There are thus $N_\nu$ emergent unit cells for a given $\nu$, with their sizes being $p_\nu$.
As such, one subscript $j$ or $q$ labeling lattice unit cells for quantum many-body spin systems or strongly correlated itinerant electron systems is now replaced by two subscripts $(l_\nu,\alpha_{l_\nu})$, with $\alpha_{l_\nu}$ labeling lattice unit cells inside an emergent unit cell labeled by $l_\nu$.

We thus need to define Green parafermions on any emergent unit cells,  regardless of partitions being atypical or typical.
Physically,  this amounts to introducing different species of Green parafermions. Following Greenberg and Messiah~\cite{greenberg}, each pair of parafields may be divided into four subsets: B for the relative bilinear commutation relations between two different fields; F for the relative bilinear anticommutation relations between two different fields; PB for the relative trilinear anticommutation relations between two different fields;  PF for the relative trilinear commutation relations between two different fields.  Indeed, any two fields belonging to the same subset have equal order and any two parafields belonging to different subsets obey ordinary relative (bilinear) commutation relations. Or equivalently, any two parafields of different orders must belong to different subsets and  any two parafields obeying the relative trilinear commutation relations must belong to the same subset~\cite{greenberg} (cf.~Appendix~\ref{tcr} for a brief summary).  As a consequence, any two Green parafermion fields from two different species may be classified into the $B$ subset, because their creation and annihilation operators obey ordinary relative (bilinear) commutation relations.

As a result, we have to deal with different species of Green parafermion fields that coexist on a (one-dimensional) lattice, if auxiliary Majorana fermions  are introduced on  periodic emergent unit cells (split into two or more emergent unit sub-cells) for an atypical partition and on non-periodic emergent unit cells for a typical partition. In particular, we note that a periodic arrangement of auxiliary Majorana fermions on emergent unit cells (split into two or more emergent unit sub-cells) for an atypical partition remains the same, with the only difference being the reduction of an internal symmetry group from ${\rm U}(p)$ to ${\rm U}(p') \times {\rm U}(p-p')$ for two sub-cells, to ${\rm U}(p') \times {\rm U}(p'-p'') \times {\rm U}(p-p'')$ for three sub-cells and so on.

For a generic partition,  we may define Green parafermions on periodic or non-periodic emergent unit cells in terms of the Green components, according to the Green Ansatz introduced for one single parafermion field~\cite{green} and its variant for two and more parafermion fields~\cite{greenberg}. Note that the Green components are neither bosons nor fermions.  It is thus necessary to introduce auxiliary Majorana fermions. For our purpose, auxiliary Majorana fermions, denoted as $\gamma_{\alpha_\nu}$ ($\alpha_\nu=0,1,\ldots,p_\nu-1$), are  attached to a lattice unit cell labeled by $\alpha_\nu$ inside an emergent unit cell labeled by $l_\nu$. Here we have assumed that a one-to-one correspondence between $j$ or $q$ and  $l_\nu$ and $\alpha_\nu$ is established for a specific set $\{ p_\nu\}$.
As a result,  external degrees of freedom are labeled by $l_\nu$  and internal degrees of freedom labeled by $\alpha_\nu$, so the dichotomy between external and internal degrees of freedom is defined with respect to  emergent unit cells of different sizes $p_\nu-1$.
In this setting, the  creation and annihilation operators $a^*_{\mu,\nu, p_\nu, l_\nu}$ and $a_{\mu,\nu, p_\nu, l_\nu}$  are constructed in terms of the Green components  $b^*_{\mu,\nu, p_\nu, l_\nu,\alpha_\nu}$ and $b_{\mu,\nu, p_\nu, l_\nu,\alpha_\nu}$. By definition, we have  $a^*_{\mu,\nu, p_\nu, l_\nu} = \sum_{\alpha_\nu} b^*_{\mu,\nu, p_\nu, l_\nu,\alpha_\nu}$ and $a_{\mu,\nu, p_\nu, l_\nu}=\sum_{\alpha_\nu} b_{\mu,\nu, p_\nu, l_\nu,\alpha_\nu}$.
The operators $a^*_{\mu,\nu, p_\nu, l_\nu}$ and $a_{\mu,\nu, p_\nu, l_\nu}$ defined on different emergent unit cells of different sizes $p_\nu$  thus represent the creation and annihilation operators for different  Green parafermions, with their orders being $p_\nu$. Accordingly, one may define Green parafermion number operators $n_{\mu,\nu, p_\nu, l_\nu}$, which are Hermitian for an Hermitian realization and  non-Hermitian for a non-Hermitian realization.
Here we have introduced an extra subscript $\nu$ to denote different species of Green parfermion fields $\phi_{\mu,\nu}$, in addition to $\mu$ ($\mu = 1,2,\ldots,n$) that labels different Green parafermions from the same species. In particular, we have assumed that only one species of  Green parafermions is introduced for emergent unit cells of the same size, although this is not necessary.  The total number of parafields needed is thus $nQ$. We remark that the relative commutation relations between two  Green parfermion fields $\phi_{\mu,\nu}$
and  $\phi_{\mu',\nu'}$ are trilinear if $\nu=\nu'$ and bilinear if  $\nu \neq \nu'$ (cf.~Appendix~\ref{tcr}).

In exactly the same way as $a^*_{\mu, p, l}$ and $a_{\mu, p, l}$ defined on an emergent unit cell consisting of $p$ adjacent lattice unit cells for an atypical partition, $a^*_{\mu,\nu, p_\nu, l_\nu}$ and $a_{\mu,\nu, p_\nu, l_\nu}$ may be defined in the constrained Hilbert space $V_1$ specified by the projection operator $\Pi_1$ or the original unconstrained Hilbert space $V_0$, depending on the symmetry group $G$, the dimensionality of the local Hilbert space and the number of Green parafermion fields involved. Accordingly, Green parafermions live in the constrained Hilbert space $V_1$ or the original unconstrained Hilbert space $V_0$, depending on what system is under investigation.

It follows that a hierarchical structure at a Hamiltonian level may be extended to any generic partition, irrespective of being atypical or typical. As a result, 
one may construct the eigenstates  $a^*_{\mu, p_{\mu,\nu_1}, l_{\nu_1}}a^*_{\mu,p_{\nu_2}, l_{\nu_2}} \ldots a^*_{\mu,p_{\nu_m},l_{\nu_m}}\vert \Omega_0 \rangle$ ($m=1,\ldots,M$) of the total Green parafermion number operator $\sum_{\mu,\nu} N_{\mu,\nu, p_\nu}$ for an Hermitian realization and the right eigenstates  $a^*_{\mu, p_{\nu_1}, l_{\nu_1}}a^*_{\mu,p_{\nu_2}, l_{\nu_2}} \ldots a^*_{p_{\mu,\nu_m},l_{\nu_m}}\vert \Omega_0 \rangle$ ($m=1,\ldots,M$) of the total Green parafermion number operator $N_{\mu,\nu, p_\nu}$ and the right eigenstates  $a^\dagger_{\mu, p_{\nu_1}, l_{\nu_1}}a^\dagger_{\mu,p_{\nu_2}, l_{\nu_2}} \ldots a^\dagger_{p_{\mu,\nu_m},l_{\nu_m}}\vert \Omega_0 \rangle$ ($m=1,\ldots,M$) of the Hermitian conjugated form $\sum_{\mu,\nu} N^\dagger_{\mu,\nu, p_\nu}$ for a non-Hermitian realization,
where $N_{\mu,\nu, p_\nu} = \sum_{l_\nu} n_{\mu,\nu, p_\nu, l_\nu}$. They play the same role as their counterparts for an atypical partition, when an emergent unit cell consists of $p$ adjacent lattice unit cells, as discussed in Subsection~\ref{greenpara-tcr}.
Note that $M$ denotes the maximum occupation number of  all species of Green parafermions of order $p_{\nu}$, each of which includes $n$  Green parafermion fields. Hence $M=n \sum_\nu p_\nu N_\nu$, as a result of the Pauli exclusion principle for Green parafermions (cf.~Appendix~\ref{tcr}).

Here we remark that, although a periodic arrangement of emergent unit cells, each of which consists of $p$ adjacent lattice unit cells, on a (one-dimensional) lattice does not exhaust all possible atypical partitions, we always restrict to this arrangement when we deal with an atypical partition. As a result, one may restrict to one species of Green parafermions, with the order being identical to $p$. For a typical partition, we choose to introduce  one species of Green parafermions for each $p_\nu$, although more than one species are allowed. That amounts to stating that  we need to introduce one set of auxiliary Majorana fermions on emergent unit cells of the same size $p_\nu$. Here one species  always accommodates $n$  
Green parafermion fields. 

As illustrative examples, we shall construct two concrete realizations of Green parafermions in 
the  ferromagnetic spin-1 biquadratic model and the ferromagnetic $\rm {SU}(2)$ flat-band Tasaki model by introducing auxiliary Majorana fermions in emergent unit cells. Further examples, including the ferromagnetic spin-orbital model, the quantum spin-1 system with competing dimer and trimer interactions and the ferromagnetic $\rm {SU}(n)$ flat-band Tasaki model, will be discussed in the forthcoming articles~\cite{shi-so,shi-dtmodel,shiqq}. 

\subsection{ Two specific realizations}~\label{biquadratic-tasaki}

\subsubsection{The ferromagnetic spin-1 biquadratic model}

For the ferromagnetic spin-1 biquadratic model ( \ref{hambq}), only one single Green parafermion field is needed for an atypical partition. One may drop off the subscript $\mu$ for brevity, so the creation and annihilation operators  are denoted as $a^*_{p, l}$ and $a_{p, l}$.
In order to establish the connection with one single Green parafermion field, we resort to the Green ansatz:  
\begin{align*}
a^*_{p, l} &= \sum_{\alpha =0}^{p-1} b^*_{p,l,\alpha}, \cr 
a_{p, l} &= \sum_{\alpha =0}^{p-1} b_{p,l,\alpha},
\end{align*} 
where  $b^*_{p,l,\alpha}$ and $b_{p,l,\alpha}$  are the Green components (cf. Appendix~\ref{tcr} for the details about the Green components). They take the form
\begin{align}
	b^*_{p,l,\alpha} &= - \gamma_\alpha \gamma_{l,\alpha} \sigma _{p,l,\alpha}^-, \cr 
	b_{p,l,\alpha} &=  \gamma_\alpha \gamma_{l,\alpha} \sigma _{p,l,\alpha}^+,            
	\label{aobiquadratic}
\end{align}
where  $\sigma _{p,l, \alpha}^\pm$  represent $\Pi_1 S_{pl+\alpha}^\pm \Pi_1$, with $S_{pl+\alpha}^+$ and  $S_{pl+\alpha}^-$ being the raising and lowering operators, defined as $S_{pl+\alpha}^\pm = (S_{pl+\alpha}^x \pm iS_{pl+\alpha}^y)/{\sqrt 2}$, respectively, and $\gamma_\alpha$ and $\gamma_{l,\alpha}$ denote auxiliary Majorana fermions: $\gamma_\alpha \gamma_{\alpha'}+ \gamma_{\alpha'} \gamma_\alpha = 2 \delta _{\alpha \alpha'}$, $\gamma_{l,\alpha}\gamma_{l',\alpha'}+ \gamma_{l',\alpha'} \gamma_{l,\alpha} =2 \delta_{ll'} \delta_{\alpha\alpha'}$ and  $\gamma_{l,\alpha}\gamma_{\alpha'}+ \gamma_{\alpha'} \gamma_{l,\alpha} =0$, in addition to the reality conditions $\gamma^\dagger_\alpha = \gamma_\alpha$ and $\gamma^\dagger_{l,\alpha} = \gamma_{l,\alpha}$~\cite{franz}.
Note that  $\gamma_{l,\alpha} \sigma _{p,l,\alpha}^+$  may be  replaced by $\sum _{\alpha'=0} ^{p-1} U_{\alpha \alpha'} \gamma_{l,\alpha'} \sigma _{p,l,\alpha'}^+$
where $U_{\alpha \alpha'}$ are entries of any $p \times p$ unitary matrix $U$. This is an internal symmetry group ${\rm U}(p)$ acting on internal degrees of freedom inside an emergent unit cell labeled by $l$, which are identified as local spin degrees of freedom and auxiliary Majorana fermions $\gamma_{l,\alpha}$ inside an emergent unit cell. Physically, this implies that there are different ways to organize local spin degrees of freedom inside an emergent unit cell so that they communicate with their counterparts on different emergent unit cells, with auxiliary Majorana fermions $\gamma_\alpha$ acting as a medium. 
As a special choice, we may choose  $U_{\alpha \alpha'} = (1/{\sqrt p}) \sum _{\alpha'=0} ^{p-1} e^{i2\pi \alpha \alpha'/p}$, formally identical to the Fourier transformation, in addition to another choice that one may restrict ${\rm U}(p)$ to be a subgroup ${\rm S}_p$ - the permutation group acting on the $p$ adjacent lattice unit cells inside an emergent unit cell. 
In fact,
the creation  and annihilation operators $a^*_{p, l}$ and $a_{p, l}$ constructed from the above Green components satisfy the defining trilinear commutation relations for one single Green parafermion field (cf. Eqs.\;(\ref{tcr2}) and (\ref{jacobi}) in Appendix~\ref{tcr}). Since the  operators $\sigma _{p,l, \alpha}^\pm$, acting on the basis states in the $l$-th emergent unit cell, involves the projection operator $\Pi_1$, they constitute a realization of Green parafermions in the constrained Hilbert space $V_1$. 
Meanwhile, we stress that the operation ``*" is identical to the adjoint operation $``\dagger"$, so this realization of the  annihilation and creation operators $a_{p, l}$  and  $a^*_{p, l}$ is Hermitian.

According to Green~\cite{green}, one may define the Green parafermion number operators $n_{p,l} =\sum_\alpha \sigma _{p,l, \alpha}^- \sigma _{p,l, \alpha}^+$ for one single  Green parafermion field for fixed $p$ (cf. Eq.~(\ref{nomu}) in Appendix~\ref{tcr}).   As already described above, the requirement that the projected Hamiltonian $\mathscr{\bar H}$ admits the Green parafermion number operators $n_{p, l}$ as a commutative  set of  conserved operators for fixed $p$ in the constrained Hilbert space $V$ results in the introduction of the projection operator $\Pi_2$. In other words, we are capable of expressing a commutative set of Green parafermion number operators in terms of spin degrees of freedom for the ferromagnetic spin-1 biquadratic model, so the  Green parafermion number operators $n_{p, l}$ do not depend on auxiliary Majorana fermions. Here we stress that the  Green parafermion number operators $n_{\sigma,p, l}$ are Hermitian. 

As a result, we may construct Green parafermion states  $a^\dagger_{p, l_1}a^\dagger_{p, l_2} \ldots a^\dagger_{p,l_m} \vert \Omega_0 \rangle$ ($m=1,2,\ldots,M$) as the eigenstates of $\sum_l n_{p, l}$, where $m$  denotes the number of Green parafermions of order $p$ and $\vert \Omega_0 \rangle$  denotes the unique parafermion Fock vacuum, namely the unique no-particle state, satisfying $a_{p, l}\vert \Omega_0 \rangle = 0$. Here $M$ is the maximum occupation number for Green parafermions of order $p$.
For this specific realization,   $\vert \Omega_0 \rangle$ is identified as the fully polarized state $\ket{\psi_0} \equiv  \otimes_{j=1}^L  \vert +_j \rangle$, namely the highest weight state for the uniform ${\rm SU(2)}$ (sub)group. It is readily seen that
$\vert \Omega_0 \rangle$  satisfies the no-particle conditions (cf. Eq.\;(\ref{noparticle2}) in Appendix~\ref{tcr}). As follows from the Pauli exclusion principle, we have $M=L$, since only one single Green parafermion field is involved, namely $n=1$.

\subsubsection{The ferromagnetic $\rm {SU}(2)$ flat-band Tasaki model}
For the ferromagnetic $\rm {SU}(2)$ flat-band Tasaki model (\ref{hamtasaki}), two Green parafermion fields are involved for an atypical partition, and their creation and annihilation operators are denoted as $a^*_{\sigma,p, l}$ and $a_{\sigma,p, l}$, where $\mu$ is replaced by $\sigma$ representing spin up or down: $\sigma = \uparrow$ or $\sigma = \downarrow$. 
In order to establish the connection with two Green parafermion fields, we resort to the Green ansatz for the creation and annihilation operators $a^*_{\sigma,p, l}$ and $a_{\sigma,p, l}$:
\begin{align*}  
a^*_{\sigma,p, l} &= \sum_{\alpha =0}^{p-1} b^*_{\sigma,p,l,\alpha}, \cr
a_{\sigma,p, l} &= \sum_{\alpha =0}^{p-1} b_{\sigma,p,l,\alpha}, 
\end{align*}
where  $b^*_{\sigma,p, l,\alpha}$ and $b_{\sigma,p, l,\alpha}$ are the Green components (cf.~Appendix~\ref{tcr} for the details about the Green components). They take the form
\begin{align}
	b_{\sigma,p, l,\alpha}^* &= - \gamma_\alpha  \hat{a}^\dagger_{\sigma,p,l,\alpha},    \cr
	b_{\sigma,p, l,\alpha} &= \gamma_\alpha    \hat{c}_{\sigma,p,l,\alpha}. 
	\label{aotasaki}
\end{align}
Here $\gamma_\alpha$ denote auxiliary Majorana fermions:  $\gamma_\alpha \gamma_{\alpha'}+ \gamma_{\alpha'} \gamma_\alpha =2 \delta _{\alpha \alpha'}$ and $\gamma^\dagger_\alpha = \gamma_\alpha$~\cite{franz}.
Note that  $ \hat{a}_{\sigma,p,l,\alpha}$ and $\hat{c}^\dagger_{\sigma,p,l,\alpha}$ may be  replaced by $\sum _{\alpha'=0} ^{p-1} U_{\alpha \alpha'} \hat{a}_{\sigma,p,l,\alpha'}$ and $\sum  _{\alpha' =0}^{p-1} U^*_{\alpha' \alpha}   \hat{c}^\dagger_{\sigma,p,l,\alpha'}$, respectively,
where $U_{\alpha \alpha'}$ are entries of any $p \times p$ unitary matrix $U$. This unitary group ${\rm U}(p)$ is an internal symmetry group, which is necessary for local fermion degrees of freedom inside an emergent unit cell to be interpreted as internal degrees of freedom.
Physically, this implies that there are different ways to organize local fermion degrees of freedom inside an emergent unit cell so that they communicate with their counterparts on different emergent unit cells, with auxiliary Majorana fermions  $\gamma_\alpha$ acting as a medium. 
As a special choice, we may choose  $U_{\alpha \alpha'} = (1/{\sqrt p}) \sum _{\alpha'=0} ^{p-1} e^{i2\pi \alpha \alpha'/p}$, formally identical to the Fourier transformation, in addition to another choice that one may restrict ${\rm U}(p)$ to be a subgroup ${\rm S}_p$ - the permutation group  acting on the $p$ adjacent lattice unit cells inside an emergent unit cell. 
Note that $\gamma_\alpha$ have been assumed to anti-commute with $\hat{a}_{\sigma,p,l,\alpha'}$ and $\hat{a}^\dagger_{\sigma,p,l,\alpha'}$ for any $\alpha$ and $\alpha'$.  In fact,
the creation  and annihilation operators $a^*_{\sigma,p, l}$ and $a_{\sigma,p, l}$, constructed from the above Green components, satisfy the defining trilinear and relative trilinear commutation relations for two Green parafermion fields, as presented in Eqs.(\ref{tcr2}), (\ref{jacobi}), (\ref{tcr3}), and (\ref{tcr4}) in Subsection~\ref{greenpara-tcr} (also cf.~Eq.(\ref{tcr5}) in Appendix~\ref{tcr}). Since the operation $*$ here is not identical to the adjoint operation $\dagger$, so they constitute a non-Hermitian realization of  Green parafermions, which live in the original unconstrained Hilbert space $V_0$.
Hence we need to introduce the Hermitian conjugated operators $(a^*_{\sigma,p, l})^\dagger$ and $a^\dagger_{\sigma,p, l}$ of the creation and annihilation operators $a^*_{\sigma,p, l}$ and $a_{\sigma,p, l}$. 

An extension to a non-Hermitian realization of the trilinear and relative trilinear commutation relations for two Green parafermion fields, derived by Greenberg and Messiah~\cite{greenberg}, is thus necessary. We refer more details to Appendix~\ref{tcr}. One may define the Green parafermion number operators $n_{\sigma,p, l}= \sum_\alpha  \hat{c}^\dagger_{\sigma,p,l,\alpha} \hat{a}_{\sigma,p,l,\alpha}$ ($\sigma =\uparrow$ and $\downarrow$) for two  Green parafermion fields (cf. Eq.~(\ref{nomu}) in Appendix~\ref{tcr}). 
Since the Green parafermion number operators $n_{\sigma,p, l}$ are not Hermitian,  we also need the Hermitian conjugated variants $n^\dagger_{\sigma,p, l}= \sum_\alpha  \hat{a}^\dagger_{\sigma,p,l,\alpha} \hat{c}_{\sigma,p,l,\alpha}$. In other words, we are capable of expressing a commutative set of Green parafermion number operators $n_{\sigma,p, l}$ and the Hermitian conjugated variants $n_{\sigma,p, l}^\dagger$ in terms of fermion degrees of freedom for the  ferromagnetic $\rm {SU}(2)$ flat-band Tasaki model.  As such, there emerge two Hermitian conjugated copies of the creation and annihilation operators for two Green parafermion fields: one copy consists of the creation and annihilation operators $a^*_{\sigma,p, l}$ and $a_{\sigma,p, l}$ and the other copy consists of the creation and annihilation operators $a^\dagger_{\sigma,p, l}$  and $(a^*_{\sigma,p, l})^\dagger$.
These two Green parafermion fields may be classified into the PF subset~\cite{greenberg} (also cf.~Appendix~\ref{tcr}). Indeed, 
a detailed calculation shows that the commutativity between $\mathscr{\bar H}$ and $n_{\sigma,p, l}$ is guaranteed if the projection operator $\Pi_2$ is implemented, since the projected Hamiltonian becomes the zero operator under this projection. Note that the  Green parafermion number operators $n_{\sigma,p, l}$ do not depend on auxiliary Majorana fermions. However, they are non-Hermitian. 

The non-Hermitian nature of Green parafermions makes it necessary to tackle both left and right eigenstates of the Green parafermion number operators $n_{\sigma,p, l}$ for fixed $p$. Mathematically,  the creation and annihilation operators $a^*_{\sigma,p, l}$ and $a_{\sigma,p, l}$ are exploited to construct the right eigenstates of the Green parafermion number operators $n_{\sigma,p, l}$ and their Hermitian conjugated variants are exploited to construct the right eigenstates of the Green parafermion number operators $n^\dagger_{\sigma,p, l}$. Note that the latter are equivalent to the left eigenstates of the Green parafermion number operators $n_{\sigma,p,l}$.

For this specific realization, the unique parafermion vacuum $\vert \Omega_0 \rangle$ is identified as the fermionic Fock vacuum $\vert \otimes_{x \in \Lambda} 0_x\rangle$, with $\langle \Omega_0 \vert$ being its Hermitian conjugation. 
Indeed, we have $a_{\sigma, p,l} \vert \Omega_0 \rangle=0$ and  $(a^*_{\sigma, p,l})^\dagger \vert \Omega_0 \rangle=0$ for all $\sigma$ and $l$, when $p$ is fixed. This is necessary for $a_{\sigma, p,l}$ and  $(a^*_{\sigma, p,l})^\dagger$ to act as the annihilation operators in the two copies conjugated to each other. Or equivalently, we have $\langle \Omega_0 \vert a^\dagger_{\sigma, p,l}=0$ and $\langle \Omega_0 \vert a^*_{\sigma, p,l}=0$ for all $\sigma$ and $l$, when $p$ is fixed. We are thus led to the no-particle conditions for  the creation and annihilation operators $a^*_{\sigma, p,l}$ and $a_{\sigma, p,l}$ (cf. Eq.(\ref{noparticle2}), in addition to the no-particle conditions for  the creation and annihilation operators $a^\dagger_{\sigma, p,l}$ and $(a^*_{\sigma, p,l})^\dagger$ (cf. Eq.(\ref{noparticle22}). Here the order $p_\sigma =p$, as follows from
$p_{\sigma}=\langle \Omega_0 \vert a_{\sigma, l} a_{\sigma, l}^* \vert \Omega_0 \rangle$, or equivalently, 
$p_\sigma =\langle \Omega_0 \vert (a^*_{\sigma,p,l})^\dagger a_{\sigma,p,l}^\dagger \vert \Omega_0 \rangle$. Note that $p_{\sigma}$ is independent of $\sigma$, since the two Green parafermion fields are time-reversed counterparts to each other.

As a result, we may construct Green parafermion states  $a^\dagger_{\sigma_1, p, l_1}a^\dagger_{\sigma_2, p, l_2} \ldots a^\dagger_{\sigma_m, p,l_m} \vert \Omega_0 \rangle$  as the right eigenstates of $\sum_{\sigma,l} n^\dagger_{\sigma,p, l}$  and 
Green parafermion states  $a^*_{\sigma_1, p, l_1}a^*_{\sigma_2, p, l_2} \ldots a^*_{\sigma_m, p,l_m} \vert \Omega_0 \rangle$  as the right eigenstates of $\sum_{\sigma,l} n_{\sigma,p, l}$,  where $m=1,2,\ldots,M$ denotes the number of Green parafermions of order $p$. Here
$M$ is the maximum occupation number of the same field or different fields for one species of Green parafermions of order $p$, as a result of the Pauli exclusion principle, and $\vert \Omega_0 \rangle$  denotes the unique parafermion Fock vacuum, namely the unique no-particle state, satisfying $a_{\sigma,p, l}\vert \Omega_0 \rangle = 0$. 
For a generic spin configuration $\{\sigma_1,\sigma_2,\ldots,\sigma_m \}$, there are $m_\uparrow$ spins up and  $m_\downarrow$ spins down, so $m=m_\uparrow+m_\downarrow$. Mathematically, this is equivalent to the statement that these Green parafermion states are defined in the sectors labeled by  $m_\uparrow$  and  $m_\downarrow$. As follows from the Pauli exclusion principle, the maximum occupation number $M$ is $M=2L$, since two Green parafermion fields are involved, namely $n=2$.

\vspace{10mm}

In summary, for a condensed matter system undergoing SSB with type-B GMs, Green parafermions are formed from auxiliary Majorana fermions and spin or fermion degrees of freedom as composite particles.
Actually, auxiliary Majorana fermions $\gamma_\alpha$ have been introduced to ensure that the Green components with the same $\alpha$ satisfy the anti-commutation relations and those with different $\alpha$'s satisfy the commutation relations. Moreover, this explains why it is necessary to
introduce auxiliary Majorana fermions $\gamma_{\mu,l,\alpha}$ for quantum many-body spin systems, given that spin degrees of freedom located at different lattice sites are always commutative.
Generically, for a given realization of Green parafermions in a  condensed matter system, one may perform a $p \times p$ unitary transformation $U$ for local spin or fermion degrees of freedom inside an emergent unit cell to yield an equivalent realization. In other words, an internal symmetry group ${\rm U}(p)$ emerges, which is necessary for local spin or fermion degrees of freedom inside an emergent unit cell to be interpreted as internal degrees of freedom (cf.~Eqs.\;(\ref{aobiquadratic}) and (\ref{aotasaki}) below for concrete examples).  An unusual feature is that this internal symmetry group ${\rm U}(p)$ contains  the symmetric group ${\rm S}_p$  as a subgroup, which consists of all the permutations with respect to $p$ lattice unit cells  inside an emergent unit cell.
Meanwhile, we note that  the trilinear commutation relations for the same field and the relative trilinear commutation relations for different fields are invariant under a $N \times N$ unitary transformation $U_N$ of the annihilation operators $a_{\mu, p, l}$, if an accompanied  $N \times N$ unitary transformation $U^\dagger_N$ is performed for the  creation operators $a^*_{\mu, p, l}$ simultaneously.

\section{Green parafermion states realized in a scenario that SSB is a limit of explicit symmetry breaking}~\label{ssbscenario}

Our aim is to connect a condensed matter system described by the model Hamiltonian $\mathscr{H}$ undergoing SSB with type-B GMs with a set of Green parafermion fields in the scenario that SSB is regarded as a limit of explicit symmetry breaking~\cite{zuber}. Here we stress that some notable features arise here, if the ground state degeneracies are exponential, which are missing in the conventional description of the scenario (also cf.~Appendix~{\ref{ssblimit} for more details).

In this scenario, an extra term  $-h \;\Sigma_{c}$ is introduced, where $h$ is an external control parameter and the subscript $c$ indicates that this term is defined as a sum over the lattice unit cells for both quantum many-body spin systems or strongly correlated itinerant electron systems.  The conventional description works for 
the ferromagnetic $\rm {SU}(2)$ spin-$1/2$ Heisenberg model - a paradigmatic example for SSB with type-B GMs. Note that it only yields polynomially many degenerate ground states. In this case,  we may take $\Sigma_{c}$ to be $\sum_j S^z_j$. The introduction of such an external magnetic field explicitly breaks the symmetry group $\rm {SU}(2)$ to the residual symmetry group $\rm {U}(1)$ generated by $\sum_j S^z_j$. However, as briefly discussed in Appendix~\ref{lemma}, one may choose $\Pi_1$ to be the projection operator that projects out all the  states containing $S^-_j \vert \uparrow~\rangle_{j}$ and $\Pi_2$ to be the identity operator, so $\Pi = \Pi_1$. As a result, the constrained Hilbert space $V$ is a one-dimensional subspace spanned by the fully polarized state with all spins up, namely the highest weight state for the symmetry group ${\rm SU(2)}$, 
consistent with the fact that this fully polarized  state is selected as the unique ground state, if $h$ is assumed to be positive. Hence
the ferromagnetic $\rm {SU}(2)$ spin-$1/2$ Heisenberg model is not a proper candidate to realize Green parafermion states as low-lying excitations in a condensed matter system. 

Mathematically, this stems from the fact that, for the ferromagnetic $\rm {SU}(2)$ spin-$1/2$ Heisenberg model, all degenerate ground states constitute an orthonormal basis for an irreducible representation of the symmetry group $\rm {SU}(2)$,  with the dimension of the representation space being $L+1$. In particular, they are permutation-invariant, in contrast to Green parafermion states that are not permutation-invariant when the order $p>1$. Indeed, this statement is valid for any condensed matter system undergoing SSB with type-B GMs, if the ground state degeneracy is only polynomial with system size. As we shall show in Section~\ref{fictitious}, the identification of a state $\vert \eta \rangle$ with Green parafermion states always  produces exponentially instead of polynomially many degenerate ground  states, if auxiliary Majorana fermions are fictitious. In this way, we are led to the conclusion that Green parafermion states are only realized (up to a projection operation) in condensed matter systems undergoing SSB with type-B GMs, subject to the condition that the ground state degeneracies are exponential.

\subsection{An extra term  $-h \;\Sigma_{c}$ connecting with Green parafermions}
Our discussion below is presented for a non-Hermitian realization of Green parafermions, since an Hermitian realization may be treated as a special case when  $N^\dagger_{\mu,p} =N_{\mu,p}$ (cf.~Appendix~\ref{ssblimit} for a brief summary about an Hermitian realization). One needs to distinguish two cases. First, if Green parafermions are realized in the constrained Hilbert space $V_1$, then $\Sigma_{c}$ is chosen to satisfy  
\begin{equation}
	\Pi_1 \Sigma_{c} \Pi_1 = \frac{1}{2} \sum_\mu (N_{\mu,p}+N^\dagger_{\mu,p}). \label{sigma1} 
\end{equation}
Second, if Green parafermions are realized in the unconstrained Hilbert space, then $\Sigma_{c}$  is chosen to satisfy 
\begin{equation}
	\Sigma_{c} =  \frac{1}{2} \sum_\mu (N_{\mu,p}+N^\dagger_{\mu,p}). \label{sigma2} 
\end{equation}
Here $ \sum_\mu N_{\mu,p}$ is the total Green parafermion number operator, with $N_{\mu,p} = \sum _l n_{\mu, p, l}$, where $n_{\mu,p,l}$ are the  Green parafermion number operators: $n_{\mu,p,l} = \sum _{\alpha} n_{\mu,p,l,\alpha}$ in the real space representation. 
Remarkably, the subscript $p$ appears in the right-hand side of  Eq.(\ref{sigma1}) or Eq.(\ref{sigma2}), but not in the left-hand side. As we shall argue below, 
this reflects a hidden form of explicit symmetry breaking, which is manifested {\it only} in  a Green parafermion system.
We remark that Eq.(\ref{sigma1}) and Eq.(\ref{sigma2}) constitute the main results in the scenario that SSB is a limit of explicit symmetry breaking. Actually, a solution to either of them is necessary to identify an extra term $-h \;\Sigma_{c}$, which is fundamental to realize Green parafermions (up to a projection operator) in condensed matter.

As a result, an internal symmetry group ${\rm U}(p)$ may be interpreted as a subsystem gauge group for $-h \sum_\mu N_{\mu,p}$, since any ${\rm U}(p)$ unitary transformation may be carried out independently in each emergent unit cell.  In this sense, this should be understood as a subsystem gauge theory (cf. Appendix~\ref{elitzur1}), so $ \sum_\mu (N_{\mu,p}+N^\dagger_{\mu,p})/2$  is a proper choice that is Hermitian and invariant under subsystem gauge transformations.
In both cases,
the total Hamiltonian $\mathscr{H}_t \equiv \mathscr{H}-h \;\Sigma_{c}$ is projected to $\mathscr{\bar H} -h  \sum_\mu \Pi_1 (N_{\mu,p}+N^\dagger_{\mu,p})\Pi_1/2$.  Afterwards,  the projection operator $\Pi_2$ is implemented such that the projected Hamiltonian $\mathscr{\bar H} -h  \sum_\mu \Pi_1 (N_{\mu,p}+N^\dagger_{\mu,p})\Pi_1/2$
becomes $-h  \sum_\mu \Pi (N_{\mu,p}+N^\dagger_{\mu,p})\Pi/2$ if $\mathscr{H}$ is in a canonical form. Indeed, $\Pi_2 \mathscr{\bar H} \Pi_2$ is proportional to $\Pi$ if $\mathscr{H}$ is not in a canonical form. Obviously,  $\Pi$ commutes with $\sum_\mu \Pi (N_{\mu,p}+N^\dagger_{\mu,p})\Pi/2$. 

As a consequence, one may first focus on a Green parafermion system described by the Hamiltonian $-h \sum_\mu N_{\mu,p}$, which is invariant under subsystem gauge transformations (cf. Appendix~\ref{elitzur1}). Here we stress that this statement is {\it only} valid for a realization of Green parafermions in terms of auxiliary Majorana fermions defined on an emergent unit cell, since its validity depends on the existence of internal degrees of freedom. 
In other words, the possibility for a realization of  Green parafermions in terms of ordinary spin or fermion degrees of freedom, together with auxiliary Majorana fermions  $\gamma _{\mu,l,\alpha}$ and $ \gamma_{\alpha}$ or $ \gamma_{\alpha}$, implies that a sort of internal structure emerges in  Green parafermions. This is in an apparent contradiction with a statement by Greenberg and Messiah~\cite{greenberg} that
the Green Ansatz (\ref{greenansatz}) for the creation and annihilation operators does not introduce any composite structure,  in the sense that it is linear. A resolution to this contradiction lies in the fact that we are dealing with a novel type of interaction that is well beyond any traditional characterization of couplings between different types of particles. Indeed,  this novel type of interaction {\it only} involves communication between spin or fermion degrees of freedom located at different emergent unit cells, with auxiliary Majorana fermions acting as a medium. As we shall see below, auxiliary Majorana fermions do not manifest themselves in the Hamiltonian explicitly, both physical and fictitious.

Actually, internal degrees of freedom are local spin degrees of freedom plus auxiliary Majorana fermions $\gamma_{\mu,l,\alpha}$  or fermion degrees of freedom inside an emergent unit cell (cf. Eq.(\ref{aobiquadratic}) and Eq.(\ref{aotasaki}) in Subsection~\ref{biquadratic-tasaki}), which in turn lead to an internal symmetry group ${\rm U}(p)$ consisting of all $p \times p$ unitary transformations for each $p$, if we restrict ourselves to an atypical partition. 
In particular,  the symmetric group ${\rm S}_p$ consisting of all the permutations with respect to  $p$ lattice unit cells labeled by $\alpha$ inside an emergent unit cell labeled by $l$ forms a subgroup of ${\rm U}(p)$.  Hence for $p$ adjacent lattice unit cells inside an emergent unit cell, their ordering is irrelevant. In other words,  $p$ lattice unit cells in a given emergent unit cell may be ordered independently as a result of the symmetric group ${\rm S}_p$.  If $p=1$, then we have a {\it local} gauge group ${\rm U}(1)$; if $p=L$, then we have a {\it global} gauge group ${\rm U}(L)$. Generically, for any $p$ between 1 and $L$, a subsystem gauge group interpolates a local gauge group ${\rm U}(1)$ and  a global gauge group ${\rm U}(L)$.
We stress that  the invariance of $-h \sum_\mu N_{\mu,p}$  under subsystem gauge transformations imposes constraints on a choice of the extra term  $-h \;\Sigma_{c}$, in addition to the requirement that $\Sigma_{c}$ is Hermitian. Note that the model Hamiltonian $\mathscr{H}$ itself is not invariant under any subsystem gauge transformation. A hierarchical structure thus arises at a Hamiltonian level: at the bottom is the model Hamiltonian $\mathscr{H}$ plus the extra term $-h \;\Sigma_{c}$ and at the top is a Green parafermion system described by the Hamiltonian $-h \; \sum_\mu N_{\mu,p}$, with
a projected Hamiltonian $-h \; \sum_\mu \Pi (N_{\mu,p}+N^\dagger_{\mu,p})\Pi/2$ under a projection operator $\Pi$ lying in between.
Note that neither a condensed matter system described by  $\mathscr{H}$ plus the extra term $-h \;\Sigma_{c}$  nor its projected form $-h \;  \sum_\mu \Pi (N_{\mu,p}+N^\dagger_{\mu,p})\Pi/2$ under a projection operator $\Pi$  is invariant under subsystem gauge transformations, in contrast to  $-h  \sum_\mu N_{\mu,p}$. Mathematically, this is due to the fact that $\Pi$ does not commute with subsystem gauge transformations.   Indeed, it is this hierarchical structure that makes the meaning of a gauge as a notion restrictive to only a set of physical observables invariant under subsystem gauge transformations.

Note that the  unique parafermion Fock vacuum $\vert \Omega_0 \rangle$ itself must be one of degenerate ground states of the model Hamiltonian $\mathscr{H}$.
This stems from the requirement that this specific state must be a simultaneous eigenstate of the projected Hamiltonian $\mathscr{\bar H}$ and the parafermion number operators $n_{\mu, p, l}$ for each $p$. Meanwhile, it is a simultaneous eigenstate of the model Hamiltonian $\mathscr{H}$ and the Cartan generator(s) of the symmetry group $G$. More precisely, $\vert \Omega_0 \rangle$ is always identified as the highest weight state of  the symmetry group $G$ if $G$ is semisimple or the semisimple subgroup of $G$ if $G$ is not semisimple, as follows from the requirements that
the total Green parafermion number operator $ \sum_\mu N_{\mu,p}$ for any specific realization  always commutes with the Cartan generators of the (semi-simple) symmetry group $G$, or with the generator of ${\rm U}(1)$ and the Cartan generators of the (semi-simple) subgroup of the (non-semi-simple) symmetry group $G$ (cf.~Appendix~\ref{elitzur1}). 

Here we have assumed that the (non-semi-simple) symmetry group $G$ always contains ${\rm U}(1)$ as a subgroup, as seen in the ferromagnetic $\rm {SU}(2)$ flat-band Tasaki model (\ref{hamtasaki}). Moreover, if the symmetry group $G$ is staggered so that the model Hamiltonian  $\mathscr{H}$ exhibits the odd-even parity effect, then it is sufficient to restrict to a smaller symmetry group that is generically uniform, as we shall demonstrate for the ferromagnetic spin-1 biquadratic model (\ref{hambq}) in the next Subsection. In the real space representation, we have $l=1, 2,\ldots, N$,  where $N=L/p$ is the total number of emergent unit cells.  Equivalently, if we choose to work in the momentum space representation, then $l$ should be replaced by $k$ that may be interpreted as crystal momentum arising from the presence of emergent unit cells, if PBCs are adopted (cf. Appendix~\ref{realandmomentum} for a brief discussion about the real and momentum space representations for Green parafermions). In the momentum space representation, we have $k= 2 \pi \delta /N$ ($\delta = 0,1,2,\ldots, N-1$). Indeed, it is advantageous to work in the momentum space representation, 
since the model Hamiltonian $\mathscr{H}$ is translation-invariant under one lattice unit cell and  the Hamiltonian $-h\;  \sum_\mu N_{\mu,p}$ is translation-invariant under $p$ lattice unit cells. Note that $\Pi$ commutes with the translation symmetry operation $\tau$ under one lattice unit cell, so $ \sum_\mu \Pi (N_{\mu,p}+N^\dagger_{\mu,p})\Pi/2$ commutes with  the translation symmetry operation $\tau^p$ under $p$ lattice unit cells. 
In addition, as far as the permutation operations with respect to $\{k_1,k_2,\ldots,k_m\}$ are concerned,  it is also privileged to investigate  Green parafermion states $a^*_{\mu_1, p, k_1}a^*_{\mu_2,p, k_2} \ldots a^*_{\mu_m, p,k_m} \vert \Omega_0 \rangle$ and  $a^\dagger_{\mu_1, p, k_1}a^\dagger_{\mu_2,p, k_2} \ldots a^\dagger_{\mu_m, p,k_m} \vert \Omega_0 \rangle$ in the momentum space representation.

Since the projection operator $\Pi$ commutes with  $\Pi (N_{\mu,p}+N^\dagger_{\mu,p})\Pi/2$ and both of them are Hermitian, they must share simultaneous eigenstates  $\Pi (a^*_{\mu_1, p, k_1}a^*_{\mu_2,p, k_2} \ldots a^*_{\mu_m, p,k_m}) \vert \Omega_0 \rangle$ and $\Pi (a^\dagger_{\mu_1, p, k_1}a^\dagger_{\mu_2,p, k_2} \ldots a^\dagger_{\mu_m, p,k_m}) \vert \Omega_0 \rangle$, as long as $\Pi$ does not nullify   $a^*_{\mu_1, p, k_1}a^*_{\mu_2,p, k_2} \ldots a^*_{\mu_m, p,k_m} \vert \Omega_0 \rangle$ and $a^\dagger_{\mu_1, p, k_1}a^\dagger_{\mu_2,p, k_2} \ldots a^\dagger_{\mu_m, p,k_m} \vert \Omega_0 \rangle$, where $m=1,\ldots,M_p$. Here $M_p$ denotes the maximum occupation number of the same field or different fields for one species of Green parafermions of order $p$. As argued in~Appendix~\ref{ssblimit}, $\Pi (a^\dagger_{\mu_1, p, l_1}a^\dagger_{\mu_2,p, l_2} \ldots a^\dagger_{\mu_m, p,l_m}) \vert \Omega_0 \rangle$ may be expressed as a linear combination of $\Pi (a^*_{\mu_1, p, l_1}a^*_{\mu_2,p, l_2} \ldots a^*_{\mu_m, p,l_m}) \vert \Omega_0 \rangle$
in each sector labeled by $m$. We are thus led to conclude that any eigenstate of  $\Pi (N_{\mu,p}+N^\dagger_{\mu,p})\Pi/2$, with the eigenvalue being $m$, must be a linear combination of   $\Pi (a^*_{\mu_1, p, k_1}a^*_{\mu_2,p, k_2} \ldots a^*_{\mu_m, p,k_m}) \vert \Omega_0 \rangle$ 
in each sector labeled by $m$.
Mathematically, this holds if $\Pi (a^*_{\mu_1, p, k_1}a^*_{\mu_2,p, k_2} \ldots a^*_{\mu_m, p,k_m}) \vert \Omega_0 \rangle$ or $\Pi (a^\dagger_{\mu_1, p, k_1}a^\dagger_{\mu_2,p, k_2} \ldots a^\dagger_{\mu_m, p,_m}) \vert \Omega_0 \rangle$ for an atypical partition and their counterparts for a typical partition are sufficient to produce a basis for the constrained Hilbert space $V$. Generically, this should be the case, as confirmed for the ferromagnetic $\rm {SU}(2)$ flat-band Tasaki model - the only known model that admits a non-Hermitian realization up to the present (cf.~Subsection~\ref{biquadratic-tasaki}). 

Note that $M_p$ is different from $M$, as long as the projection operator $\Pi_2$ is non-trivial. This stems from a modified form of the Pauli exclusion principle, due to the presence of the combined projection operator $\Pi$. In order to quantify the effect of the projection operator $\Pi$  on parastatistics, it is useful to introduce a notion -- the rank for a realization of Green parafermions in condensed matter.

\subsection{The rank and a modified form of the Pauli exclusion principle}
 
As already pointed out, Green parafermions obey the Pauli exclusion principle~\cite{green,greenberg}. However, what is realized in condensed matter systems under investigation is Green parafermion states up to a projection, as revealed in a hierarchical structure at a Hamiltonian level. 
Indeed, the presence of the combined projection operator $\Pi$ does affect parastatistics, which may be quantified by the rank $r$ for a specific realization of Green parafermions in a condensed matter system.  Mathematically, the rank $r$ is defined to be an integer that satisfies $\Pi (a^\dagger_{\mu,p, l})^r \vert \Omega_0 \rangle\neq 0$, but $\Pi (a^\dagger_{\mu,p, l})^{r+1} \vert \Omega_0 \rangle=0$ for the same field, and $\Pi (a^\dagger_{\mu,p, l})^{r/2} (a^\dagger_{\mu',p, l})^{r/2})\vert \Omega_0 \rangle\neq 0$, but $\Pi (a^\dagger_{\mu,p, l})^{r/2+1}  (a^\dagger_{\mu',p, l})^{r/2})\vert \Omega_0 \rangle = 0$ and $\Pi (a^\dagger_{\mu,p, l})^{r/2}  (a^\dagger_{\mu',p, l})^{r/2+1})\vert \Omega_0 \rangle = 0$ ($\mu,\mu' = 1,2,\ldots,n$) for  any two different fields labeled by $\mu$ and $\mu'$, when $p\;{\rm mod}\;(4) =0$ or $3$ and $\Pi (a^\dagger_{\mu,p, l})^{(r+1)/2} (a^\dagger_{\mu',p, l})^{(r-1)/2}\vert \Omega_0 \rangle\neq 0$ and $\Pi (a^\dagger_{\mu,p, l})^{(r-1)/2} (a^\dagger_{\mu',p, l})^{(r+1)/2}\vert \Omega_0 \rangle\neq 0$, but $\Pi (a^\dagger_{\mu,p, l})^{(r+1)/2}  (a^\dagger_{\mu',p, l})^{(r+1)/2}  \vert \Omega_0 \rangle = 0$  ($\mu,\mu' = 1,2,\ldots,n$) for  any two different fields labeled by $\mu$ and $\mu'$, when $p\;{\rm mod}\;(4) =1$ or $2$. If one restricts to one  parafermion field, namely $n=1$, then we have $\Pi (a^\dagger_{p, l})^r \vert \Omega_0 \rangle\neq 0$, but $\Pi (a^\dagger_{p, l})^{r+1} \vert \Omega_0 \rangle=0$, where we have dropped off the subscript $\mu$. 
Note that it is usually possible to attach a discrete ${\rm Z}_n$ cyclic group to $n$ different values of $\mu$, such as the time-reversal symmetry group ${\rm Z}_2$ for the ferromagnetic $\rm {SU}(2)$ flat-band Tasaki model and the exchange symmetry group  ${\rm Z}_2$ between the spin operators $\textbf{S}_j=(S^x_j,S^y_j,S^z_j)$ and the pseudo-spin operators  $\textbf{T}_j=(T^x_j,T^y_j,T^z_j)$ for the ferromagnetic spin-orbital model~\cite{shi-so}. 

The rank $r$ is not always equal to the order $p$ for a generic realization of Green parafermions in a condensed matter system, due to the presence of the projection operator $\Pi$. In this sense, the order $p$ is {\it intrinsic}, meaning that it {\it only} depends on the defining trilinear and relative trilinear commutation relations. In contrast, the rank is {\it extrinsic}, in the sense that it also depends on the projection operator $\Pi$. Physically, the rank $r$ quantifies the maximum occupation number of Green parafermions from the same field or the  maximum occupation number of  Green parafermions from  any two different fields among $n$  Green parafermion fields, if we restrict to the projected Green parafermion states, when Green parafermions from  any two different fields are {\it only} present as equally as possible in one emergent unit cell labeled by $l$.

\subsection{Miscellaneous remarks}

Here we present a few miscellaneous remarks, relevant to some salient features manifested in our scenario that SSB is a limit of explicit symmetry breaking in the context of realizations of Green parafermions in condensed matter. 

First, as already mentioned above,  the total Green parafermion number operator $ \sum_\mu N_{\mu,p}$ for any specific realization  always commutes with the Cartan generator(s) of the (semi-simple) symmetry group $G$ for quantum many-body spin models, or with the generator of ${\rm U}(1)$ generated by the electron number and the Cartan generator(s) of the (semi-simple) subgroup of the symmetry group $G$ for strongly correlated itinerant electron systems. As a result, for any $h$, all the sectors labeled by $m$ may be re-labeled by the eigenvalue(s) of the Cartan generator(s) or by the electron number and the eigenvalue(s) the Cartan generator(s).  In particular,  
as $h$ tends to be zero, all states in these sectors become degenerate in this scenario (cf.~Appendix~\ref{elitzur1}).
In addition,  Green parafermion states in the real and momentum space representations are connected via the Fourier transformation (cf.~Appendix~\ref{realandmomentum}). Physically, this results from the equivalence between the two representations.

Second, auxiliary Majorana fermions themselves do not appear in this extra term $-h \;\Sigma_{c}$ as well as  in the model Hamiltonian $\mathscr{H}$. Moreover, they are not involved explicitly in the Hamiltonian $-h \sum_\mu N_{\mu,p}$. Nevertheless, the creation and annihilation operators of Green parafermions contain auxiliary Majorana fermions as a key ingredient. Hence they behave as if they are {\it spectators} instead of {\it participants}. Indeed, the extra term $-h \;\Sigma_{c}$ only involves spin or fermion degrees of freedom, depending on what model is under investigation. 
However, whether or not the Hilbert space should be enlarged to accommodate both the subspace for spin or fermion degrees of freedom and the subspace for auxiliary Majorana fermions depends on the nature of auxiliary Majorana fermions.  If auxiliary Majorana fermions are physical, then it is necessary to enlarge the Hilbert space to accommodate both the subspace for spin or fermion degrees of freedom and the subspace for auxiliary Majorana fermions, depending on the specifics of a condensed matter system. In contrast, 
if auxiliary Majorana fermions are fictitious,  then the Hilbert space only involves spin or fermion degrees of freedom, depending on what model is under investigation.

Third,  as we have already pointed out, the model Hamiltonian $\mathscr{H}$ and the extra term $-h \;\Sigma_{c}$ is translation-invariant under one lattice unit cell, but a Green parafermion system described by the Hamiltonian $-h \sum_\mu N_{\mu,p}$
is only translation-invariant under  $p$ lattice unit cells, as a result of the presence of auxiliary Majorana fermions $\gamma_\alpha$ ($\alpha =0,1,\ldots,p-1$) arranged periodically on emergent unit cells for an atypical partition. Physically, this leads to explicit symmetry breaking such that the translation symmetry  under one lattice unit cell is broken into the translation symmetry under $p$ lattice unit cells, thus representing a novel type of symmetry breaking -- a {\it hidden} form of explicit symmetry breaking, which is not manifested itself in the model Hamiltonian $\mathscr{H}$ as well as the extra term $-h \;\Sigma_{c}$. In contrast, this hidden form of explicit symmetry breaking is reflected in a Green parafermion system, which is connected with the model Hamiltonian $\mathscr{H}$ including the extra term $-h \;\Sigma_{c}$ in a hierarchical structure at a Hamiltonian level (also cf.~Appendix~\ref{elitzur1}). In other words, this explicit symmetry breaking is {\it hidden} in  a Green parafermion system, with Green parafermions living in the constrained 
Hilbert space $V_1$ or the original unconstrained Hilbert space $V_0$. In fact,  as already mentioned above, in Eqs.\;(\ref{sigma1}) and (\ref{sigma2}), the subscript $p$ only occurs in the total Green parafermion number operator $\sum_\mu N_{\mu,p}$.

\subsection{Two illustrative examples for an extra term  $-h \Sigma_{c}$}

\subsubsection{The ferromagnetic spin-1 biquadratic model}
For the ferromagnetic spin-1 biquadratic model (\ref{hambq}), one may introduce an extra term $-h \sum_j S^z$, where $h $ is an external control parameter. In other words, we have chosen $\Sigma_{c}$ to be the Cartan generator  $\sum_j S^z$ of the uniform ${\rm SU(2)}$ (sub)group. 
As already mentioned in Section\ref{ssbtype-b}, this model exhibits the even-odd parity effect. However, one only needs to focus on the uniform ${\rm SU(2)}$ (sub)group, as far as the construction of Green parafermion states is concerned,  irrespective of $L$ being even or odd. Mathematically,
this is due to the fact that the Green components   $b^*_{p,l,\alpha}$ and $b_{p,l,\alpha}$ in Eq.~(\ref{aobiquadratic}) only involve the generators  of the uniform ${\rm SU(2)}$ (sub)group, irrespective of  $L$ being even or odd. A physical consequence one may draw from this fact is that 
the pattern for a realization of Green parafermions remains the same for  even and odd $L$, in contrast to the symmetry groups and the SSB patterns.
It is readily seen that this choice for $\Sigma_{c}$ satisfies Eq.\;(\ref{sigma1}), since Green parafermions are realized in the constrained Hilbert space $V_1$. Indeed, this is the only choice that satisfies the requirement that the Cartan generator  $\sum_j S^z$  commutes with  the total Green parafermion number operators $N_{p}= \sum_l n_{p,l}$. This extra term also commutes with the projection operator $\Pi_1$. In fact, after the projection  operator $\Pi_1$ is implemented, $\sum_j S^z$ becomes   $-N_{p}$.

Notably, this implies that the translation symmetry under one lattice unit cell has been explicitly broken to the translation symmetry under $p$ lattice unit cells, due to the presence of auxiliary Majorana fermions, either physical or fictitious. We remark that this explicit symmetry breaking is in a {\it hidden} form, given that the extra term $-h \sum_j S^z$ as well as the model Hamiltonian (\ref{hambq}) are translation-invariant under one lattice unit cell.
In the scenario that SSB is regarded as a limit of explicit symmetry breaking, a bridge may be built to connect the model Hamiltonian (\ref{hambq})  with a Green parafermion model described by the Hamiltonian $-h N_{p}$ via the projection operators $\Pi_1$ and $\Pi_2$, thus revealing a hierarchical structure at a Hamiltonian level. Note that $-h N_{p}$ is invariant under  subsystem gauge transformations. In contrast, $\mathscr{H}$ is not invariant under any subsystem gauge transformation.
Given the realization specified by the Green components (\ref{aobiquadratic}) is Hermitian, and only one single Green parafermion field is involved, one may construct Green parafermion states $a^\dagger_{p, l_1}a^\dagger_{p, l_2} \ldots a^\dagger_{p, l_m} \vert \Omega_0 \rangle$, where we have used $\dagger$ instead of $*$ to construct eigenstates, since they are identical for an Hermitian realization. Here $\vert \Omega_0 \rangle$ represents the unique parafermion Fock vacuum, identified to be the highest weight state $\vert \otimes_{j=1}^L +_j \rangle$.
It follows that one single Green parafermion field of order $p$ occurs for each $p$, if auxiliary Majorana fermions $\gamma_\alpha$ and $\gamma_{l,\alpha}$ are introduced, with the order $p$ being identical to the emergent unit cell size (for more details, cf.~Appendix~\ref{tcr}). 
As a consequence, $\Pi (a^\dagger_{p, l_1}a^\dagger_{p, l_2} \ldots a^\dagger_{p, l_m}) \vert \Omega_0 \rangle$ are eigenstates of $\Pi N_{p}\Pi/2$, with the eigenvalue being $m$, 
in each sector labeled by $m$ for this Hermitian realization, where $m=1,2,\ldots,M_p$ (cf.~Appendix~\ref{ssblimit}).

\subsubsection{The ferromagnetic $\rm {SU}(2)$ flat-band Tasaki model}

For the ferromagnetic $\rm {SU}(2)$ flat-band Tasaki model (\ref{hamtasaki}), one may introduce an extra  term $-h\sum_{\sigma,q} (\hat{a}^\dagger_{\sigma,q} \hat{c}_{\sigma,q}+ \hat{c}^\dagger_{\sigma,q} \hat{a}_{\sigma,q})/2$, where $h$ is an external control parameter. 
In other words, we have chosen $\Sigma_{c}$ to be $\sum_{\sigma,q} (\hat{a}^\dagger_{\sigma,q} \hat{c}_{\sigma,q}+ \hat{c}^\dagger_{\sigma,q} \hat{a}_{\sigma,q})/2$. It is readily seen that this choice for $\Sigma_{c}$ satisfies Eq.\;(\ref{sigma2}), since Green parafermions are realized in the original unconstrained Hilbert space.
In addition, this choice satisfies the requirement that it commutes with both the electron number operator $\hat{N} = {\hat {N}}_\uparrow + {\hat {N}}_\downarrow$ as the generator of the ${\rm U(1)}$ group in the charge sector and the Cartan generator  $\sum_j S^z = ({\hat {N}}_\uparrow - {\hat {N}}_\downarrow)/2$ of the ${\rm SU(2)}$ group in the spin sector, where ${\hat {N}}_\sigma = \sum _x {\hat n}_{x,\sigma}$, with ${\hat n}_{x,\sigma} ={\hat c}^\dagger_{x,\sigma} {\hat c}_{x,\sigma}$ ($\sigma = \;\uparrow$ and $\downarrow$).  Note that  $\sum_{\sigma,q}  (\hat{a}^\dagger_{\sigma,q} \hat{c}_{\sigma,q}+ \hat{c}^\dagger_{\sigma,q} \hat{a}_{\sigma,q})/2$  becomes 
$\sum_{\sigma} \Pi (N_{\sigma,p}+N^\dagger_{\sigma,p})\Pi/2$ after the projection operator $\Pi$ is implemented, where $N_{\sigma,p} = \sum_l n_{\sigma,p, l}$. Obviously,  $\sum_{\sigma} \Pi (N_{\sigma,p}+N^\dagger_{\sigma,p})\Pi/2$ commutes with $\Pi$. Here $\sum_{\sigma} N_{\sigma,p}$ is  the total Green parafermion number operator, namely the sum of all the Green parafermion number operators $n_{\sigma,p, l}$ for fixed $p$. Here both $N_{\sigma,p}$ and $N^\dagger_{\sigma,p}$ only involve fermion degrees of freedom.   Although   $a^*_{\sigma_1, p, l_1}a^*_{\sigma_2,p, l_2} \ldots a^*_{\sigma_m, p,l_m} \vert \Omega_0 \rangle$ are completely different from $a^\dagger_{\sigma_1, p, l_1}a^\dagger_{\sigma_2,p, l_2} \ldots a^\dagger_{\sigma_m, p,l_m} \vert \Omega_0 \rangle$, a tedious but straightforward calculation confirms that any eigenstate of  $\Pi (N_{\sigma,p}+N^\dagger_{\sigma,p})\Pi/2$, with the eigenvalue being $m$, must be a linear combination of   $\Pi (a^*_{\sigma_1, p, l_1}a^*_{\sigma_2,p, l_2} \ldots a^*_{\sigma_m, p,l_m}) \vert \Omega_0 \rangle$ 
in each sector labeled by $m$ for this non-Hermitian realization, where $m=1,2,\ldots,M_p$ (cf.~Appendix~\ref{ssblimit}). 
Mathematically, this follows from the fact that  $\Pi (N_{\sigma,p}+N^\dagger_{\sigma,p})\Pi/2$ is identical to $\Pi {\hat {N}}_\sigma \Pi$, where $\Pi {\hat {N}}_\sigma \Pi$ are related with
the electron number operator $\hat{N} = {\hat {N}}_\uparrow + {\hat {N}}_\downarrow$ as the generator of the ${\rm U(1)}$ group in the charge sector and the Cartan generator  $\sum_j S^z = ({\hat {N}}_\uparrow - {\hat {N}}_\downarrow)/2$ of the ${\rm SU(2)}$ group in the spin sector (cf.~Section~\ref{ssbtype-b}).  Alternatively, it is possible to show this identification by noting that 
 $\hat{c}^\dagger_{x,\sigma}$ ($x \in \Lambda$) may be expressed linearly in terms of  $\hat{a}^\dagger_{q,\sigma}$ ($q \in \mathscr{E}$) and $\hat{b}^\dagger_{u,\sigma}$ ($u \in \mathscr{I}$).

\vspace{10mm}

The construction above is restricted to an atypical partition. However, an extension to a typical partition is straightforward, as follows from our discussion in Subsection~\ref{non-periodic}.  As a result, 
the projected Green parafermion states  $\Pi(a^*_{ p_{\nu_1}, l_{\nu_1}}a^*_{p_{\nu_2}, l_{\nu_2}} \ldots a^*_{p_{\nu_m},l_{\nu_m}})\vert \Omega_0 \rangle$ for an Hermitian realization and $\Pi(a^*_{ p_{\nu_1}, l_{\nu_1}}a^*_{p_{\nu_2}, l_{\nu_2}} \ldots a^*_{p_{\nu_m},l_{\nu_m}})\vert \Omega_0 \rangle$ or  $\Pi(a^\dagger_{ p_{\nu_1}, l_{\nu_1}}a^\dagger_{p_{\nu_2}, l_{\nu_2}} \ldots a^\dagger_{p_{\nu_m},l_{\nu_m}})\vert \Omega_0 \rangle$ for a non-Hermitian realization play the same role as their counterparts for an atypical partition, when an emergent unit cell consists of $p$ adjacent lattice unit cells.

As we shall argue below, if auxiliary Majorana fermions are physical, then Green parafermions  emerge as {\it real} composite particles only in one specific partition.  For an atypical partition, Green parafermion states  (in the momentum space representation) (up to a projection operator) are realized as flat-band excitations in a condensed matter system  undergoing SSB with type-B GMs, given that $\mathscr{H}$ is translation-invariant under one lattice unit cell. In contrast, if auxiliary Majorana fermions are fictitious, then Green parafermions  emerge as {\it imaginary} composite particles in every possible partition. For an atypical partition,  Green parafermion states  (in the real and momentum space representations) (up to a projection operator) enable us to demystify many puzzling features in exponentially many degenerate ground states for a condensed matter system  undergoing SSB with type-B GMs.

\section{Real Green parafermions from auxiliary (physical) Majorana fermions}~\label{real}

Now we are ready to elaborate on the emergence of a Green parafermion state as a flat-band excitation in a condensed matter system described by the model Hamiltonian $\mathscr{H}$,  if auxiliary Majorana fermions are physical. In this case,  Green parafermions are {\it real} so that the Hilbert space needs to be augmented. The underlying physics involves SSB with type-B GMs from the symmetry group $G$ to the residual symmetry group $H$, which may be understood in the scenario that SSB is regarded as a limit of explicit symmetry breaking, when an external control parameter $h$ vanishes.
In this scenario, the extra term  $-h \;\Sigma_{c}$ is formally translation-invariant under one lattice unit cell, if it is expressed in terms of spin or fermion degrees of freedom in the original unconstrained Hilbert space $V_0$. However, it explicitly breaks the  translation symmetry  under one lattice unit cell into the translation symmetry under $p$ lattice unit cells if it is rewritten in terms of Green parafermion number operators $n_{\mu, p,k}$ in the constrained Hilbert space $V$, as a result of the presence of auxiliary (physical) Majorana fermions that are arranged  periodically on emergent unit cells for an atypical partition -- a hidden form of explicit symmetry breaking.

\subsection{The projected Green parafermion states as flat-band excitations}

For an atypical partition, Green parafermion states of order $p$ (up to a projection operator $\Pi$) in the momentum space representation, denoted as $\Pi(a^*_{\mu_1, p, k_1}a^*_{\mu_2,p, k_2} \ldots a^*_{\mu_m, p,k_m})\vert \Omega_0 \rangle$ or $\Pi(a^\dagger_{\mu_1, p, k_1}a^\dagger_{\mu_2,p, k_2} \ldots a^\dagger_{\mu_m, p,k_m})\vert \Omega_0 \rangle$,  are realized in a sector labeled by the number of Green parafermion number $m$,  due to the presence of auxiliary (physical) Majorana fermions on periodic emergent unit cells. Note that $m$ may be re-labeled by the eigenvalue(s) of the Cartan generator(s) or by the electron number and the eigenvalue(s) the Cartan generator(s) of the symmetry group $G$. This construction is valid for any $h$, even when $h$ vanishes.  As a consequence,  one may interpret the projected Green parafermion states $\Pi(a^*_{\mu_1, p, k_1}a^*_{\mu_2,p, k_2} \ldots a^*_{\mu_m, p,k_m})\vert \Omega_0 \rangle$ or $\Pi(a^\dagger_{\mu_1, p, k_1}a^\dagger_{\mu_2,p, k_2} \ldots a^\dagger_{\mu_m, p,k_m})\vert \Omega_0 \rangle$ (up to a unitary transformation in each sector labeled by $m$) as flat-band excitations in a condensed matter system described by the model Hamiltonian $\mathscr{H}$, as $h$ tends to be zero (cf.~Appendix~\ref{ssblimit}).  Mathematically, this amounts to stating that the excitation energy $\omega_{\mu, p, k}$ of a Green parafermion in a state labeled by $\mu$, $p$ and $k$ becomes zero, namely we have $\omega_{\mu,p, k}=0$.

We are thus led to conclude that the projected Green parafermion states in the momentum space representation, namely $\Pi(a^*_{\mu_1, p, k_1}a^*_{\mu_2,p, k_2} \ldots a^*_{\mu_m, p,k_m})\vert \Omega_0 \rangle$ or $\Pi (a^\dagger_{\mu_1, p, k_1}a^\dagger_{\mu_2,p, k_2} \ldots a^\dagger_{\mu_m, p,k_m}) \vert \Omega_0 \rangle$, appear as flat-band excitations in a condensed matter system for an Hermitian realization, where $*$ is identical to $\dagger$. In contrast, for a non-Hermitian realization, what is realized in a specific condensed matter system is a flat-band excitation that represents a linear combination of the projected Green parafermion states  $\Pi (a^*_{\sigma_1, p, k_1}a^*_{\sigma_2,p, k_2} \ldots a^*_{\sigma_m, p,k_m}) \vert \Omega_0 \rangle$ 
in each sector labeled by $m$. They are connected to the projected Green parafermion states $\Pi(a^*_{\mu_1, p, l_1}a^*_{\mu_2,p, l_2} \ldots a^*_{\mu_m, p,l_m})\vert \Omega_0 \rangle$ or $\Pi (a^\dagger_{\mu_1, p, l_1}a^\dagger_{\mu_2,p, l_2} \ldots a^\dagger_{\mu_m, p,l_m}) \vert \Omega_0 \rangle$ in the real space representation via the Fourier transformations (cf.~Appendix~\ref{realandmomentum}).

We refer to Appendix~\ref{ssblimit} for a more detailed discussion about the difference between Hermitian and non-Herrmitian realizations. Here we stress that the introduction of such an extra term  $-h \;\Sigma_{c}$ makes it possible to understand how  Green parafermion states emerge in the scenario that SSB is regarded as a limit of explicit symmetry breaking. Physically, the projected Green parafermion states explicitly break the translation symmetry under one lattice unit cell into the translation symmetry under $p$ lattice unit cells for an atypical partition, in addition to continuous SSB from $G$ to $H$. Note that, as far as  the translation symmetry is concerned, this explicit symmetry breaking is in a {\it hidden} form, given that the extra term as well as the model Hamiltonian are translation-invariant under one lattice unit cell.
Here we remark that specifying a model Hamiltonian is not sufficient to specify a condensed matter system. Our discussion above shows that both a model Hamiltonian and its associated Hilbert space for all particles involved are necessary, including auxiliary but physical Majorana fermions.

In this setting, degeneracies arising from the symmetry group $G$ are lifted when $h$ is non-zero,  since $G$ is explicitly broken to the residual symmetry group $H$. This explicit symmetry breaking becomes SSB as $h$ tends to be zero. In contrast, the hidden form of explicit symmetry breaking from the translation symmetry under one lattice unit cell to the translation symmetry under $p$ lattice unit cells survives, even when $h$ vanishes. Physically, this is due to the fact that auxiliary (physical) Majorana fermions remain to be in the Hilbert space, which are periodically arranged on emergent unit cells and act as a medium for local spin or fermion degrees of freedom located at different emergent unit cells to communicate with each other, irrespective of whether or not $h$ is zero.  In addition,
this is consistent with the fact that the introduction of auxiliary (physical) Majorana fermions defined on emergent unit cells for two (atypical or typical) partitions into a condensed matter system described by the same model Hamiltonian $\mathscr{H}$ results in two different systems, even if the extra term $-h \;\Sigma_{c}$ expressed in terms of spin or fermion degrees of freedom takes an identical form.

The presence of auxiliary (physical) Majorana fermions  provides a mechanism that blocks accessibility to any local measurements performed on lattice sites inside an emergent unit cell, in the sense that auxiliary  Majorana fermions $\gamma_\alpha$, together with spin degrees of freedom and auxiliary  Majorana fermions $\gamma_{l,\alpha}$ or fermion degrees of freedom in an emergent unit cell, constitute a {\it composite} particle so that any local measurement performed on any local spin or fermion degrees of freedom inside an emergent unit cell is forbidden. This is due to the fact that this composite particle should be treated as a whole, namely that any local degree of freedom inside a specific emergent unit cell is inaccessible, since any local measurement destroys such a composite particle itself. As such, any local degree of freedom inside a specific emergent unit cell may be interpreted as internal degrees of freedom. In particular, internal symmetry groups may be implemented on emergent unit cells independently, thus leading to subsystem gauge groups. We stress that such an internal symmetry group is drastically different from the conventional internal symmetry groups, because they act on local degrees of freedom located at different lattice unit cells inside an emergent unit cell.   This in turn is relevant to a hierarchical structure at a Hamiltonian level, which reveals a Green parafermion system described by the Hamiltonian $-h \sum_\mu N_{\mu,p}$ behind a condensed matter system undergoing SSB with type-B GMs, if the ground state degeneracies are exponential with system size.
Indeed, there are different ways to organize local spin or fermion degrees of freedom inside an emergent unit cell; one way corresponds to one choice of gauge for a Green parafermion system, and different ways are connected via a gauge transformation, under the condition that {\it only} observables invariant under subsystem gauge transformations are allowed. Otherwise, inaccessibility to internal degrees of freedom is violated.

We now turn to a brief discussion of a possible way to realize Green parafermion states (up to a projection operator) in the real space representation, which are realized in a condensed matter system described by the model Hamiltonian $\mathscr{H}$, if auxiliary Majorana fermions are physical. More precisely, a linear combination of $\Pi(a^*_{\mu_1, p, l_1}a^*_{\mu_2,p, l_2} \ldots a^*_{\mu_m, p,l_m})\vert \Omega_0 \rangle$ or $\Pi (a^\dagger_{\mu_1, p, l_1}a^\dagger_{\mu_2,p, l_2} \ldots a^\dagger_{\mu_m, p,l_m}) \vert \Omega_0 \rangle$ for chosen $\{ \mu \}$ and $\{ l \}$ when $p$ and $m$ is fixed may be realized as a ground state, where $\{ \mu \}$ and $\{ l \}$ denote $\{ \mu_1,\ldots, \mu_n \}$ and $\{ l_1,l_2, \ldots,l_m \}$, with $n$ being the number of Green parafermion fields. One way to achieve this is to introduce inhomogeneities in an external control parameter $h$ such that $h$ becomes $h_{\mu,l}$, since $\Pi(a^*_{\mu_1, p, l_1}a^*_{\mu_2,p, l_2} \ldots a^*_{\mu_m, p,l_m})\vert \Omega_0 \rangle$ or $\Pi (a^\dagger_{\sigma_1, p, l_1}a^\dagger_{\sigma_2,p, l_2} \ldots a^\dagger_{\sigma_m, p,l_m}) \vert \Omega_0 \rangle$ appear as  eigenstates of  $\Pi (n_{\mu,p,l}+n^\dagger_{\mu,p,l})\Pi/2$ for any $l$, not just as  eigenstates of $\Pi(N_{\mu,p}+N^\dagger_{\mu,p})\Pi/2$. As such, one only needs to choose $h_{\mu,l}$ to ensure that a linear combination of $\Pi(a^*_{\mu_1, p, l_1}a^*_{\mu_2,p, l_2} \ldots a^*_{\mu_m, p,l_m})\vert \Omega_0 \rangle$ or $\Pi (a^\dagger_{\sigma_1, p, l_1}a^\dagger_{\sigma_2,p, l_2} \ldots a^\dagger_{\sigma_m, p,l_m}) \vert \Omega_0 \rangle$ for chosen $\{ \mu \}$ and $\{ l \}$ is an eigenstate of  $\Pi (n_{\mu,p,l}+n^\dagger_{\mu,p,l})\Pi/2$ for any $l$ when $p$ and $m$ are fixed. Once this is done,  $h_{\mu,l}$ are slowly varied to ensure that the system remains in the same state, which becomes a ground state of the model Hamiltonian, as $h_{\mu,l}$ vanish for all $\{ \mu \}$ and $\{ l \}$.

\subsection{Flat-band excitations in the ferromagnetic spin-1 biquadratic model and the ferromagnetic $\rm {SU}(2)$ flat-band Tasaki model}

For the ferromagnetic spin-1 biquadratic model  (\ref{hambq}), if auxiliary Majorana fermions are physical, then the projected Green parafermion states  in the momentum space representation, namely $\Pi(a^*_{ p, k_1}a^*_{p, k_2} \ldots a^*_{p,k_m})\vert \Omega_0 \rangle$ ($m=1,2,\ldots,M_p$) emerge as flat-band excitations. We stress that the Hilbert space is now enlarged to be the tensor product of both the subspace for spin degrees of freedom and the subspace for auxiliary (physical) Majorana fermions $\gamma_\alpha$ and $\gamma_{l,\alpha}$, although the model Hamiltonian $\mathscr{H}$ and the extra term  $-h \sum_j S^z$ only involve spin degrees of freedom.

For the ferromagnetic $\rm {SU}(2)$ flat-band Tasaki model (\ref{hamtasaki}), if auxiliary Majorana fermions are physical, then a linear combination of the projected Green parafermion states  in the momentum space representation, namely $\Pi(a^*_{\sigma_1, p, k_1}a^*_{\sigma_2,p, k_2} \ldots a^*_{\sigma_m, p,k_m})\vert \Omega_0 \rangle$ or $\Pi (a^\dagger_{\sigma_1, p, k_1}a^\dagger_{\sigma_2,p, k_2} \ldots a^\dagger_{\sigma_m, p,k_m}) \vert \Omega_0 \rangle$  ($m=1,2,\ldots,M_p$) appear as flat-band excitations. We stress that the Hilbert space is enlarged to be the tensor product of both the subspace for fermion degrees of freedom and the subspace for auxiliary Majorana fermions $\gamma_\alpha$, although the model Hamiltonian (\ref{hamtasaki}) and the extra term $-h\sum_{\sigma,q} (\hat{a}^\dagger_{\sigma,q} \hat{c}_{\sigma,q}+ \hat{c}^\dagger_{\sigma,q} \hat{a}_{\sigma,q})/2$ only involves fermion degrees of freedom. 

In both cases, $M_p$ is different from $M$. This stems from a modified form of the Pauli exclusion principle. Note that this difference is also embodied in the fact that the rank $r$ is different from the order $p$. 
Specifically, for the realization of Green parafermions, defined by the Green components (\ref{aobiquadratic}), in the spin-1 ferromagnetic biquadratic model, we have $r= (p+1)/2$ for odd $p$ and $r= p/2$ for even $p$. For the realization of Green parafermions, defined by the Green components (\ref{aotasaki}), in the ferromagnetic $\rm {SU}(2)$ flat-band Tasaki model, we have $r = p$ if only one Green parafermion field $\phi_\sigma$ ($\sigma = \uparrow$ or $\downarrow$) appears and $r =  (p+1)/2$ for odd $p$ and $r= p/2$ for even $p$ if  two Green parafermion fields $\phi_\sigma$ ($\sigma = \uparrow$ and $\downarrow$) are involved as equally as possible.
We remark that the rank $r$ really depends on the explicit form of this projection operator $\Pi$ (cf. Ref.~\cite{shi-dtmodel} for an example that retains $r$ to be always identical to $p$).

We stress that the rank $r$ does not depend on the nature of auxiliary Majorana fermions for a specific realization in condensed matter. As a result, the ranks for the two illustrative models remain the same for imaginary Green parafermions discussed in Section~\ref{fictitious},  which are formed from auxiliary (fictitious) Majorana fermions and local 
spin  degrees of freedom plus auxiliary Majorana fermions $\gamma_{\mu,l,\alpha}$ or fermion degrees of freedom inside an emergent unit cell. This enables us to construct all highest and generalized highest weight states of a condensed matter system undergoing SSB with type-B GMs, if the ground state degeneracies are exponential under PBCs and OBCs.

\section{Imaginary Green parafermions from auxiliary (fictitious) Majorana fermions}~\label{fictitious}

\subsection{Imaginary Green parafermion states on emergent unit cells}
For imaginary Green parafermions formed from auxiliary (fictitious) Majorana fermions and local spin or fermion degrees of freedom inside emergent unit cells,  no auxiliary (physical) Majorana fermions are involved. The Hilbert space thus only involves spin or fermion degrees of freedom, depending on what model is under investigation. 
As a consequence, auxiliary but {\it fictitious} Majorana fermions are introduced {\it only} as a mathematical device, which are imagined to be defined on emergent unit cells of any sizes for all possible partitions, both atypical and typical, with the only condition that they must be removed eventually from the Hilbert space. This leads to a notable but counter-intuitive conclusion that
the absence of auxiliary (physical) Majorana fermions defined on an emergent unit cell of {\it fixed} size for a specific partition is equivalent to the presence of auxiliary (fictitious) Majorana fermions defined on emergent unit cells of any different sizes for all possible partitions, given the Hilbert space {\it only} include spin or fermion degrees of freedom.  As a result, a (one-dimensional) lattice can be partitioned into a union of emergent unit cells of different sizes in all possible ways, with none of the emergent unit cells intersecting with others. We emphasize that all possible partitions are on the {\it same} footing, regardless of being atypical or typical.  In particular, imaginary Green parafermions of any orders may be introduced on emergent unit cells of any sizes for a typical partition, as described in Subsection~\ref{non-periodic} below.
In a sense, the introduction of auxiliary (fictitious) Majorana fermions into a condensed matter system amounts to introducing  auxiliary (physical) Majorana fermions on an emergent unit cell for every possible partition, but they must be removed from one partition before one is switched to another
partition, in order to ensure that no auxiliary Majorana fermions are left in the Hilbert space.

This counter-intuitive conclusion may be understood from a basic physical requirement for the formation of Green parafermions - internal degrees of freedom are inaccessible to local measurements performed on any lattice sites inside an emergent unit cell.  Here we encounter a subtle situation. For a specific partition, it is legitimate to say that {\it imaginary} Green parafermions are formed from internal spin or fermion degrees of freedom inside an emergent unit cell and accompanied auxiliary  ({\it fictitious}) Majorana fermions.   We stress that it only makes sense to view local degrees of freedom inside an emergent unit cell as internal degrees of freedom in one single specific partition. Thus a given lattice unit cell belongs to different emergent unit cells of different sizes for different partitions, given all possible partitions are on the same footing. 
This stems from the statement that auxiliary (fictitious) Majorana fermions introduced on emergent unit cells for one partition must be removed from the Hilbert space, before one switches to another partition. In other words,  although all possible partitions are allowed for auxiliary (fictitious) Majorana fermions, it only makes sense to speak of internal degrees of freedom for one specific partition. Otherwise, no local spin or fermion degrees of freedom could be regarded as internal degrees of freedom inaccessible to local measurements,  since imaginary Green parafermions formed in one specific partition are destroyed in another partition.

Put differently, auxiliary (fictitious) Majorana fermions are introduced as a mathematical device to facilitate the construction of exponentially many degenerate ground states of the model Hamiltonian $\mathscr{H}$, under the condition that any physical consequence to be drawn from this construction does not explicitly depend on auxiliary (fictitious) Majorana fermions themselves. More precisely, auxiliary (fictitious) Majorana fermions {\it only} offer a mathematical means to construct exponentially many degenerate ground states in such a way that exponentially many highest and generalized highest weight states are well reorganized into  Green parafermion states of any orders (up to a projection operator $\Pi$). This amounts to stating that auxiliary (fictitious) Majorana fermions  may be regarded as a mathematical device to construct (a subset of) highest and generalized highest weight states, which in turn makes it possible to yield all exponentially many degenerate ground states arising from SSB with type-B GMs.

Mathematically, auxiliary (fictitious) Majorana fermions are not included in the Hilbert space for any possible partition, so the projected Green parafermion states $\Pi(a^*_{\mu_1, p, l_1}a^*_{\mu_2,p, l_2} \ldots a^*_{\mu_m, p,l_m})\vert \Omega_0 \rangle$ or $\Pi (a^\dagger_{\sigma_1, p, l_1}a^\dagger_{\sigma_2,p, l_2} \ldots a^\dagger_{\sigma_m, p,l_m}) \vert \Omega_0 \rangle$ may be expanded in terms of monomials constructed from auxiliary Majorana fermions, namely $\gamma_\alpha$ and $\gamma_{l,\alpha}$ for quantum many-body spin systems and $\gamma_\alpha$ for strongly correlated itinerant electron models, if we resort to a specific realization of Green parafermions. Physically, the possibility for this expansion relies on the fact that local degrees of freedom inside an emergent unit cell for any partition are indeed accessible to local measurements,  when one takes all possible partitions into account. In this way no degrees of freedom beyond one lattice unit cell can be regarded as internal degrees of freedom such that any imaginary Green parafermions defined on emergent unit cells of any sizes are destroyed. As a consequence, the coefficients in front of exponentially many (linearly independent)  monomials yield exponentially many highest and generalized highest weight states. That is, they appear to be degenerate ground states of the model Hamiltonian $\mathscr{H}$, once one moves back to the original unconstrained Hilbert space $V_0$.

In addition, a subtle difference between real and imaginary Green parafermions arises, as far as explicit symmetry breaking of the translation symmetry under one lattice unit cell into the translation symmetry under $p$ lattice unit cells is concerned. For auxiliary (fictitious) Majorana fermions, this hidden form of explicit symmetry breaking becomes SSB when an external control parameter $h$ vanishes. This is in sharp contrast to auxiliary (physical) Majorana fermions. For the latter, this explicit symmetry breaking  still survives even when an external control parameter $h$ vanishes,
already mentioned in the previous Subsection.  
As a consequence, we are led to conclude that it is not sufficient to distinguish SSB from explicit symmetry breaking at a Hamiltonian level solely - a novel feature missing in the conventional scenario that SSB is regarded as  a limit of explicit symmetry breaking. Note that, in both cases, explicit symmetry breaking of the symmetry group $G$ becomes continuous SSB, as $h$ tends to be zero.

\subsection{The identification of a state $\vert \eta \rangle$ with imaginary Green parafermion states}

This partial SSB of the translation symmetry under one lattice unit cell accounts for the emergence of atypical  (degenerate) ground states of the model Hamiltonian $\mathscr{H}$ that remain periodic, with period $p$, as long as $p$ divides $L$.  However, not all degenerate ground states constructed from atypical partitions appear to be atypical degenerate ground states, since not all of them are periodic, though all atypical generalized highest weight states are constructed from the projected Green parafermion states for atypical partitions. This in turn justifies why it is necessary to introduce both atypical and typical partitions as different notions from atypical and typical degenerate ground states.

As such, we are led to the identification of a state $\vert \eta \rangle$ with imaginary Green parafermion states   
$a^*_{\sigma_1, p, k_1}a^*_{\sigma_2,p, k_2} \ldots a^*_{\sigma_m, p,k_m} \vert \Omega_0 \rangle$ or $a^\dagger_{\sigma_1, p, k_1}a^\dagger_{\sigma_2,p, k_2} \ldots a^\dagger_{\sigma_m, p,k_m} \vert \Omega_0 \rangle$ for an atypical partition and their counterparts for a typical partition in the momentum space representation. Indeed, it also works to identify $\vert \eta \rangle$ with
imaginary Green parafermion states   
$a^*_{\sigma_1, p, l_1}a^*_{\sigma_2,p, l_2} \ldots a^*_{\sigma_m, p,l_m} \vert \Omega_0 \rangle$ or $a^\dagger_{\sigma_1, p, l_1}a^\dagger_{\sigma_2,p, l_2} \ldots a^\dagger_{\sigma_m, p,l_m} \vert \Omega_0 \rangle$ for an atypical partition and their counterparts for a typical partition in the real space representation. 
This identification reveals a hierarchical structure behind exponentially many highest and generalized highest weight states that appear as degenerate ground states, since it does not necessarily produce all generalized highest weight states. 
Hence it is conceptually demanding to develop some notions to elaborate on different families, into which exponentially many degenerate ground states fall.

Indeed, the usefulness of the identification of a state $\vert \eta \rangle$ with imaginary Green parafermion states for both atypical and typical partitions, in either the momentum or the real space representation, lies in the fact that it attaches a physical meaning to generalized  highest weight states -- a mathematical notion introduced in Ref.~\cite{goldensu3}, in the sense that generalized  highest weight states, which appear as (a subset of) fully factorized degenerate ground states, may be re-interpreted as projected Green parafermion states in the constrained Hilbert space $V$. In addition,
this identification {\it not only} justifies the introduction of another notion -- primary generalized highest weight states, if Green parafermions live in the constrained Hilbert space $V_1$ (cf.~Appendix~\ref{primary}), {\it but also} reveals a hierarchical structure hidden behind fully factorized degenerate ground states, which in turn provides a basis for understanding a hierarchical structure in the ground state subspace. They are defined as a subset of generalized highest weight states derivable from this identification, in addition to all possible highest weight states. In other words, one may define the primary family as  all fully factorized degenerate ground states derivable from  the identification of $\vert \eta \rangle$ with imaginary Green parafermion states, constructed from  imaginary Green parafermions defined on emergent unit cells of any sizes for all possible partitions, both atypical and typical.
In particular, all atypical generalized highest weight states follow from this identification for atypical partitions, which in turn explains the origin of partial SSB of the translation symmetry under one lattice unit cell.

\vspace{10mm}

A few remarks are in order. First, for Green parafermions living in the constrained Hilbert space $V_1$,  not all  generalized highest weight states are primary for quantum many-body spin systems.  Mathematically, this is due to the fact that some local spin states must be projected out by introducing the projection operator $\Pi_1$ when the dimension of the local Hilbert space is not $2^n$  so that it is impossible to accommodate $n$ Green parafermion fields, due to the requirement that  local spin degrees of freedom associated with  one Green parafermion field must anti-commute with each other.
Here we recall that the Pauli matrices anti-commute on the same lattice site, thus explaining why  the dimension of the local Hilbert space must be $2^n$, if one attempts to introduce $n$ Green parafermion fields by resorting to a unitary equivalence between a $2^n$-dimensional local Hilbert space and a vector space accommodating $n$ sets of the Pauli matrices. An example for $n=2$ is seen in the ferromagnetic spin-orbital model~\cite{shi-so}.
This requirement is necessary for realizing Green components in terms of local spin degrees of freedom and auxiliary Majorana fermions. In addition,
generalized highest weight states themselves are defined in the unconstrained Hilbert space $V_0$ so that they are not exhausted by the identification of $\vert \eta \rangle$ with imaginary Green parafermion states, thus leading to a notion --  an emergent subsystem non-invertible symmetry operation tailored to a specific primary family member, which appears as a degenerate ground state. In Appendix~\ref{elitzur}, we have defined emergent subsystem invertible and non-invertible symmetries tailored  to a specific degenerate ground state for a condensed matter system undergoing SSB with type-B GMs.
Indeed, they stems from conceptual developments regarding emergent subsystem invertible  and non-invertible subsystem symmetries, which in turn are connected with an equivalent restatement of the Elitzur theorem (cf.~Appendix~\ref{elitzur}).
As a result, one has to divide  all generalized highest weight states into primary generalized highest weight states and secondary ones, in the sense that the latter are derived from the former via an  emergent subsystem non-invertible symmetry. Second, if Green parafermions live in the constrained Hilbert space $V_1$  for quantum many-body spin systems, then the primary family consists of  primary generalized highest weight states,  in addition to one unique highest weight state. In contrast, if Green parafermions live in the original unconstrained Hilbert space $V_0$, then all highest and generalized highest weight states follow from this identification, in addition to all lowest and generalized lowest weight states. However, a distinction needs to be made between quantum many-body spin systems and strongly correlated itinerant electron systems.  For quantum many-body spin systems, 
primary family consists of only the unique highest weight state.
For strongly correlated itinerant electron systems,
primary family consists of all exponentially many highest weight states.  In both cases, all generalized highest weight states are secondary family members derivable from primary family.  Note that the ferromagnetic spin-orbital model offers a specific example for the former, which  shall be investigated in detail in a forthcoming article~\cite{shi-so}.  Further, there are at least one  highest weight state in each sector by the eigenvalues of the electron number operator and the eigenvalue(s) of the Cartan generators of the semi-simple subgroup of the (non-semi-simple) symmetry group $G$ for a strongly correlated itinerant electron system. A detailed discussion is relegated to Appendix~\ref{primary}.  Third, once all generalized highest weight states are derived, one may produce all degenerate ground states from the action of the generators of the symmetry group $G$ on them.
Fourth, in the thermodynamic limit, atypical degenerate ground states appear as monomerized, dimerized, trimerized and tetramerized states and so on, which appear to be a characteristic feature for this novel type of quantum state of matter~\cite{FMGM,hqzhou,goldensu3,spinorbitalsu4,TypeBtasaki,2dtypeb}.

Here we stress that the remarks above should be regarded as some rules of thumb that may be used to draw a unifying picture behind concrete examples for specific realizations of Green parafermions in condensed matter.
Generically,  the identification of a state  $\vert \eta \rangle$ with imaginary Green parafermion states in the real or momentum space representation enables to classify exponentially many highest and generalized highest weight states into two different families, with only one family being primary  and the other family secondary. This distinction naturally explains the origin of a hierarchical structure in the ground state subspace of a condensed matter system undergoing SSB with type-B GMs, as long as the ground state degeneracies are exponential under PBCs and OBCs. In particular, the secondary family 
members are derived from the primary family members  via emergent subsystem non-invertible symmetries, which may be understood as a result of a novel phenomenon -- the time-reversal symmetry fragmentation, which represents the time-reversal symmetry operation that {\it only} acts on spin or fermion states in a block consisting of  adjacent lattice unit cells. We speculate that the time-reversal symmetry fragmentation
is relevant to the so-called Hilbert space fragmentation, as revealed in Ref.~\cite{moudgalya} for the ferromagnetic spin-1 biquadratic model.

\subsection{A hierarchical structure in the ground state subspace: two illustrative examples}~\label{hierarchical}

Here we mainly focus on Green parafermion states for atypical partitions. However, an  extension to Green parafermion states for typical partitions, as far as their mathematical expressions are concerned. Note that they are necessary for constructing all of the primary family members for condensed matter systems undergoing SSB with type-B GMs, as already mentioned in Subsection~\ref{non-periodic}.

Note that there are two common features for any specific realization of Green parafermions in a condensed matter system undergoing SSB with type-B GMs. One is that the order of Green parafermions involved is identical to the (periodic) emergent unit cell size for an atypical partition. Here we have assumed that $p$ always divides $L$, subject to $p < L$ (measured in terms of the lattice unit cells). The other is that, for a specific realization, the projected Green parafermion states always involve the projection operator $\Pi$.  As already emphasized above, its presence does affect parastatistics, thus leading to a modified form of the Pauli exclusion principle.  Hence it is convenient to introduce the rank $r$, a key ingredient to formalize a modified form of the Pauli exclusion principle for a specific realization of Green parafermions in a condensed matter system (cf. Appendix~\ref{tcr} for more details). 

As already argued in Section~\ref{gscheme}, the absence of auxiliary (physical) Majorana fermions defined on an emergent unit cell of {\it fixed} size for a specific partition is equivalent to the presence of auxiliary (fictitious) Majorana fermions defined on any emergent unit cells of different sizes for all possible partitions.
As a result, the entire (one-dimensional) lattice is partitioned into a union of emergent unit cells of different sizes in all possible ways. Indeed, all possible partitions are on the {\it same} footing, regardless of emergent unit cells being periodic or non-periodic, namely  atypical and typical partitions. Hence even if only spin or fermion degrees of freedom already contained in the Hamiltonian are involved, it is still possible to introduce extra degrees of freedom into the Hilbert space, as long as they are eventually removed.
In other words, the model Hamiltonian (\ref{hambq}) or (\ref{hambq}) remains the same, and the Hilbert space only involves spin or fermion degrees of freedom for auxiliary (fictitious) Majorana fermions, which are defined on any emergent unit cells for a specific partition.

\subsubsection{The ferromagnetic spin-1 biquadratic model}

Recall that the Green parafermion states $a^\dagger_{p, l_1}a^\dagger_{p, l_2} \ldots a^\dagger_{p, l_m} \vert \Omega_0 \rangle$ ($m=1,\ldots,M$)
appear as simultaneous eigenstates of the Green parafermion number operators  $n_{p, l}$ for any $l$, when $p$ is fixed.
Here the creation and annihilation operators $a^*_{p, l}$ and $a_{p, l}$ are realized via the Green Ansatz, with the Green components  $b^*_{p, l,\alpha}$ and $b_{p, l,\alpha}$ in (\ref{aobiquadratic}). As argued in Appendix~\ref{lemma},  the commutativity of  $\Pi n_{p, l} \Pi$  with $\Pi$ leads us to conclude that the projected Green parafermion states 
$ \Pi (a^\dagger_{p, l_1}a^\dagger_{p, l_2} \ldots a^\dagger_{p, l_m}) \vert \Omega_0 \rangle$ ($m=1,\ldots,M_p$) constitute degenerate ground states of the model Hamiltonian (\ref{hambq}). Here the maximum occupation number $M$ for the Green parafermion states becomes the maximum occupation number $M_p$ for the projected Green parafermion states. In fact, this reflects how the presence of the projection operator $\Pi$ affects parastatistics.
Note that  the fully polarized state $\ket{\psi_0} \equiv  \otimes_{j=1}^L \vert +_j \rangle$ becomes the fully polarized state $\ket{\psi_0} \equiv  \otimes_{j=1}^L \vert -_j \rangle$ under the time-reversal symmetry operation $K$, which is the lowest weight state for the uniform ${\rm SU(2)}$ (sub)group. In fact, our construction may be reformulated for the lowest weight state $\ket{\psi_0} \equiv  \otimes_{j=1}^L \vert -_j \rangle$, if we choose  $\Pi_1$ to project out all the states containing $\vert +\rangle_j$  ($j=1,\ldots,L$).
 
All the sectors labeled by $m$ ($m=1,\ldots,M_p$) may be equivalently labeled by the eigenvalue of the Cartan generator $\sum_j S^z$ of the uniform ${\rm SU(2)}$ (sub)group, since it commutes with the total Green parafermion number operator $N_{p}$. So once the identification of $\eta$ with a Green parafermion state of any order $p$ is made, one may move back to the original unconstrained Hilbert space $V_0$.  We are thus led to a set of fully factorized degenerate ground states that are a subset of generalized highest weight states, namely  $|\psi_m^{j_1,j_2,\ldots,j_m}\rangle \equiv S_{j_1}^-S_{j_2}^- \ldots S_{j_m}^- \ket{\psi_0}$,  where $j_1$ is not less than 1, $j_{\beta+1}$ not less than $j_{\beta}+2$ ($\beta =1,2,\ldots,m-2$), and $j_m$  not less than $j_{m-1}+2$, but less than $L$. Note that the conditions  above for $j_{\alpha}$ are nothing but the requirement from a modified form of the Pauli exclusion principle for Green parafermions, due to the presence of the projection operator $\Pi$ (for a detailed discussion on this subtle point, we refer to Appendix~\ref{tcr}). 

In fact, this modified form manifests itself in the rank $r$, which is always equal to the order $p$ for Green parafermions, but not for a specific realization of Green parafermions, due to the presence of a projection operator $\Pi$. Specifically, for this specific realization via the Green Ansatz with the Green components (\ref{aobiquadratic}), we have $r= (p+1)/2$ for odd $p$ and $r= p/2$ for even $p$. 
In particular, for any system size $L$,  Green parafermions of order $p$ appear, as long as $p$ divides $L$, with ordinary fermions corresponding to the order $p=1$ as a special case.  Note that the maximum occupation number $M_p$ depends on the rank $r$, since both of them are relevant to a modified form of the Pauli exclusion principle.
Generically, $p$ is not equal to 1. However, the presence of the projection operator $\Pi$ does affect parastatistics. Even if $L=2$, we have two choices for $p$: $p=1$ or $p=2$. In both cases, we have $r=1$. As a result,  two Green parafermions of order 1 are not allowed to occur in the same state, due to the presence of the projection operator $\Pi$. Hence they behave as constrained ordinary fermions when $p=1$. Physically, this simply follows from the fact that the constrained Hilbert space $V$ specified by the projection operator $\Pi$ consists of three states $\vert ++ \rangle$,  $\vert +0 \rangle$ and $\vert 0+ \rangle$, whereas four states are needed to host two ordinary fermions.  Generically, the presence of the projection operator $\Pi$ does affect parastatistics, as discussed in Ref.~\cite{shi-dtmodel}, where a concrete realization of Green parafermion in condensed matter is constructed, with the projection operator $\Pi_2$ being trivial, namely the identity operator.

As discussed in Appendix~\ref{primary}, we are capable of identifying the primary family as a subset of  highest and generalized highest weight states, which appear to be fully factorized degenerate ground states. More precisely, the primary family includes all  primary generalized highest weight states, in addition to the unique highest weight state,  namely the fully polarized state  $\ket{\psi_0} \equiv  \otimes_{j=1}^L \vert +_j \rangle$. By definition, 
primary generalized highest weight states are those derived from the identification of a state $\vert \eta \rangle$ with imaginary Green parafermion states. As it turns out, the  primary generalized highest weight states consists of all fully factorized (degenerate) ground states $|\Psi_0\rangle$  consisting of $\vert + \rangle_j$ and $\vert 0 \rangle_j$, as long as those states containing the local states $\vert 00 \rangle_{j,j+1}$ on the two adjacent lattice sites are excluded. In other words, not all generalized highest weight states are necessarily  the primary family members, thus leading to a notion - the primary generalized highest weight states. In fact, all the secondary generalized highest weight states follow from emergent subsystem non-invertible symmetries tailored to a specific primary generalized highest weight state, which in turn stems from the Elitzur theorem~\cite{elitzur} stating that no local gauge symmetry is spontaneously broken. For a detailed discussion on a connection between an emergent subsystem (invertible and non-invertible) symmetry operation tailored to a specific degenerate ground state and the Elitzur theorem, we refer to Appendix~\ref{elitzur}.
As a consequence, all generalized highest weight states, which appear as fully factorized  degenerate ground states, are derived. This in turn leads to all degenerate ground states, if the action of the generator(s) of the symmetry group $G$ on  all generalized highest weight states is carried out~\cite{goldensu3,dimertrimer,jesse}. Here $G$ is the staggered ${\rm SU(3)}$ symmetry if $L$ is even and  the uniform ${\rm SU(2)}$ symmetry if $L$ is odd.

Here we emphasize that  an emergent subsystem non-invertible symmetry tailored to a specific  primary generalized highest weight state may be regarded as a {\it fragmented}  time-reversal symmetry operation,  which {\it only} acts on all spins between any two lattice sites in the local states $\vert 0 \rangle_j$. In a sense, this fragmented  time-reversal symmetry operation might provide a novel perspective for understanding the so-called Hilbert space fragmentation, as revealed in Ref.~\cite{moudgalya} for the ferromagnetic spin-1 biquadratic model. In addition, there are also  other emergent subsystem  non-invertible symmetries tailored to a subset of the entire ground state subspace. They may be interpreted as the ladder operators, including the raising and lowering operators, thus offering an alternative means to construct all generalized highest weight states successively, as explained in Appendix~\ref{primary}.
 
\subsubsection{The ferromagnetic $\rm {SU}(2)$ flat-band Tasaki model}

Recall that the Green parafermion states  $a^\dagger_{\sigma_1, p, l_1}a^\dagger_{\sigma_2, p, l_2} \ldots a^\dagger_{\sigma_m, p,l_m} \vert \Omega_0 \rangle$ ($m=1,\ldots,M$) appear as simultaneous right eigenstates of $n^\dagger_{\sigma,p, l}$  and the
Green parafermion states  $a^*_{\sigma_1, p, l_1}a^*_{\sigma_2, p, l_2} \ldots a^*_{\sigma_m, p,l_m} \vert \Omega_0 \rangle$ ($m=1,\ldots,M$) as simultaneous right eigenstates of $n_{\sigma,p, l}$ for any $\sigma$ and $l$, when $p$ is fixed.
Here the creation and annihilation operators $a^*_{\sigma,p, l}$ and $a_{\sigma,p, l}$ are realized via the Green Ansatz, with the Green components  $b^*_{\sigma,p, l,\alpha}$ and $b_{\sigma,p, l,\alpha}$ in (\ref{aotasaki}).
As argued in Appendix~\ref{lemma},  the commutativity of  $\Pi n_{\sigma,p, l} \Pi$  with $\Pi$ leads us to conclude that the projected Green parafermion states 
$\Pi (a^\dagger_{\sigma_1, p, l_1}a^\dagger_{\sigma_2, p, l_2} \ldots a^\dagger_{\sigma_m, p,l_m})\vert \Omega_0  \rangle$ or 
$\Pi (a^*_{\sigma_1, p, l_1}a^*_{\sigma_2, p, l_2} \ldots a^*_{\sigma_m, p,l_m}) \vert \Omega_0 \rangle$ 
($m=1,\ldots,M_p$) constitute fully factorized degenerate ground states of the model Hamiltonian (\ref{hamtasaki}). Here the maximum occupation number $M$ for the Green parafermion states becomes the maximum occupation number $M_p$ for the projected Green parafermion states. In fact, this reflects how the presence of the projection operator $\Pi$ affects parastatistics.

All the sectors labeled by $m_\uparrow$ and $m_\downarrow$ ($m=m_\uparrow + m_\downarrow$) are equivalently labeled by  the eigenvalues of the electron number operator $\hat{N} = {\hat {N}}_\uparrow + {\hat {N}}_\downarrow$ of the Abelian subgroup ${\rm U(1)}$ and the Cartan generator  $\sum_j S^z$ of the simple subgroup ${\rm SU(2)}$ of the non-semi-simple symmetry group ${\rm U(1)} \times {\rm SU(2)}$, since they commute with the Green parafermion number operators $N_{\sigma,p}$ ($\sigma = \uparrow$ and $\downarrow$).  Once the identification of a state $\eta$ with a Green parafermion state of any order $p$ is made, one may move
back to the original unconstrained Hilbert space $V_0$. We are thus led to a set of fully factorized degenerate ground states
$\hat{a}_{q_1,\sigma_1}^\dagger  \hat{a}_{q_2,\sigma_2}^\dagger \ldots \hat{a}_{q_m,\sigma_m}^\dagger |\otimes_{x \in \Lambda} 0_x\rangle$,
where $(q_1,\sigma_1), \ldots, (q_m,\sigma_m)$ are subject to the condition that $\sigma_\beta \neq {\bar \sigma}_\gamma$, if $|q_\gamma - q_\beta| =0$  or 1 for any $\beta, \gamma \in \{1,\ldots,m\}$.
They act as the highest weight states in the sector when all spins are up, namely $\sigma_\beta = \uparrow$ for any $\beta \in \{1,\ldots,m\}$, as the lowest weight states in the sector  when all spins are down, namely $\sigma_\beta = \downarrow$ for any $\beta \in \{1,\ldots,m\}$, and as generalized highest weight states in the sector when $m_\uparrow$ spins are up and $m_\downarrow$ spins are down. In other words, any local states with
both spin up and spin down are not allowed on the same site and on the two adjacent sites in the external sublattice. In particular, the conditions  above for $q_{\beta}$ are nothing but the requirement from a modified form of the Pauli exclusion principle for Green parafermions, due to the presence of the projection operator $\Pi$ (for a detailed discussion on this subtle point, we refer to Appendix~\ref{tcr}).

In fact, this modified form manifests itself in the rank $r$. Indeed, the rank $r$ is always equal to the order $p$ for Green parafermions, but not for a specific realization of Green parafermions, due to the presence of the projection operator $\Pi$. Specifically,  for the realization of Green paragermions via the Green Ansatz, with the Green components (\ref{aotasaki}), we have $r = p$ if only one of the two Green parafermion fields $\phi_\sigma$ ($\sigma = \uparrow$ or $\downarrow$) appears and $r =  (p+1)/2$ for odd $p$ and $r= p/2$ for even $p$ if  two Green parafermion fields $\phi_\uparrow$ and $\phi_\downarrow$ are involved. Note that $m$ ranges from 1 to $M_p$, where the maximum occupation number $M_p$ depends on the rank $r$.

Here we remark that the creation and annihilation operators  $a^*_{k, \downarrow}$ and $a_{k,\downarrow}$  are time-reversed counterparts of 
the creation and annihilation operators $a^*_{k, \uparrow}$ and $a_{k,\uparrow}$, since they  map into each other under the time-reversal symmetry operation $K$.    Indeed, the emergence of such a subsystem symmetry operation tailored to a specific degenerate ground state in turn stems from an equivalent restatement of the Elitzur theorem~\cite{elitzur}, as discussed in Appendix~\ref{elitzur}.

In Appendix~\ref{primary}, it is found that the primary family consists of all the highest weight states for the ferromagnetic $\rm {SU}(2)$ flat-band Tasaki model. The secondary family follow from the primary family members by acting emergent subsystem non-invertible symmetries on them, thus yielding all fully factorized  degenerate ground state. Once this is done, all exponentially many degenerate ground states may be generated from all highest and generalized highest weight states, if one takes the symmetry group into account~\cite{TypeBtasaki}. Actually, the total number of degenerate ground states has been counted. As a result, the ground state degeneracies behave asymptotically as the golden spiral, which in turn reflects an intrinsic abstract fractal underlying the ground state subspace~\cite{hqzhou,2dtypeb}.
In particular, the unique ground state (modulo those generated from the action of the lowering operator of the symmetry group $\rm{SU(2)}$ in the spin sector) at quarter filling is reached when $m=L$. This unique ground state is nothing but a Green parafermion state with the maximum occupation number allowed by a modified form of the Pauli exclusion principle for Green parafermions (cf. Appendix~\ref{tcr}). In this sense, the saturated ferromagnetism at quarter filling~\cite{tasakibook} arises from  a modified form of the Pauli exclusion principle for Green parafermions, due to the presence of the projection operator $\Pi$ (cf. Appendix~\ref{tcr}).  Here we remark that 
$\hat{a}_{q_1,\sigma_1}^\dagger  \hat{a}_{q_2,\sigma_2}^\dagger \ldots \hat{a}_{q_m,\sigma_m}^\dagger |\otimes_{x \in \Lambda} 0_x\rangle$ are formally fully factorized, though they are generically entangled. This is particularly so at quarter filling when $m=L$. Meanwhile, a sophisticated understanding of the rich physics underlying the model Hamiltonian (\ref{hamtasaki}) requires to go beyond quarter filling.
 
In addition, one may identify two emergent subsystem  non-invertible symmetries that act as the  ladder operators, namely raising and lowering operators, to generate all  highest and generalized highest weight states successively. However, the lowering  operators are not the Hermitian conjugation of the lowering operators (cf. Appendix~\ref{primary}). This reflects the fact that a realization of Green parafermions in the ferromagnetic $\rm {SU}(2)$ flat-band Tasaki model is non-Hermitian.

 \vspace{10mm}

Our generic scheme offers a natural explanation for the fact that  both exponentially many generalized highest weight states in quantum many-body spin systems and
exponentially many  highest and generalized highest weight states in strongly correlated itinerant electron models stem from their connection with imaginary Green parafermion states, given that this connection holds for both atypical and typical partition (cf.~Subsection~\ref{non-periodic} below).   In other words, our construction offers a physical interpretation for the origin of generalized highest weight states, in the sense that primary generalized highest weight states originate from the identification of a state $\vert \eta \rangle$ with imaginary Green parafermion states.
As already stressed above, all auxiliary ( fictitious) Majorana fermions have to be removed from the Hilbert space eventually, this implies that imaginary Green parafermions {\it only} manifest themselves in some theoretical predictions, including non-zero residual entropy, emergence of atypical degenerate ground states that are translation-invariant under $p$ adjacent lattice unit cells and some restrictive conditions for a fully factorized state to be a degenerate ground state, which may be traced back to a modified form of the Pauli exclusion principle for Green parafermions.

Although we have focused on Green parafermions on periodic emergent unit cells for an atypical partition, we regularly refer to Green parafermions on non-periodic emergent unit cells in various places. Indeed, in order to reveal a hierarchical structure behind exponentially many (fully factorized) degenerate ground states, it is required to extend  the identification of a state $\vert \eta \rangle$ with imaginary Green parafermion states to any generic partition, irrespective of being atypical or typical. As a result, for a typical partition,
the projected Green parafermion states  $\Pi(a^*_{\mu, p_{\nu_1}, l_{\nu_1}}a^*_{p_{\mu,\nu_2}, l_{\nu_2}} \ldots a^*_{\mu,p_{\nu_m},l_{\nu_m}})\vert \Omega_0 \rangle$ for an Hermitian realization and $\Pi(a^*_{\mu, p_{\nu_1}, l_{\nu_1}}a^*_{\mu,p_{\nu_2}, l_{\nu_2}} \ldots a^*_{\mu,p_{\nu_m},l_{\nu_m}})\vert \Omega_0 \rangle$ or  $\Pi(a^\dagger_{\mu, p_{\nu_1}, l_{\nu_1}}a^\dagger_{\mu,p_{\nu_2}, l_{\nu_2}} \ldots a^\dagger_{\mu,p_{\nu_m},l_{\nu_m}})\vert \Omega_0 \rangle$ for a non-Hermitian realization play the same role as their counterparts for an atypical partition. The only difference lies in the fact that Green parafermions on non-periodic emergent unit cells concern different species of Green parafermions, which fall into the B subset according to a classification scheme of Green parafields by Greenberg and Messiah~\cite{greenberg}. 
 
In addition, the ground state degeneracies under both PBCs and OBCs behave asymptotically as the golden spiral for both the ferromagnetic spin-1  biquadratic model~\cite{goldensu3} and the ferromagnetic $\rm {SU}(2)$ flat-band Tasaki model~\cite{TypeBtasaki}, which in turn reflects an intrinsic abstract fractal underlying the ground state subspace~\cite{FMGM,hqzhou}. Actually, the golden ratio manifests itself in the total number of the primary family members for the ferromagnetic spin-1 biquadratic model~\cite{shiqq}. In general, the ground state degeneracies under both PBCs and OBCs behave asymptotically as self-similar geometric objects, with the golden spiral as the simplest example, for many condensed matter systems undergoing SSB with type-B GMs~\cite{goldensu3,spinorbitalsu4,dimertrimer,finitesize,TypeBtasaki,jesse}.   As a consequence, a surprising correspondence between Green parafermions and  self-similar geometric objects, such as the golden spiral, is revealed.

\section{Concluding remarks}~\label{conclusion}

We have shown that Green parafermion states may be realized in a broad class of  condensed matter systems (up to a projection operator), if they undergo SSB  with type-B GMs, as long as the ground state degeneracies are exponential with system size, irrespective of the type of boundary condition adopted. However, a realization of Green parabosons as low-lying excitations remains elusive.

We have chosen the ferromagnetic spin-1  biquadratic model and the ferromagnetic $\rm {SU}(2)$ flat-band Tasaki model as two typical examples to realize real and imaginary Green parafermion states (subject to a projection), if auxiliary Majorana fermions introduced are physical and fictitious.  The construction has been carried out in the scenario that SSB is regarded as a limit of explicit symmetry breaking, with a notable feature that it is not sufficient to distinguish SSB from explicit symmetry breaking at a Hamiltonian level solely. Specifically, an extra term is introduced into the model Hamiltonian so that it explicitly breaks the symmetry group $G$ into the residual symmetry group $H$, but it still keeps the  translation symmetry  under one lattice unit cell. However, once we introduce auxiliary Majorana fermions on emergent unit cells for an atypical partition, then it explicitly breaks the translation symmetry  under one lattice unit cell  into the translation symmetry under $p$ lattice unit cells. 
Physically, the latter arises from the presence of auxiliary Majorana fermions on an emergent unit cell, thus representing a {\it hidden} form of explicit symmetry breaking. It manifests itself in a periodic arrangement of auxiliary (physical or fictitious) Majorana fermions on emergent unit cells, which act as a medium for local spin or fermion degrees of freedom to communicate with each other for an atypical partition, though it does not appear in the model Hamiltonian plus the extra term. 

Indeed,  the construction challenges a folklore that a condensed matter system is completely specified, once the model Hamiltonian is given. In fact, {\it not only} the Hamiltonian {\it but also} the accompanied Hilbert space for all fields involved are necessary for specifying such a condensed matter system under investigation. This salient feature stems from a novel type of interaction that is well beyond any traditional characterization of couplings between different types of particles. Indeed,  this novel type of interaction {\it only} involves communication between spin or fermion degrees of freedom located at different emergent unit cells, with auxiliary Majorana fermions acting as a medium. It is  this {\it weird} nature of auxiliary Majorana fermions that enables us to resolve an apparent contradiction with
a statement by Greenberg and Messiah~\cite{greenberg} that the Green Ansatz for the creation and annihilation operators does not introduce any composite structure,  in the sense that it is linear. Instead, that real and imaginary Green parafermions  are formed from auxiliary (physical and fictitious) Majorana fermion and local spin or fermion degrees of freedom and imaginary Green parafermions.
As a result, instead of one single Hamiltonian in the conventional scenario,  we have to deal with a hierarchical structure at a Hamiltonian level in the scenario that SSB is regarded as a limit of explicit symmetry breaking. Indeed, a Green parafermion system arises, which is invariant under subsystem gauge transformations acting on internal degrees of freedom inside emergent unit cells for a specific partition, if some proper constraints are imposed on the Hilbert space. Here the constraints amounts to introducing projection operators, thus leading to a connection between a condensed matter system undergoing SSB with type-B GMs and a  Green parafermion system.

If auxiliary Majorana fermions are physical, then Green parafermion states or their linear combinations in the momentum space representation  emerge as flat-band excitations (subject to a projection operator). In this case, a hierarchical structure at a Hamiltonian level has been revealed that involves a Green parafermion system and the model Hamiltonian plus the extra term, with the projected Hamiltonian in between. As a result, for a Green parafermion system that is invariant under subsystem gauge transformations, one chosen gauge corresponds to a specific way to organize local spin or fermion degrees of freedom that communicate with each other, with auxiliary (physical) Majorana fermions as a medium.  Once we move back to the model Hamiltonian defined in the  original unconstrained  Hilbert space $V_0$, the invariance under subsystem gauge transformations is lost. Hence different ways to organize local spin or fermion degrees of freedom are physically discernible. In addition, the symmetry group $G$ is spontaneously broken into the residual symmetry group $H$, as an external control parameter $h$ tends to be zero.  Physically, this (continuous) SSB happens in exactly the same way as in the ferromagnetic spin-$1/2$ $\rm {SU}(2)$ Heisenberg model.

On the other hand, if auxiliary Majorana fermions are fictitious, then our construction offers a nice explanation for many puzzling features arising from exponentially many degenerate ground states in condensed matter systems undergoing SSB with type-B GMs:
(i) Type-B GMs as a result of SSB in these condensed matter systems are not {\it fundamental}, in the sense that they may be reduced to exponentially many Green flat bands that emerge from  a realization of Green parafermions. Actually, one may decompose Goldstone flat-band multi-magnon excitations into a linear combination of exponentially many degenerate ground states that are recognized as generalized highest weight states and others derived from them via the symmetry generators for the ferromagnetic spin-1 biquadratic model~\cite{jesse}. In fact, this observation is not surprising, since type-B GMs in such a condensed matter system violate the basic assumption made in Refs.~\cite{watanabe,watanabe1} that low-energy effective field theories only rely on the symmetry group, as already emphasized in Refs.~\cite{goldensu3,spinorbitalsu4,dimertrimer}.
(ii) The even-odd parity effect casts a doubt on the existence of the thermodynamic limit in the ferromagnetic spin-1 biquadratic model. Physically, this stems from the fact that the symmetry group is the staggered $\rm{SU(3)}$  group  for even $L$, in contrast to uniform $\rm {SU}(2)$ for odd $L$. As a consequence, the number of type-B GMs for even $L$ is two, but only one for odd $L$. In contrast, the pattern for exponentially many flat bands from  realizations in terms of Green parafermions remains the same for both even and odd $L$. This implies that the even-odd parity effect as a result of the different numbers of type-B GMs for even and odd $L$ is irrelevant, as far as the existence of the thermodynamic limit is concerned.
(iii) From realizations of Green parafermion in terms of  auxiliary (fictitious) Majorana fermions,  the translation symmetry under one lattice unit cell is explicitly broken to the translation symmetry under $p$ lattice unit cells, if $p$ divides $L$, when an extra term is introduced. However, this explicit symmetry breaking becomes SSB when the external control parameter vanishes. This explains why the translation symmetry under one lattice unit cell is partially spontaneously broken. Given that any emergent unit cells of different sizes coexist, there are atypical degenerate ground states that are monomerized,  dimerized, trimerized, tetramerized and so on in the thermodynamic limit~\cite{goldensu3,spinorbitalsu4,dimertrimer,finitesize,TypeBtasaki,jesse}. 
(iv) Exponentially many degenerate ground states may be traced back to exponentially many highest and generalized highest weight states, which in turn stem from  the identification between $\vert \eta \rangle$ and all the projected Green parafermion states. This identification leads to the primary family among highest and generalized highest weight states.  All other secondary families follow from  the emergent subsystem  non-invertible symmetries (if Green parafermions live in the constrained Hilbert space $V_1$) and the symmetry group  (cf. Appendix~\ref{elitzur} for details about emergent subsystem (invertible and non-invertible) symmetries). As a result, we are capable of reproducing all exponentially many degenerate ground states for a specific model undergoing SSB with type-B GMs. In this sense, our construction offers a physical explanation for the appearance of generalized highest weight states in condensed matter systems undergoing SSB with type-B GMs, if the ground state degeneracies are exponential.
(v) A modified form of the Pauli exclusion principle for Green parafermions accounts for the constraints imposed on  degenerate ground states arising from SSB with type-B GMs. This justifies the introduction of a notion -- the rank for a specific realization of Green parafermions. In particular, the unique ground state up to a  multiplicity due to the $\rm{SU(2)}$ symmetry in the spin sector, constructed by Tasaki~\cite{tasaki,tasakibook} at quarter filling for the ferromagnetic $\rm {SU}(2)$ flat-band Tasaki model, results from the maximum occupation allowed by a modified form of the Pauli exclusion principle for Green parafermions.  An intriguing point, relevant to this modified form  of the Pauli exclusion principle for Green parafermions, is whether or not the presence of a projection operator $\Pi$ always changes  Green  parastatistics of parafermions realized in condensed matter systems. The answer relies on the explicit forms of the two projection operators $\Pi_1$ and $\Pi_2$ for a specific model. As it has been found in Ref.~\cite{shi-dtmodel}, the uniform $\rm {SU}(3)$ spin-1 trimer model  offers an example to demonstrate that this is not always the case, given that the projection operator $\Pi_2$ is the identity operator.

We now turn to the equivalence between the first and second quantization formalisms for Green paraparticles. As shown in Ref.~\cite{stolt},  the first and second quantization theories of Green paraparticles are equivalent, if the order $p$ is finite. Consequently, 
for a specific realization of real Green parafermions in a condensed matter system, the first and second quantization formalisms are  equivalent, as long as the emergent unit cell size $p$ is finite, given that the order $p$ is identical to the emergent unit cell size for an atypical partition.  As a result, real Green parafermions
follow from a high-dimensional representation of the symmetric group $S_N$ ($N=L/p$). 
However,  for a specific realization of imaginary Green parafermions in a condensed matter system,  auxiliary (fictitious) Majorana fermions are not included in the Hilbert space, in contrast to auxiliary (physical) Majorana fermions (cf. Section~\ref{gscheme}).  Hence imaginary Green parafermions are not regarded as composite particles, in contrast to real Green parafermions. An alternative approach to imaginary Green parafermions in the first quantization formalism is briefly discussed in Appendix~\ref{symmetricgroup}. As demonstrated for the ferromagnetic spin-1  biquadratic model and  the ferromagnetic $\rm {SU}(2)$ flat-band Tasaki model,
imaginary Green parafermion states for all possible partitions produce a sequence of wave functions that constitute a high-dimensional representation of the symmetric group $S_L$. Generically, the presence of the projection operator $\Pi$ implies that we have to tackle a subgroup of  the symmetric group $S_L$. Consequently, a modified form of the Pauli exclusion principle is associated with a subgroup of  the symmetric group $S_L$.

For condensed matter systems undergoing SSB with type-B GMs,  the three properties, namely, (1) the exponential ground state degeneracies with system size under both PBCs and OBCs, (2) the existence of emergent (local) symmetry operations tailored to  degenerate ground states, and (3) the emergence of  Goldstone flat bands, are equivalent in the context of SSB with type-B GMs~\cite{jesse}. Here we have implicitly assumed that Goldstone flat bands are equivalent to exponentially many flat bands from the projected Green parferrmiuons. As a consequence, an unexpected connection is revealed between Green parafermions and self-similar geometric objects -- the logarithmic spirals, with the celebrated golden spiral as the simplest example. This stems from the observation that the ground state degeneracies under both PBCs and OBCs behave asymptotically as the golden spiral in the two models under investigation, as long as the system sizes are sufficiently large. Indeed, the golden ratio appears in the non-zero residual entropy for both  the ferromagnetic spin-1 biquadratic model~\cite{goldensu3, saleur, dtmodel,jesse} and the ferromagnetic $\rm {SU}(2)$ flat-band Tasaki model~\cite{TypeBtasaki} (also cf. Refs.~\cite{tasakidegeneracy,tasakidegeneracy1}). As pointed out in Refs.~\cite{goldensu3,TypeBtasaki}, the non-zero residual entropy measures some kind of disorder present in degenerate ground states via a pattern in local spin or fermion states, thus ensuring self-similarities reflecting an abstract fractal underlying the ground state subspace for a condensed matter system under investigation.

Here we have focused on condensed matter systems undergoing SSB with type-B GMs in one spatial dimension. However, it is possible to extend to condensed matter systems on any lattices in two and higher spatial dimensions.  Even in one spatial dimension, there are many other examples to realize Green parafermions. As far as quantum many-body spin models are concerned, we are mainly concerned with a realization of the  creation and annihilation operators for one single Green parafermion field in terms of the spin-1 operators, as appears in the ferromagnetic spin-1 biquadratic model. There is a realization in terms of the spin-$s$ operators in the ferromagnetic spin-s model (biquadratic in the $\rm {SO}(2s+1)$ generators), which undergoes SSB with $2s$ type-B GMs. As shown in Ref.~\cite{barber1}, the model constitutes a representation of the Temperley-Lieb algebra~\cite{tla,baxterbook,martin}. In particular, when $s=3/2$, the local Hilbert space is four-dimensional, so it is unitarily equivalent to the ferromagnetic spin-orbital model, as discussed in Refs.~\cite{spinorbitalsu4,jesse}. This spin-orbital model offers another example for a specific realization of the  creation and annihilation operators for two Green parafermion fields. Moreover, various realizations of  the trilinear commutation relations between the  creation and annihilation operators for the same field and the relative trilinear commutation relations between  the  creation and annihilation operators for different fields naturally appear in condensed matter systems, which involve more than two Green parafermion fields. Here we mention two examples that realize  $n$ Green parafermion fields. One is the  extended spin-orbital model, which possesses staggered  ${\rm SU}(2^n)$ symmetry for even $L$ under PBCs or for any $L$ under OBCs, but only uniform  ${\rm SO}(2^n)$ for odd $L$ under PBCs, which is unitarily equivalent to the ferromagnetic spin-$s$ model with $s=(2^n-1)/2$. This model constitutes a representation of the Temperley-Lieb algebra, so it is exactly solvable by means of the Bethe Ansatz. The other is the ferromagnetic ${\rm SU}(n)$ flat-band  Tasaki model~\cite{wzhang,tamura,tamura1}, which is the variant of the  ${\rm SU}(n)$ Hubbard model on the decorated lattices. 

Last but not least, a natural question concerns the necessity for a condensed matter system to undergo SSB with type-B GMs,  as far as a realization of Green parafermions  is concerned. It appears that a key notion in our conceptual framework is an emergent unit cell as a result of partial SSB of the translation symmetry under one lattice unit cell. An immediate consequence of this partial SSB is the occurrence of exponential many degenerate ground states for a condensed matter system, as long as it is translation-invariant under  one lattice unit cell. Conceptually, this partial (discrete) SSB accompanies continuous SSB from the symmetry group $G$ to the residual symmetry group $H$ when type-B GMs emerge. However, there is no {\it a priori} reason that makes it impossible to separate this partial (discrete) SSB from any continuous SSB. In particular, one may anticipate that  it is possible for this type of partial (discrete) SSB to occur if a condensed matter system does not possess any continuous symmetry group. If so, one may expect that Green parafermions are realizable in such a system. Consequently, if a (translation-invariant) condensed matter system yields exponentially many degenerate ground states, but it does not possess any continuous symmetry group, then it  is a natural candidate for realizing Green parafermions (up to a projection operator). A nontrivial example for such a quantum many-body spin system  is the Kitaev-AKLT model introduced in Ref.~\cite{ganesh}. We shall address this intriguing issue in a forthcoming article~\cite{hqz}.

{\it Acknowledgment.-}  We thank John O. Fj{\ae}restad for his enlightening discussions. We also thank Jesse J. Osborne and Qian-Qian Shi for their collaboration  on a related project~\cite{jesse}.

\newpage
\onecolumngrid
\section*{Appendices}
\twocolumngrid
\setcounter{equation}{0}
\setcounter{section}{0}
\renewcommand{\theequation}{A\arabic{equation}}

\subsection{A model Hamiltonian undergoing SSB with type-B GMs is frustration-free}~\label{frustration-free}

Here we outline a mathematical proof for the statement that a model Hamiltonian $\mathscr{H}$ undergoing SSB  from $G$ to $H$ with type-B GMs is frustration-free, under a reasonable assumption that there is a unique highest weight state $\vert \psi_0 \rangle$ of the symmetry group $G$, which is translation-invariant under one lattice unit cell. Here we have assumed that the symmetry group $G$ is a semi-simple. If the symmetry group $G$ is not semi-simple, then one may restrict to its semi-simple subgroup, as far as SSB with type-B GMs is concerned. 

Before proceeding, we note that the presence of (gapless) type-B GMs stems from the fact that the repeated action of the lowering operators of the symmetry group $G$ on this highest weight state $\vert \psi_0 \rangle$ yields a sequence of degenerate ground states, and all of these degenerate ground states constitute an irreducible representation of the symmetry group $G$. Note that $G$ always contains a uniform subgroup ${\rm SU}(2)$, whose generators are denoted as  $S^\pm$ and $S^z$, with $S^\pm=\sum_j S_j^\pm$ and $S^z=\sum_j S_j^z$, irrespective of $G$ being uniform or staggered, as follows from the Peter-Weyl theorem that asserts the complete reducibility  of unitary representations of any compact Lie group~\cite{hall}. Note that we have applied this theorem to the uniform subgroup ${\rm SU}(2) \subset G$. Hence $\vert \psi_0 \rangle$ must be the highest weight state of the subgroup ${\rm SU}(2)$, with the eigenvalue of $S^z$ being the maximum. In addition, any irreducible representation for the symmetry group $G$, as a unitary representation, is completely reducible for the subgroup ${\rm SU}(2)$, since any compact Lie group is isomorphic to a matrix group~\cite{knapp}. Hence all other degenerate ground states are eigenvectors of $S^z$, with the eigenvalues less than the maximum eigenvalue. We remark that, for a generic symmetry group $G$, some degenerate ground states also act as the highest weight states for irreducible representations of the subgroup ${\rm SU}(2)$, but the eigenvalues of $S^z$ are strictly smaller than the maximum.

The proof consists of the following steps. First,  we note that a model Hamiltonian $\mathscr{H}$ is bounded from below. Hence the introduction of an additive constant in the Hamiltonian ensures that the ground state energy is zero. 
Second, $\mathscr{H} \vert \psi_0 \rangle =0$ is equivalent to $\langle \psi_0 \vert \mathscr{H} \vert \psi_0 \rangle=0$, and  $\langle \psi_0 \vert \mathscr{H} \vert \psi_0 \rangle=0$ is equivalent to  $\langle \psi_0 \vert \mathscr{H}_{j,j+1} \vert \psi_0 \rangle=0$ for all $j$ under PBCs, given that $\vert \psi_0 \rangle$ is translation-invariant.
Third,  $\langle \psi_0 \vert \mathscr{H}_{j,j+1} \vert \psi_0 \rangle=0$ is equivalent to  $\mathscr{H}_{j,j+1} \vert \psi_0 \rangle=0$ for all $j$.
Obviously,   $\langle \psi_0 \vert \mathscr{H}_{j,j+1} \vert \psi_0 \rangle=0$ follows from  $\mathscr{H}_{j,j+1} \vert \psi_0 \rangle=0$ for all $j$. 
However, the converse is a bit involved.
Given $\langle \psi_0 \vert \mathscr{H}_{j,j+1} \vert \psi_0 \rangle=0$, it is readily seen that  $\mathscr{H}_{j,j+1} \vert \psi_0 \rangle$ for all $j$ must be orthogonal to $\vert \psi_0 \rangle$. However, $\mathscr{H}_{j,j+1}$ commute with $S^\pm$ and $S^z$ for all $j$, so  $\mathscr{H}_{j,j+1} \vert \psi_0 \rangle$  is also an eigenvector of $S^z$, with the same maximum eigenvalue. That is, $\mathscr{H}_{j,j+1} \vert \psi_0 \rangle$ is essentially $\vert \psi_0 \rangle$ for all $j$, as a result of the fact that $\vert \psi_0 \rangle=0$ is assumed to be unique as the highest weight state, with the maximum eigenvalue of $S^z$. This is in a contradiction with the fact that $\mathscr{H}_{j,j+1} \vert \psi_0 \rangle$ must be orthogonal to $\vert \psi_0 \rangle$, unless  $\mathscr{H}_{j,j+1} \vert \psi_0 \rangle=0$ for all $j$.
In other words,  there is a unique highest weight state  $\vert \psi_0 \rangle$ for the symmetry group $G$ such that  $\mathscr{H}_{j,j+1} \vert \psi_0 \rangle=0$ for all $j$. This is nothing but the conditions for $\mathscr{H}$ to be frustration-free~\cite{tasakibook}.

Four remarks are in order. First, for a frustration-free Hamiltonian $\mathscr{H}$, it is always possible to decompose it into a sum of the Hamiltonian densities $\mathscr{H}_{j,j+1}$ that are positive semi-definite, namely $\mathscr{H}_{j,j+1} \ge 0$~\cite{tasakibook}. As a result, one may replace $\mathscr{H}_{j,j+1} \equiv \mathscr{H}_{j,j+1}$ by $\tilde {\mathscr{H}}_{j,j+1} \equiv \xi_j \mathscr{H}_{j,j+1} + \epsilon_j$, with $\xi_j$ being positive and  $\epsilon_j$ any real number. Under this replacement, the conditions for $\tilde {\mathscr{H}} = \sum _j \tilde {\mathscr{H}}_{j,j+1}$ to be frustration-free become $\tilde {\mathscr{H}}_{j,j+1} \vert \psi_0 \rangle=\epsilon_j$ for all $j$. This fact has been exploited when the Hilbert space fragmentation was discussed for the ferromagnetic spin-1  biquadratic model~\cite{moudgalya}. 
Second,  for simplicity, the notations we have adopted here are suitable for quantum many-body spin systems, with the lattice unit cell contains only one lattice site.  However, it is straightforward to extend our discussion to more general situations when the lattice unit cell contains more than one lattice site, as long as all the interactions involved are short-ranged. Meanwhile, the argument above may be adapted to strongly correlated electron systems. Third, the assumption regarding the unique translation-invariant highest weight state $\vert \psi_0 \rangle$ of a semi-simple symmetry group $G$ or a semi-simple subgroup of a non-semi-simple symmetry group $G$
is valid for all known condensed matter systems undergoing SSB with type-B GMs, regardless of the ground state degeneracies being polynomial or exponential with system size~\cite{FMGM,hqzhou,2dtypeb,goldensu3,spinorbitalsu4,dimertrimer,finitesize,TypeBtasaki,jesse}.  For the ferromagnetic spin-1  biquadratic model, $\vert \psi_0 \rangle$ is the highest weight state of $\rm {SU}(2)$, namely $\vert \psi_0 \rangle = \vert ++ \ldots + \rangle$ for both odd and even $L$'s, given that the symmetry group $G$ is the uniform $\rm {SU}(2)$ for odd $L$ and the staggered $\rm {SU}(3)$ for even $L$, which contains the uniform $\rm {SU}(2)$ as a subgroup (cf. Section~\ref{ssbtype-b}). For the ferromagnetic $\rm {SU}(2)$ flat-band Tasaki model,  $\vert \psi_0 \rangle$ is the highest weight state of $\rm {SU}(2)$ in the spin sector -- the semi-simple subgroup of the symmetry group $G= \rm {U}(1) \times \rm {SU}(2)$, namely $\vert \psi_0 \rangle = \hat{a}_{q_1,\sigma_1}^\dagger  \hat{a}_{q_2,\sigma_2}^\dagger \ldots \hat{a}_{q_L,\sigma_L}^\dagger |\otimes_{x \in \Lambda} 0_x\rangle$, which is the ground state at quarter filling~\cite{tasakibook}.
Fourth, our argument has been focused on condensed matter systems under PBCs. However, for condensed matter systems under OBCs, one {\it only} needs to replace the condition for the translation invariance of $\vert \psi_0 \rangle$  under one lattice unit cell by the condition that it is invariant under the permutation operation $P_{12} P_{23} \ldots P_{L-1L}$, where $P_{j j+1}$ denote the permutation operators acting on two adjacent lattice unit cells, if the lattice unit cell contains one lattice site. Note that the situation becomes more complicated if the lattice unit cell contains more than one lattice site, in which a more sophisticated treatment is necessary, since there are more than one ways to associate a condensed matter systems under PBCs to the same system under PBCs, as done for the ferromagnetic $\rm {SU}(2)$ flat-band Tasaki model~\cite{TypeBtasaki}.  This condition is valid for all known condensed matter systems undergoing SSB with type-B GMs, if OBCs are adopted~\cite{FMGM,hqzhou,2dtypeb,goldensu3,spinorbitalsu4,dimertrimer,finitesize,TypeBtasaki,jesse}, and has been exploited to construct the matrix-product state representations for degenerate ground states under PBCs and OBCs~\cite{exactmps,TypeBtasaki}. 

\subsection{ A mathematical lemma regarding the combined projection operator $\Pi$}~\label{lemma}

Here we first formalize a mathematical lemma regarding the combined projection operator $\Pi$ for a condensed matter system described by a model Hamiltonian $\mathscr{H}$. Here it only involves spin or fermion degrees of freedom, but undergoes SSB with type-B GMs. This mathematical lemma points a way towards constructing  degenerate ground states arising from SSB with type-B GMs through the explicit construction of the two projection operators $\Pi_1$ and $\Pi_2$. In particular, it offers an efficient means for constructing exponentially many degenerate ground states, although it is also applicable to polynomially many degenerate ground states arising from SSB with type-B GMs.

Consider a specific condensed matter system undergoing SSB with type-B GMs, described by a  Hamiltonian $\mathscr{H}$ that is frustration-free (cf. Appendix~\ref{frustration-free}). As it turns out, the  Hamiltonian $\mathscr{H}$ itself is a sum of local projection operators, up to a multiplicative constant and an additive constant. As a convention, we set the multiplicative constant to be one and the additive constant to be zero. According to this convention, the Hamiltonian $\mathscr{H}$ is positive semi-definite, with the ground state energy being zero. In this sense, we say that $\mathscr{H}$ is in a canonical form. As defined in Section~\ref{projection}, there is a  projection operator $\Pi_1$ that maps the Hamiltonian $\mathscr{H}$ into a projected Hamiltonian $\mathscr{\bar H}$, in the sense that it is defined as a product of local projection operators acting on local Hilbert spaces in a region, which is not necessarily identical to the lattice unit cell. Mathematically, we have
\begin{equation*}
	\mathscr{\bar H} = \Pi _1\mathscr {H} \Pi_1. 
\end{equation*}
The projected Hamiltonian $\mathscr{\bar H}$ is simply a sum of local projection operators in the constrained Hilbert space $V_1$. As a result, another projection operator $\Pi_2$ is introduced such that projection of $\mathscr{\bar H}$ by $\Pi_2$ gives the zero operator.  Mathematically, we have
\begin{equation*}
	\Pi_2 \mathscr{\bar H} \Pi_2 = 0. 
\end{equation*}
Note that $\Pi_2$, by construction, commutes with $\Pi_1$: $\Pi_1 \Pi_2 = \Pi_2 \Pi_1$. Note that the projection operator $\Pi_2$ is simply the identity operator if $\mathscr{\bar H}$ is already the zero operator. Generically, the projection operator $\Pi_2$ is defined to be factorized as a product of local projection operators acting on a local Hilbert space in a region involving more than one lattice unit cell. Generically, both $\Pi_1$ and $\Pi_2$ depend on what types of boundary conditions are adopted, as seen from the ferromagnetic $\rm {SU}(2)$ flat-band Tasaki model in Section~\ref{projection}.
If we define the combined projection operator $\Pi = \Pi_2 \Pi_1$, then the above equation  becomes
\begin{equation}
	\Pi \mathscr {H} \Pi  = 0. \label{comm}
\end{equation}
An intriguing fact is that (\ref{comm}) holds if and only if one of the following two equivalent conditions is valid
\begin{equation}
	\Pi \mathscr {H} =0 \;\;\; {\rm or} \;\;\;\mathscr{H} \Pi = 0. \label{comm1}
\end{equation}
Here we note that $\mathscr{H} \Pi = 0$ is equivalent to $\Pi \mathscr {H} =0$, since they follow from each other by performing the Hermitian conjugation, given both $\Pi$ and $\mathscr {H}$ are Hermitian, namely $\Pi^\dagger= \Pi$ and $\mathscr {H}^\dagger=\mathscr {H}$.
One may thus introduce the constrained Hilbert space $V$ specified by the projection operator $\Pi$, which consists of all vectors $\vert v \rangle$  that satisfy $\Pi\;\vert v \rangle=\vert v \rangle$. The commutativity between the combined projection operator $\Pi$ and $\mathscr {H}$ implies that any state  $\vert v \rangle$  in the constrained Hilbert space $V$ is a degenerate ground state of the model Hamiltonian $\mathscr{H}$, with the ground state energy being zero. In other words, we have $\mathscr{H} \vert v \rangle=0$ for any vector $\vert v \rangle$  in the constrained Hilbert space $V$. 

We turn to the equivalence between (\ref{comm}) and (\ref{comm1}). It is obvious to see that  (\ref{comm}) follows from (\ref{comm1}).
To show the converse, we assume that the opposite is true. That is, $\Pi \mathscr {H} \neq 0$ or equivalently, $\mathscr{H} \Pi \neq 0$, since $\Pi \mathscr {H}$ and $\mathscr{H} \Pi$ are Hermitian conjugated to each other. This implies that there is at least one vector $\vert v_0 \rangle$ such that $\mathscr {H} \Pi \vert v_0 \rangle \neq 0$. In other words, $\Pi \vert v_0 \rangle$ is not a (degenerate) ground state, because $\mathscr {H}$ is in a canonical form. As a consequence, we have $\langle v_0 \vert \Pi \mathscr {H} \Pi \vert v_0 \rangle \neq 0$, thus leading to a contradiction with (\ref{comm}). In addition, the equivalence between $\mathscr{H} \Pi = 0$  and $\Pi \mathscr {H} =0$ implies that $\Pi$ commutes with $ \mathscr {H}$. Hence the fact that any state  $\vert v \rangle$  in the constrained Hilbert space $V$ is a degenerate ground state simply follows from an observation that $\Pi \mathscr{H} \Pi\vert v \rangle =  \mathscr{H} \Pi \vert v \rangle =  \mathscr{H} \vert v \rangle$ and $\Pi \mathscr{H} \Pi = \Pi \mathscr{H} = \mathscr{H} \Pi=0$. Mathematically, we have $\mathscr{H} \vert v \rangle=0$ for any  $\vert v \rangle \in V$.

As a common feature for a condensed matter system undergoing SSB with type-B GMs, the combined projection operator $\Pi$ always maps the model Hamiltonian $\mathscr {H}$ into the zero operator, after the projection operation $\Pi$ acts from either the left-hand side or the right-hand side. Mathematically, any state in the form $\vert v \rangle = \Pi \vert \eta \rangle$ is a degenerate ground state of $\mathscr {H}$, with $ \vert \eta \rangle$ being any vector in the original unconstrained Hilbert space $V_0$. Here we note that if a model Hamiltonian $\mathscr{H}$ is not in a canonical form, then it is proportional to the projection operator $\Pi$ after the two projection operators $\Pi_1$ and $\Pi_2$ are implemented.
Generically, $\Pi \vert \eta \rangle$ yields a nontrivial degenerate ground state of the model Hamiltonian $\mathscr{H} $ if and only if  $ \vert \eta \rangle$ is not orthogonal to the ground state subspace.

As a warm-up exercise, we examine what happens if we exploit this mathematical lemma to investigate polynomially many degenerate ground states arising from SSB with type-B GMs. For this purpose, consider the spin-$1/2$ $\rm {SU}(2)$ ferromagnetic Heisenberg model described by the Hamiltonian
\begin{equation*}
	\mathscr{H}=\sum_{j}\left( \frac{1}{4} - \textbf{S}_j \cdot \textbf{S}_{j+1}\right). 
\end{equation*}
Here $\textbf{S}_j=(S^x_j,S^y_j,S^z_j)$ is the vector of the spin-$1/2$ spin operators at lattice site $j$, respectively. The sum over $j$ is taken from $1$ to $L$ under PBCs. The symmetry group is uniform ${\rm SU(2)}$ group, which has a total of 3 generators.
The SSB pattern is from ${\rm SU(2)}$ to ${\rm U(1)}$, with one type-B GM: $N_B=1$. Indeed, the ground state degeneracy is $L+1$. 
Note that we have included an additive constant $1/4$ in this Hamiltonian to ensure that it is a sum of local projection operators, so it is positive semi-definite. In fact, this model is the simplest example for representations of the Temperley-Lieb algebra with a continuous symmetry group, it is thus exactly solvable by means of the Bethe Ansatz~\cite{barber}. Here we mention that it is not always obvious to identify a frustration-free Hamiltonian as a sum of local projection operators, as it is readily seen from an extension of the ferromagnetic spin-$1/2$ $\rm {SU}(2)$ Heisenberg model  to arbitrary spin $s$.

If we choose $\Pi_1$ to be a projection operator that projects out all the states containing $S^-_j \vert \uparrow~\rangle_{j}$ defined on a region consisting of one lattice site labeled by $j$, then the projected Hamiltonian $\mathscr{\bar H}$ becomes the zero operator. One may simply choose $\Pi_2$ to be the identity operator, we thus have $\Pi = \Pi_1$. As a result, the constrained Hilbert space $V$ is a one-dimensional subspace spanned by the fully polarized state with all spins up, namely the highest weight state for the symmetry group ${\rm SU(2)}$.

An important lesson we have learned from this simple example is that it is impossible to realize Green parafermion states as low-lying excitations in the ferromagnetic quantum spin-$1/2$ $\rm {SU}(2)$  Heisenberg model. Mathematically, this stems from the fact that $L+1$ degenerate ground states are permutation-invariant, in contrast to Green parafermions of order $p$ when $p >1$. Indeed,  if one identifies $\vert \eta \rangle$ as a Green parafermion state, then exponentially many degenerate ground states are produced. This explains why it is necessary to investigate condensed matter systems with exponentially many degenerate ground states. In the main text, we focus on the ferromagnetic spin-1 biquadratic model and the ferromagnetic $\rm {SU}(2)$ flat-band Tasaki model as two illustrative examples. A detailed discussion on an explicit construction of the projection operators $\Pi_1$ and $\Pi_2$ for a few other condensed matter systems may be found in Refs.~\cite{shi-so,shi-dtmodel,shiqq}.

Generically, if we randomly choose  $\vert \eta \rangle$, then $\Pi \vert \eta \rangle$ is a linear combination of many degenerate ground states, given  the dimension of the ground state subspace for such a condensed matter system is exponential with system size. Hence it is theoretically challenging to sort out exponentially many degenerate ground states, even if we take the symmetry group $G$ into account. In this sense, it is highly desirable to develop a mathematical method that enables one to classify exponentially many degenerate ground states into different families. As the first step, one needs to get rid of any extra complication arising from continuous SSB from $G$ to $H$, thus leading to (exponentially many) fully factorized degenerate ground states. Afterwards, a hierarchical structure behind  all fully factorized degenerate ground states may be revealed by resorting to emergent subsystem non-invertible symmetries,  with only one family being primary  and all other fully factorized degenerate ground states secondary. As such, it is sufficient to focus on the primary family. In this regard, both the symmetry group and emergent subsystem (invertible and non-invertible) symmetries play crucial roles, in the sense that they enable us to identify the primary family as a subset of fully factorized degenerate ground states, thus leading to a notion - primary generalized highest weight states (a detailed discussion about primary generalized highest weight states, cf. Appendix~\ref{primary}).

The above mathematical lemma {\it only} relies on the combined projection operator $\Pi$, which is split into  two  projection operators $\Pi_1$ and $\Pi_2$. The necessity for introducing them may be attributed to the fact that the projection operator $\Pi_1$ defines the projected Hamiltonian  $\mathscr{\bar H}$, which in turn leads to the projection operator $\Pi_2$, as a result of the requirement of the commutativity of the projected Hamiltonian  $\mathscr{\bar H}$ with the Green parafermion number operators $n_{\mu, p, k}$  for an Hermitian realization and with the Green parafermion number operators $n_{\mu, p, k}$ or  $n^\dagger_{\mu, p, k}$ for a non-Hermitian realization in the constrained Hilbert space $V$.  It is this physical requirement that leads to a connection between a state  $\vert \eta \rangle$ and Green parafermion states, as established in the scenario that SSB is regarded as a limit of explicit symmetry breaking (cf.~Appendix~\ref{ssblimit} for more details). 

\subsection{ The projection operators $\Pi_1$ and $\Pi_2$ for the two illustrative models: explicit expressions}~\label{pipi}

Our aim is to explicitly construct the projection operators $\Pi_1$ and $\Pi_2$ for the  two illustrative models under investigation. Recall that $\Pi_1$ is, by definition, the projection operator  that decomposes the original (unconstrained) Hilbert space $V_0$  into the direct sum of the constraint Hilbert space $V_1$ and its (orthogonal) complement $V_1^c$ in $V_0$, namely $V_0 = V_1 \oplus V_1^c$. Similarly,  $\Pi_2$ is the projection operator that decomposes $V_1$ into the direct sum of $V$ and its (orthogonal) complement $V^c$ in $V_1$, namely $V_1 = V \oplus V^c$. As a result,  explicit mathematical expressions for the projection operators $\Pi_1$ and $\Pi_2$ may be constructed if all the orthonormal basis states in $V_0$, $V_1$ and $V$ are enumerated. In particular,  they do not depend on any specific choice of the orthonormal basis states, due to their invariance under any unitary transformation from one set of the orthonormal basis states to another set.

\subsubsection{The ferromagnetic spin-1 biquadratic model}

For the ferromagnetic spin-1 biquadratic model,  a set of the orthonormal basis states in the original (unconstrained) Hilbert space $V_0$ may be constructed from the action of $ S^-_j$ ($j=1,2,...,L$) on the highest weight state $\vert \psi_0 \rangle = \vert ++ \ldots + \rangle$. Formally,  the orthonormal basis states take the form $\prod_{j \in A_0} S^-_j \prod_{j \in A_{-1}} (S^-_j)^2 \vert \psi_0 \rangle$, where $ A_0$ and $A_{-1}$ represent two subsets of $\{1,2,\ldots,L\}$, subject to the condition that the intersection of $ A_0$ and $A_{-1}$ is empty, namely $ A_0 \cap A_{-1} = \phi$, where $\phi$ denotes the empty set. The number of these orthonormal basis states may be counted as $\sum_{|A_{-1}|} C_L^{|A_{-1}|} 2^{L-|A_{-1}|}$, where $|A_{-1}|$ denotes the number of elements in $A_{-1}$ and $C_L^{|A_{-1}|}$ denote the binomial coefficients. We thus have
$\sum_{|A_{-1}|} C_L^{|A_{-1}|} 2^{L-|A_{-1}|} = 2^L \sum_{|A_{-1}|} C_L^{|A_{-1}|} 2^{-|A_{-1}|}= 2^L (1+1/2)^L=3^L$, identical to the dimension of the  original (unconstrained) Hilbert space $V_0$. Since $\Pi_1$ is defined to project out all the (orthonormal) states $\prod_{j \in A_0} S^-_j \prod_{j \in A_{-1}} (S^-_j)^2 \vert \psi_0 \rangle$ when $A_{-1} = \phi$, $V_1$ is spanned by  $\prod_{j \in A_0} S^-_j \vert \psi_0 \rangle$, where $A_0$ is any subset of $\{1,2,\ldots,L\}$. Note that the number of these basis states in $V_1$ is $2^L$.  Equivalently, $V_1^c$ is spanned by $\prod_{j \in A_0} S^-_j \prod_{j \in A_{-1}} (S^-_j)^2 \vert \psi_0 \rangle$ as long as  $A_{-1}$ is not empty, with the dimension being $3^L - 2^L$.

For the orthonormal basis states $\prod_{j \in A_0} S^-_j \prod_{j \in A_{-1}} (S^-_j)^2 \vert \psi_0 \rangle$ in the original (unconstrained) Hilbert space $V_0$, the resolution of the identity implies that
\begin{equation}
\sum _{A_0,\; A_{-1}}\prod_{j \in A_0} S^-_j \prod_{j \in A_{-1}} (S^-_j)^2 \vert \psi_0 \rangle \langle \psi_0 \vert  \prod_{j \in A_{-1}} (S^+_j)^2 \prod_{j \in A_0} S^+_j =I_{V_0},\label{unity}
\end{equation}
where the sum  is taken over all possible $A_0$ and $A_{-1}$, subject to the condition that the intersection of $A_0$ and $A_{-1}$ is empty, namely $A_0 \cap A_{-1} = \phi$. Note that the subscript $V_0$ in ``$I_{V_0}$" is used to indicate the identity operator in $V_0$. 

The projection operator $\Pi_1$ takes the form
\begin{equation*}
	\Pi_1 = \sum _{A_0}\prod_{j \in A_0} S^-_j  \vert \psi_0 \rangle \langle \psi_0 \vert \prod_{j \in A_0} S^+_j.
\end{equation*}
As follows from the resolution of the identity (\ref{unity}), we have
\begin{equation*}
	\Pi_1 = I_{V_0} -\sum _{A_0,\; A_{-1} \neq \phi}\prod_{j \in A_0} S^-_j \prod_{j \in A_{-1}} (S^-_j)^2 \vert \psi_0 \rangle \langle \psi_0 \vert  \prod_{j \in A_{-1}} (S^+_j)^2 \prod_{j \in A_0} S^+_j.
\end{equation*}
The above expression for $\Pi_1$ may be rewritten as follows
\begin{equation}
	\Pi_1 = \prod _{A_0,\; A_{-1} \neq \phi} \Pi_{A_0, A_{-1}},
\end{equation}
where
\begin{equation*}
	\Pi_{A_0,\; A_{-1}}=I_{V_0} -\prod_{j \in A_0} S^-_j \prod_{j \in A_{-1}} (S^-_j)^2 \vert \psi_0 \rangle \langle \psi_0 \vert  \prod_{j \in A_{-1}} (S^+_j)^2 \prod_{j \in A_0} S^+_j.
\end{equation*}
The projection operator $\Pi_1$ may thus be expressed as the product of a sequence of the projection operators $\Pi_{A_0,\; A_{-1}}$, with $A_{-1} \neq \phi$, so that all the (orthonormal) basis states in $V_1^c$  are projected out.

As for the projection operator $\Pi_2$ that decomposes $V_1$ into the direct sum of $V$ and its (orthogonal) complement $V^c$ in $V_1$, we need to construct a set of the orthonormal basis states in $V$. Note that $V$ is spanned by  $\prod_{j \in A_0} S^-_j \vert \psi_0 \rangle$, where $A_0$ is any subset of $\{1,2,\ldots,L\}$, but subject to the condition that no two elements in $A_0$ are adjacent to each other. Hence $V$ is decomposed into the direct sum of the sectors, which are labeled by $|A_0|$. The number of the basis states in the sector labeled by $|A_0|$ is $C_{L-|A_0|}^{|A_0|}+C_{L-|A_0|-1}^{|A_0|-1}$~\cite{jesse}.  Thus $\Pi_2$ takes the form
\begin{equation*}
	\Pi_2 = {\sum}' _{A_0}\prod_{j \in A_0} S^-_j  \vert \psi_0 \rangle \langle \psi_0 \vert \prod_{j \in A_0} S^+_j.
\end{equation*}
Here the sum $\sum'$ is taken over all possible $A_0$'s, subject to the condition that no two elements in $A_0$ represent two lattice sites adjacent to each other. As follows from the resolution of the identity in $V_1$, we have
\begin{equation*}
	\Pi_2 = I_{V_1} -{\sum}^{''} _{A_0}\prod_{j \in A_0} S^-_j  \vert \psi_0 \rangle \langle \psi_0 \vert \prod_{j \in A_0} S^+_j,
\end{equation*}
where $I_{V_1}$ denotes the identity operator in $V_1$, and the sum $\sum^{''}$ is taken over all possible $A_0$'s, subject to the condition that at least there are  two elements in $A_0$  representing two lattice sites adjacent to each other.
The above expression for $\Pi_1$ may be rewritten as follows
\begin{equation}
	\Pi_2 = {\prod}^{''} _{A_0} \Pi_{A_0},
\end{equation}
where the product ${\prod}^{''}$ is taken over all possible $A_0$'s, subject to the condition that at least there are  two elements in $A_0$ representing two lattice sites adjacent to each other
\begin{equation*}
	\Pi_{A_0}=  I_{V_1} -\prod_{j \in A_0} S^-_j  \vert \psi_0 \rangle \langle \psi_0 \vert \prod_{j \in A_0} S^+_j.
\end{equation*}
In other words,  $\Pi_2$ may be expressed as the product of a sequence of the projection operators $\Pi_{A_0}$ that project out all the (orthonormal) basis states in $V^c$.

\subsubsection{The ferromagnetic $\rm {SU}(2)$ flat-band Tasaki model}
For the ferromagnetic $\rm {SU}(2)$ flat-band Tasaki model, the situation becomes a bit cumbersome, since the basis states constructed from the action of  ${\hat a}_{q,\sigma}^\dagger$ and  ${\hat b}_{u,\sigma}^\dagger$  on the fermionic Fock vacuum state $|\otimes_{y \in \Lambda} 0_y\rangle$ are not orthogonal, though they are  linearly independent.  Formally,  the basis states $\vert  A_{\uparrow}, A_{\downarrow}, B_{\uparrow}, B_{\downarrow} \rangle$ take the form~\cite{tasakibook}
\begin{eqnarray}
\vert  A_{\uparrow}, A_{\downarrow}, B_{\uparrow}, B_{\downarrow} \rangle &=& \frac{1}{N( A_{\uparrow}, A_{\downarrow}, B_{\uparrow}, B_{\downarrow})} 
\prod_{q \in A_{\uparrow}}  {\hat a}_{q,\uparrow}^\dagger  \prod_{q \in A_{\downarrow}}  {\hat a}_{q,\downarrow}^\dagger \times \nonumber \\
&& \prod_{u \in B_{\uparrow}}  {\hat b}_{u,\uparrow}^\dagger  \prod_{u \in B_{\downarrow}}  {\hat b}_{u,\downarrow}^\dagger |\otimes_{y \in \Lambda} 0_y\rangle, \label{basis-tasaki}
\end{eqnarray}
where $A_{\uparrow}$ and  $A_{\downarrow}$ represent two subsets of $\{ 1,2,\ldots,L\}$ and $B_{\uparrow}$ and  $B_{\downarrow}$ represent two subsets of $\{1/2,3/2,\ldots,L+1/2\}$, and $N( A_{\uparrow}, A_{\downarrow}, B_{\uparrow}, B_{\downarrow})$ is a normalization factor. In contrast to the ferromagnetic spin-1  biquadratic model, there is no constraint imposed on $A_{\uparrow}$  $A_{\downarrow}$ $B_{\uparrow}$ and  $B_{\downarrow}$, in the sense that they are allowed to intersect with each other. The number of these orthonormal basis states may be counted as $\sum_{|A_{\uparrow}|,|A_{\downarrow}|,|B_{\uparrow}|,|B_{\downarrow}|} C_L^{|A_{\uparrow}|} C_L^{|A_{\downarrow}|}C_L^{|B_{\uparrow}|} C_L^{|B_{\downarrow}|} = 2^L \times 2^L \times 2^L \times 2^L= 16^L$, identical to the dimension of the  original (unconstrained) Hilbert space $V_0$.
Given that $\Pi_1$ is defined to project out all the basis states (\ref{basis-tasaki}) when $B_{\uparrow}\neq \phi$ or $B_{\downarrow}\neq \phi$, $V_1$ is spanned by  
\begin{equation}
	\vert  A_{\uparrow}, A_{\downarrow} \rangle = \frac{1}{N( A_{\uparrow}, A_{\downarrow})} 	\prod_{q \in A_{\uparrow}}  {\hat a}_{q,\uparrow}^\dagger  \prod_{q \in A_{\downarrow}}  {\hat a}_{q,\downarrow}^\dagger 
|\otimes_{y \in \Lambda} 0_y\rangle. \label{basis-tasaki-V1}
\end{equation}
Note that the number of these basis states in $V_1$ is $2^L \times 2^L=4^L$.  Equivalently, $V_1^c$ is spanned by all the basis states (\ref{basis-tasaki}) when $B_{\uparrow}\neq \phi$ or $B_{\downarrow}\neq \phi$, with the dimension being $16^L - 4^L$.

The basis states (\ref{basis-tasaki}) and  (\ref{basis-tasaki-V1}) in both $V_0$ and $V_1$ are not always orthogonal to each other, though linearly independent. In this regard, we stress that the basis states (\ref{basis-tasaki}) fall into distinct sectors labeled by $|A_\uparrow|
+|B_\uparrow|$ and $|A_\downarrow|+|B_\downarrow|$, since both ${\hat N}_{\uparrow}$ and ${\hat N}_{\downarrow}$  are conserved. This amounts to decomposing $V_0$ into  the direct sum of the sectors labeled by the eigenvalues $m$ and $\sigma/2$ ($\sigma =\sum_\sigma \sigma_\alpha $) of the electron number operator ${\hat N}= {\hat N}_{\uparrow} +{\hat N}_{\downarrow}$ and the $z$-projection of the total spin $S^z=({\hat N}_{\uparrow} - {\hat N}_{\downarrow})/2$. As a result,  the basis states (\ref{basis-tasaki}) are  orthogonal to each other, if $|A_\uparrow|
+|B_\uparrow|$ or $|A_\downarrow|+|B_\downarrow|$ is different. In particular, the basis states (\ref{basis-tasaki-V1}) in $V_1$ are orthogonal to the basis states $\vert  A_{\uparrow}, A_{\downarrow}, B_{\uparrow}, B_{\downarrow} \rangle$ in $V_1^c$, where either $B_{\uparrow} \neq 0$ or $B_{\downarrow} \neq 0$. Note that the basis states  (\ref{basis-tasaki-V1}) in $V_1$ may be labeled by
$|A_\uparrow|$ and $|A_\downarrow|$, meaning that $V_1$ is decomposed into the direct sum of the sectors labeled by $m$ and $\sigma/2$ ($\sigma =\sum_\sigma \sigma_\alpha $), with $m=|A_\uparrow|+ |A_\downarrow|$ and $\sigma=|A_\uparrow|- |A_\downarrow|$, since both $B_{\uparrow}= \phi$ and $B_{\downarrow} =\phi$.

As we have learned from the ferromagnetic spin-1 biquadratic model, if a set of the orthonormal basis states in $V_0$ are available, then $\Pi_1$ may be expressed as the product of a sequence of the projection operators that project out all the (orthonormal) basis states in $V_1^c$. As a consequence, it is necessary to construct a set of the orthonormal basis states in $V_0$. To this end,
we assume that $\vert \kappa \rangle _{|A_\uparrow|+|B_\uparrow|, |A_\downarrow|+|B_\downarrow|}$ represent a set of the orthonormal basis states in a sector labeled by  $|A_\uparrow|+|B_\uparrow|$ and $|A_\downarrow|+|B_\downarrow|$, where $\kappa=1,2,\ldots, d_{|A_\uparrow|+|B_\uparrow|, |A_\downarrow|+|B_\downarrow|}$, with  $d_{|A_\uparrow|+|B_\uparrow|, |A_\downarrow|+|B_\downarrow|}$ being the dimension of this sector: $d_{|A_\uparrow|+|B_\uparrow|, |A_\downarrow|+|B_\downarrow|}= \sum _{|A_\uparrow|+|B_\uparrow|, |A_\downarrow|+|B_\downarrow|} C_L^{|A_{\uparrow}|} C_L^{|A_{\downarrow}|}C_L^{|B_{\uparrow}|} C_L^{|B_{\downarrow}|}$.  Here the sum $\sum _{A_\uparrow|+|B_\uparrow|, |A_\downarrow|+|B_\downarrow|}$ is restricted to fixed $|A_\uparrow|+|B_\uparrow|$ and $|A_\downarrow|+|B_\downarrow|$. 

The resolution of the identity in $V_0$ thus takes the form
\begin{equation}
	\sum  _{|A_\uparrow|+|B_\uparrow|, |A_\downarrow|+|B_\downarrow|}   \vert \kappa \rangle _{|A_\uparrow|+|B_\uparrow|, |A_\downarrow|+|B_\downarrow|} \;{}_{|A_\uparrow|+|B_\uparrow|, |A_\downarrow|+|B_\downarrow|} \langle \kappa \vert =I_{V_0}. \label{unity-tasaki}
\end{equation}
Since the basis states (\ref{basis-tasaki})  are linearly independent, we have
\begin{equation}
\vert \kappa \rangle _{\{A_\uparrow|+|B_\uparrow|, |A_\downarrow|+|B_\downarrow|\}}= {\sum}'_{A_{\uparrow}, A_{\downarrow}, B_{\uparrow}, B_{\downarrow}}
g^\kappa_{A_{\uparrow}, A_{\downarrow}, B_{\uparrow}, B_{\downarrow}}
\vert  A_{\uparrow}, A_{\downarrow}, B_{\uparrow}, B_{\downarrow} \rangle , \label{basis-tasaki-sector}
\end{equation}
where $g^\kappa_{A_{\uparrow}, A_{\downarrow}, B_{\uparrow}, B_{\downarrow}}$ denote the coefficients yet to be determined and the sum ${\sum}'$ is taken over all possible $A_{\uparrow}, A_{\downarrow}, B_{\uparrow}$ and $B_{\downarrow}$, subject to the condition that both $|A_\uparrow|+|B_\uparrow|$ and $|A_\downarrow|+|B_\downarrow|$ are fixed. In particular, $\vert \kappa \rangle _{\{A_\uparrow|, |A_\downarrow|\}}$, as a special case of $\vert \kappa \rangle _{\{A_\uparrow|+|B_\uparrow|, |A_\downarrow|+|B_\downarrow|\}}$, when $B_\uparrow = \phi$ and $B_\downarrow = \phi$, appear to be a set of the orthogonal basis states in a sector in the decomposition of $V_1$.

The projection operator $\Pi_1$ takes the form 
\begin{equation*}
	\Pi_1 = \sum _{\{|A_\uparrow|,\; |A_\downarrow|\}}\Pi_{\{|A_\uparrow|, |A_\downarrow|\}},
\end{equation*}
where $\Pi_{\{|A_\uparrow|,\;|A_\downarrow|\}}$ are defined as 
\begin{equation*}
	\Pi_{\{|A_\uparrow|,\; |A_\downarrow|\}} = \sum _\kappa \vert \kappa \rangle _{\{|A_\uparrow|, |A_\downarrow|\}} {}_{\{|A_\uparrow|, |A_\downarrow|\}} \langle \kappa \vert.
\end{equation*}
As follows from the resolution of the identity (\ref{unity-tasaki}), we have
\begin{equation*}
	\Pi_1 = I_{V_0} - {\sum}'_{\{A_\uparrow|+|B_\uparrow|, |A_\downarrow|+|B_\downarrow| \}}\Pi_{\{A_\uparrow|+|B_\uparrow|, |A_\downarrow|+|B_\downarrow| \}},
\end{equation*}
with
\begin{equation*}
	\Pi_{\{|A_\uparrow|+|B_\uparrow|, |A_\downarrow|+|B_\downarrow| \}} = \sum _\kappa \vert \kappa \rangle _{\{|A_\uparrow|+|B_\uparrow|, |A_\downarrow|+|B_\downarrow| \}} {}_{\{|A_\uparrow|+|B_\uparrow|, |A_\downarrow|+|B_\downarrow| \}} \langle \kappa \vert.
\end{equation*}
Here the sum ${\sum}'$ is taken over all the sectors satisfying the condition that either $B_\uparrow \neq \phi$ or $B_\downarrow \neq \phi$.

The above expression for $\Pi_1$ may be rewritten as follows
\begin{equation}
	\Pi_1 = {\prod}'_{\{A_\uparrow|+|B_\uparrow|, |A_\downarrow|+|B_\downarrow| \}} \left( I_{V_0} - \Pi_{\{A_\uparrow|+|B_\uparrow|, |A_\downarrow|+|B_\downarrow| \}} \right),
\end{equation}
where the product ${\prod}'$ is taken over all the sectors satisfying the condition that either $B_\uparrow \neq \phi$ or $B_\downarrow \neq \phi$.
In other words, $\Pi_1$ may be expressed as the product of a sequence of the projection operators so that all the (orthonormal) basis states in $V_1^c$
are projected out.

Taking (\ref{basis-tasaki-sector}) into account, $\Pi_{\{|A_\uparrow|+|B_\uparrow|, |A_\downarrow|+|B_\downarrow| \}}$ may be expressed as follows
\begin{eqnarray}
\Pi_{\{A_\uparrow|+|B_\uparrow|, |A_\downarrow|+|B_\downarrow| \}} &=&  {\sum}' _{A_{\uparrow}, A_{\downarrow}, B_{\uparrow}, B_{\downarrow},
A'_{\uparrow}, A'_{\downarrow}, B'_{\uparrow}, B'_{\downarrow}} \times \nonumber\\
&&  w_{A_{\uparrow}, A_{\downarrow}, B_{\uparrow}, B_{\downarrow};A'_{\uparrow}, A'_{\downarrow}, B'_{\uparrow}, B'_{\downarrow}} \times \nonumber\\
&& \vert  A_{\uparrow}, A_{\downarrow}, B_{\uparrow}, B_{\downarrow} \rangle   \langle A'_{\uparrow}, A'_{\downarrow}, B'_{\uparrow}, B'_{\downarrow} \vert,
\end{eqnarray}
with
\begin{equation*}
w_{A_{\uparrow}, A_{\downarrow}, B_{\uparrow}, B_{\downarrow};A'_{\uparrow}, A'_{\downarrow}, B'_{\uparrow}, B'_{\downarrow}} = \sum_\kappa
g^\kappa_{A_{\uparrow}, A_{\downarrow}, B_{\uparrow}, B_{\downarrow}} (g^\kappa_{A'_{\uparrow}, A'_{\downarrow}, B'_{\uparrow}, B'_{\downarrow}})^*,
\end{equation*}
where the sum ${\sum}'$ is taken over all possible $A_{\uparrow}, A_{\downarrow}, B_{\uparrow}$ and $B_{\downarrow}$ and $A'_{\uparrow}, A'_{\downarrow}, B'_{\uparrow}$ and $B'_{\downarrow}$, subject to the conditions that $|A_\uparrow|+|B_\uparrow|= |A'_\uparrow|+|B'_\uparrow|$ and $|A_\downarrow|+|B_\downarrow|=|A'_\downarrow|+|B'_\downarrow|$ are satisfied.

The remaining task is to determine the coefficients $g^\kappa_{A_{\uparrow}, A_{\downarrow}, B_{\uparrow}, B_{\downarrow}}$ in (\ref{basis-tasaki-sector}). For this purpose, it is convenient to reshape $g^\kappa_{A_{\uparrow}, A_{\downarrow}, B_{\uparrow}, B_{\downarrow}}$  into a matrix $\mathscr{G}$, with $\kappa$ labeling the rows and $A_{\uparrow}, A_{\downarrow}, B_{\uparrow}, B_{\downarrow}$ labeling the columns. After reshaping, $w_{A_{\uparrow}, A_{\downarrow}, B_{\uparrow}, B_{\downarrow};A'_{\uparrow}, A'_{\downarrow}, B'_{\uparrow}, B'_{\downarrow}}$ becomes the entries of a matrix $\mathscr{W}$: $\mathscr{W}= \mathscr{G}^\dagger \mathscr{G}$. Performing a singular value decomposition for the coefficient matrix $\mathscr{G}$~\cite{svd}, we have 
\begin{equation*}
\mathscr{G} = \mathscr{U}\mathscr{D}\mathscr{V},
\end{equation*}
where $\mathscr{U}$ is a unitary matrix, with rows and columns labeled by $\kappa$ and $\kappa'$, $\mathscr{V}$ is a unitary matrix, with rows and columns labeled by $A_{\uparrow}, A_{\downarrow}, B_{\uparrow}, B_{\downarrow}$ and $A'_{\uparrow}, A'_{\downarrow}, B'_{\uparrow}, B'_{\downarrow}$, and $\mathscr{D}$ denotes the singular value matrix. We thus have $\mathscr{W}= \mathscr{V}^\dagger \mathscr{D}^2 \mathscr{V}$. Note that the coefficient matrix $\mathscr{G}$ is invertible, so $\mathscr{D}$ is invertible.

In addition, for any state $\vert  A{''}_{\uparrow}, A^{''}_{\downarrow}, B^{''}_{\uparrow}, B^{''}_{\downarrow} \rangle$, we have  $\Pi_{\{|A_\uparrow|+|B_\uparrow|, |A_\downarrow|+|B_\downarrow| \}}\vert  A{''}_{\uparrow}, A^{''}_{\downarrow}, B^{''}_{\uparrow}, B^{''}_{\downarrow} \rangle = \vert  A{''}_{\uparrow}, A^{''}_{\downarrow}, B^{''}_{\uparrow}, B^{''}_{\downarrow} \rangle$, as long as  the conditions $|A^{''}_\uparrow|+|B^{''}_\uparrow|= |A_\uparrow|+|B_\uparrow|$ and $|A^{''}_\downarrow|+|B^{''}_\downarrow|=|A_\downarrow|+|B_\downarrow|$ are satisfied. Mathematically, we have
\begin{eqnarray*}
 &&{\sum}' _{A_{\uparrow}, A_{\downarrow}, B_{\uparrow}, B_{\downarrow},A'_{\uparrow}, A'_{\downarrow}, B'_{\uparrow}, B'_{\downarrow}} 
w_{A_{\uparrow}, A_{\downarrow}, B_{\uparrow}, B_{\downarrow};A'_{\uparrow}, A'_{\downarrow}, B'_{\uparrow}, B'_{\downarrow}} \times \nonumber\\
&&o_{A^{''}_{\uparrow}, A^{''}_{\downarrow}, B^{''}_{\uparrow}, B^{''}_{\downarrow};A'_{\uparrow}, A'_{\downarrow}, B'_{\uparrow}, B'_{\downarrow}} \vert  A_{\uparrow}, A_{\downarrow}, B_{\uparrow}, B_{\downarrow} \rangle = \nonumber\\
	&& \vert  A^{''}_{\uparrow}, A^{''}_{\downarrow}, B^{''}_{\uparrow}, B^{''}_{\downarrow} \rangle,
\end{eqnarray*}
where $o_{A^{''}_{\uparrow}, A^{''}_{\downarrow}, B^{''}_{\uparrow}, B^{''}_{\downarrow};A'_{\uparrow}, A'_{\downarrow}, B'_{\uparrow}, B'_{\downarrow}}$
denotes the overlap between $\vert  A^{''}_{\uparrow}, A^{''}_{\downarrow}, B^{''}_{\uparrow}, B^{''}_{\downarrow} \rangle$ and $\vert A'_{\uparrow}, A'_{\downarrow}, B'_{\uparrow}, B'_{\downarrow} \rangle$, namely
\begin{equation*}
	o_{A^{''}_{\uparrow}, A^{''}_{\downarrow}, B^{''}_{\uparrow}, B^{''}_{\downarrow};A'_{\uparrow}, A'_{\downarrow}, B'_{\uparrow}, B'_{\downarrow}} = \langle A'_{\uparrow}, A'_{\downarrow}, B'_{\uparrow}, B'_{\downarrow} \vert  A^{''}_{\uparrow}, A^{''}_{\downarrow}, B^{''}_{\uparrow}, B^{''}_{\downarrow} \rangle.
\end{equation*}
This implies that 
\begin{eqnarray*}
&&{\sum}' _{A'_{\uparrow}, A'_{\downarrow}, B'_{\uparrow}, B'_{\downarrow}} 
w_{A_{\uparrow}, A_{\downarrow}, B_{\uparrow}, B_{\downarrow};A'_{\uparrow}, A'_{\downarrow}, B'_{\uparrow}, B'_{\downarrow}}
o_{A^{''}_{\uparrow}, A^{''}_{\downarrow}, B^{''}_{\uparrow}, B^{''}_{\downarrow};A'_{\uparrow}, A'_{\downarrow}, B'_{\uparrow}, B'_{\downarrow}} \nonumber \\
&& = \delta _{A_{\uparrow}, A^{''}_{\uparrow}} \delta _{A_{\downarrow}, A^{''}_{\downarrow}}\delta _{B_{\uparrow}, B^{''}_{\uparrow}} \delta _{B_{\downarrow}, B^{''}_{\downarrow}}.
\end{eqnarray*}
After reshaping $	o_{A^{''}_{\uparrow}, A^{''}_{\downarrow}, B^{''}_{\uparrow}, B^{''}_{\downarrow};A'_{\uparrow}, A'_{\downarrow}, B'_{\uparrow}, B'_{\downarrow}}$ into a matrix $\mathscr{O}^\dagger$, we have
\begin{equation*}
	\mathscr{W}\mathscr{O}^\dagger =I,
\end{equation*}
meaning that $\mathscr{W}$ is nothing but the inverse of $\mathscr{O}^\dagger$. As a consequence, we have $\mathscr{O}^\dagger= \mathscr{V}^\dagger \mathscr{D}^{-2} \mathscr{V}$. We are thus led to conclude that the coefficient matrix $\mathscr{G}$ simply follows from the singular value decomposition of the overlapping matrix $\mathscr{O}$. This is due to the fact that both $\mathscr{D}$ and $\mathscr{V}$ may be read off from the singular value decomposition of the overlapping matrix $\mathscr{O}$, and  $\mathscr{U}$ may be taken as an arbitrary unitary matrix, due to the fact that  performing a unitary transformation for the orthogonal basis states  leaves the projection operator   $\Pi_{\{|A_\uparrow|+|B_\uparrow|, |A_\downarrow|+|B_\downarrow| \}}$ intact.

The projection operator $\Pi_2$ decomposes $V_1$ into the direct sum of $V$ and its (orthogonal) complement $V^c$ in $V_1$. 
We turn to the construction of a set of the orthonormal basis states in $V$. Note that $V$ is spanned by  
\begin{equation}
	\vert \vert A_{\uparrow}, A_{\downarrow} \rangle = \frac{1}{N_V( A_{\uparrow}, A_{\downarrow})} 	{\prod}' _{q \in A_{\uparrow}, q' \in A_{\downarrow}}  {\hat a}_{q,\uparrow}^\dagger   {\hat a}_{q',\downarrow}^\dagger 
	|\otimes_{y \in \Lambda} 0_y\rangle, \label{basis-V-tasaki1}
\end{equation}
where the product ${\prod}'$ is over all possible $A_{\uparrow}$ and $A_{\downarrow}$, subject to the conditions that  $A_{\uparrow} \cap A_{\downarrow} =  \phi$ so that they do not share the same lattice unit cell and that $A_{\uparrow} \cup A_{\downarrow}$  does not contain any elements labeling the two lattice unit cells adjacent to each other, and $N_V( A_{\uparrow}, A_{\downarrow})$ denotes a normalization factor. As a consequence, the orthogonal complement $V^c$ in $V_1$ is spanned by
\begin{equation}
	\vert \vert A_{\uparrow}, A_{\downarrow} \rangle = \frac{1}{N_{V^c} ( A_{\uparrow}, A_{\downarrow})} 	{\prod}^{''} _{q \in A_{\uparrow}, q' \in A_{\downarrow}}  {\hat a}_{q,\uparrow}^\dagger   {\hat a}_{q',\downarrow}^\dagger 
	|\otimes_{y \in \Lambda} 0_y\rangle, \label{basis-V-tasaki2}
\end{equation}
where the product ${\prod}^{''}$ is over all possible $A_{\uparrow}$ and $A_{\downarrow}$, if either $A_{\uparrow} \cap A_{\downarrow} \neq  \phi$ so that $A_{\uparrow}$ and $A_{\downarrow}$ share  some lattice unit cells or $A_{\uparrow} \cup A_{\downarrow}$ contains any elements labeling the two lattice unit cells adjacent to each other, and $N_{V^c}( A_{\uparrow}, A_{\downarrow})$ denotes a normalization factor.

As a result, both $V$ and $V^c$ are decomposed into the direct sum of the sectors labeled by $|A_{\uparrow}|$ and $|A_{\downarrow}|$. In other words, for  $V_1 = V \oplus V^c$, there is a finer decomposition into the direct sum of the sectors labeled by $|A_{\uparrow}|$ and $|A_{\downarrow}|$. Generically, this decomposition results in a set of the orthogonal basis states in $V_1$, denoted as $\vert \zeta \rangle _{|A_\uparrow|, |A_\downarrow|}$, which are different from   $\vert \kappa \rangle _{|A_\uparrow|, |A_\downarrow|}$ introduced above.

Assume that $\vert \zeta \rangle _{|A_\uparrow|, |A_\downarrow|}$ represent a set of the orthonormal basis states in a sector labeled by  $|A_\uparrow|$ and $|A_\downarrow|$, where $\zeta=1,2,\ldots, d_{|A_\uparrow|, |A_\downarrow|}$, with  $d_{|A_\uparrow|, |A_\downarrow|}$ being the dimension of this sector. Since the basis states (\ref{basis-V-tasaki1})  and (\ref{basis-V-tasaki2}) are linearly independent, we have
\begin{equation}
	\vert \zeta \rangle _{\{|A_\uparrow|, |A_\downarrow|\}}= {\sum}'_{A_{\uparrow}, A_{\downarrow}}
	g^\zeta_{A_{\uparrow}, A_{\downarrow}}
	\vert \vert A_{\uparrow}, A_{\downarrow} \rangle , \label{basis-tasaki-sector-1}
\end{equation}
where $g^\zeta_{A_{\uparrow}, A_{\downarrow}}$ denote the coefficients yet to be determined and the sum ${\sum}'$ is taken over all possible $A_{\uparrow}$ and $A_{\downarrow}$, subject to the condition that both $|A_\uparrow|$ and $|A_\downarrow|$ are fixed. Hence the projection operator $\Pi_2$ is the sum of the projection operators $\Pi_{\{A_\uparrow|, |A_\downarrow| \}}$ in all the sectors satisfying this condition:
\begin{equation*}
\Pi_2 = \sum_{|A_\uparrow|, |A_\downarrow|} \Pi_{\{|A_\uparrow|, |A_\downarrow| \}}, 
\end{equation*}
with 
\begin{equation*}
	\Pi_{\{|A_\uparrow|, |A_\downarrow| \}} = \sum _\zeta \vert \zeta \rangle _{\{|A_\uparrow|, |A_\downarrow| \}} {}_{\{|A_\uparrow|, |A_\downarrow|\}} \langle \zeta \vert.
\end{equation*}
Taking (\ref{basis-tasaki-sector-1}) into account, we have
\begin{equation*}
	\Pi_{\{ |A_\uparrow|, |A_\downarrow|\}} =  {\sum}' _{A_{\uparrow}, A_{\downarrow},
		A'_{\uparrow}, A'_{\downarrow}} 
     w_{A_{\uparrow}, A_{\downarrow};A'_{\uparrow}, A'_{\downarrow}}
 \vert \vert  A_{\uparrow}, A_{\downarrow} \rangle   \langle A'_{\uparrow}, A'_{\downarrow} \vert \vert,
\end{equation*}
with
\begin{equation*}
	w_{A_{\uparrow}, A_{\downarrow};A'_{\uparrow}, A'_{\downarrow}} = \sum_\zeta
	g^\zeta_{A_{\uparrow}, A_{\downarrow}} (g^\zeta_{A'_{\uparrow}, A'_{\downarrow}})^*,
\end{equation*}
where the sum ${\sum}'$ is taken over all possible $A_{\uparrow}, A_{\downarrow}$ and $A'_{\uparrow}, A'_{\downarrow}$, subject to the condition that $|A_\uparrow|= |A'_\uparrow|$ and $|A_\downarrow|=|A'_\downarrow|$.

We now turn to determining the coefficients $g^\zeta_{A_{\uparrow}, A_{\downarrow}}$ in (\ref{basis-tasaki-sector-1}). By reshaping $g^\zeta_{A_{\uparrow}, A_{\downarrow}}$  into a matrix $\mathscr{G}_A$, with $\zeta$ labeling the rows and $A_{\uparrow}, A_{\downarrow}$ labeling the columns, $w_{A_{\uparrow}, A_{\downarrow};A'_{\uparrow}, A'_{\downarrow}}$ becomes the entries of a matrix $\mathscr{W}_A$: $\mathscr{W}_A= \mathscr{G}_A^\dagger \mathscr{G}_A$. Performing a singular value decomposition for the coefficient matrix $\mathscr{G}_A$~\cite{svd}, we have 
\begin{equation*}
	\mathscr{G}_A = \mathscr{U}_A\mathscr{D}_A\mathscr{V}_A,
\end{equation*}
where $\mathscr{U}_A$ is a unitary matrix, with rows and columns labeled by $\zeta$ and $\zeta'$, $\mathscr{V}_A$ is a unitary matrix, with rows and columns labeled by $A_{\uparrow}, A_{\downarrow}$ and $A'_{\uparrow}, A'_{\downarrow}$, and $\mathscr{D}_A$ denotes the singular value matrix. We thus have $\mathscr{W}_A= \mathscr{V}_A^\dagger \mathscr{D}_A^2 \mathscr{V}_A$. Note that the coefficient matrix $\mathscr{G}_A$ is invertible, so $\mathscr{D}_A$ is invertible.

In addition, for any state $\vert \vert  A{''}_{\uparrow}, A^{''}_{\downarrow} \rangle$, we have  $\Pi_{\{|A_\uparrow|, |A_\downarrow| \}}\vert  A{''}_{\uparrow}, A^{''}_{\downarrow} \rangle = \vert  A{''}_{\uparrow}, A^{''}_{\downarrow} \rangle$, as long as  $|A^{''}_\uparrow|= |A_\uparrow|$ and $|A^{''}_\downarrow|=|A_\downarrow|$. Mathematically, we have
\begin{eqnarray*}
	{\sum}' _{A_{\uparrow}, A_{\downarrow},A'_{\uparrow}, A'_{\downarrow}} 
	w_{A_{\uparrow}, A_{\downarrow};A'_{\uparrow}, A'_{\downarrow}} 
	o_{A^{''}_{\uparrow}, A^{''}_{\downarrow};A'_{\uparrow}, A'_{\downarrow}} \vert \vert  A_{\uparrow}, A_{\downarrow} \rangle = 
	\vert \vert  A^{''}_{\uparrow}, A^{''}_{\downarrow} \rangle,
\end{eqnarray*}
with $o_{A^{''}_{\uparrow}, A^{''}_{\downarrow};A'_{\uparrow}, A'_{\downarrow}}$
denoting the overlap between $\vert \vert A^{''}_{\uparrow}, A^{''}_{\downarrow} \rangle$ and $\vert \vert A'_{\uparrow}, A'_{\downarrow} \rangle$, namely
\begin{equation*}
	o_{A^{''}_{\uparrow}, A^{''}_{\downarrow};A'_{\uparrow}, A'_{\downarrow}} = \langle A'_{\uparrow}, A'_{\downarrow} \vert \vert  A^{''}_{\uparrow}, A^{''}_{\downarrow} \rangle.
\end{equation*}
This implies that 
\begin{eqnarray*}
	{\sum}' _{A'_{\uparrow}, A'_{\downarrow}} 
	w_{A_{\uparrow}, A_{\downarrow};A'_{\uparrow}, A'_{\downarrow}}
	o_{A^{''}_{\uparrow}, A^{''}_{\downarrow};A'_{\uparrow}, A'_{\downarrow}} 
    = \delta _{A_{\uparrow}, A^{''}_{\uparrow}} \delta _{A_{\downarrow}, A^{''}_{\downarrow}}.
\end{eqnarray*}
After reshaping $	o_{A^{''}_{\uparrow}, A^{''}_{\downarrow};A'_{\uparrow}, A'_{\downarrow}}$ into a matrix $\mathscr{O}^\dagger_A$, we have
\begin{equation*}
	\mathscr{W}_A\mathscr{O}^\dagger_A =I.
\end{equation*}
In other words, $\mathscr{W}_A$ is the inverse of $\mathscr{O}^\dagger_A$. As a result, we have $\mathscr{O}^\dagger_A= \mathscr{V}_A^\dagger \mathscr{D}_A^{-2} \mathscr{V}_A$. We are thus led to conclude that the coefficient matrix $\mathscr{G}_A$ simply follows from the singular value decomposition of the overlapping matrix $\mathscr{O}_A$. This is due to the fact that both $\mathscr{D}_A$ and $\mathscr{V}_A$ may be read off from the singular value decomposition of the overlapping matrix $\mathscr{O}_A$, and  $\mathscr{U}_A$ may be taken as an arbitrary unitary matrix, since this amounts to performing a unitary transformation for the orthogonal basis states that leaves the projection operator   $\Pi_{\{|A_\uparrow|, |A_\downarrow|\}}$ intact.

\vspace{10mm}

A few remarks are in order. First, the construction for the ferromagnetic spin-1 biquadratic model  works for any condensed matter system undergoing SSB with type-B GMs, if the basis states generated from the action of the relevant lowering operator(s) on one chosen highest weight state are orthogonal. Generically, distinct sectors may be labeled by the eigenvalues of the Cartan generator(s) of the (semi-simple) symmetry group $G$. Second, the construction for the ferromagnetic $\rm {SU}(2)$ flat-band Tasaki model  works for any condensed matter system undergoing SSB with type-B GMs, if the basis states generated from the action of the relevant creation operators on any chosen highest weight state are not orthogonal. Here distinct sectors may be labeled by the eigenvalues of the electron number operator and the Cartan generator(s) of the (semi-simple) subgroup of the (non-semi-simple) symmetry group $G$.
Note that our approach developed above may be regarded as a special form of the Gram-Schmidt orthogonalization procedure~\cite{sd-orth} exclusively tailored to identifying a set of the orthonormal basis states from a set of the non-orthogonal basis states in all the sectors labeled by the eigenvalues of the electron number operator ${\hat N}$ and the Cartan generator(s) of the semi-simple subgroup of the symmetry group $G$ for strongly correlated itinerant electron models. 
In principle,  one can always turn a given set of linearly independent basis states into a set of orthonormal basis states by resorting to the Gram-Schmidt orthogonalization procedure~\cite{sd-orth}. However, our approach provides a constructive means to demonstrate the existence of the projection operators $\Pi_1$ and $\Pi_2$ in condensed matter systems undergoing SSB with type-B GMs, as long as the ground state degeneracies are exponential. Actually, {\it not only} the existence {\it but also} the uniqueness (up to an irrelevant  unitary transformation) are established for the two projection operators $\Pi_1$ and $\Pi_2$. 
Third, the notations we have exploited here is abstract. However, there is one-to-one correspondence between the notations used here and those in the rest of this article. For instance, for the spin-1 ferromagnetic biquadratic model,  the basis states $\prod_{j \in A_0} S^-_j \vert \psi_0 \rangle$, where $A_0$ is any subset of $\{1,2,\ldots,L\}$, but subject to the condition that no two elements in $A_0$ are adjacent to each other, correspond to $|\psi_m^{j_1,j_2,\ldots,j_m}\rangle \equiv S_{j_1}^-S_{j_2}^- \ldots S_{j_m}^- \ket{\psi_0}$,  where $j_1$ is not less than 1, $j_{\beta+1}$ not less than $j_{\beta}+2$ ($\beta =1,2,\ldots,m-2$), and $j_m$  not less than $j_{m-1}+2$, but less than $L$, if $|A_0|=m$. For the ferromagnetic $\rm {SU}(2)$ flat-band Tasaki model, the basis  states $\vert \vert A_{\uparrow}, A_{\downarrow} \rangle$ in (\ref{basis-V-tasaki1}) correspond to $\hat{a}_{q_1,\sigma_1}^\dagger  \hat{a}_{q_2,\sigma_2}^\dagger \ldots \hat{a}_{q_m,\sigma_m}^\dagger |\otimes_{y \in \Lambda} 0_y\rangle$, where $(q_1,\sigma_1), \ldots, (q_m,\sigma_m)$ are subject to the condition that $\sigma_\alpha \neq {\bar \sigma}_\beta$, if $|q_\gamma - q_\beta| =0$  or 1 for any $\beta,  \gamma \in \{1,\ldots,m\}$.

\subsection{ Trilinear and relative trilinear commutation relations for Green parafermions}~\label{tcr}

We restrict ourselves to Green parafermions, although almost all the discussions are applicable to Green parabosons. 
Historically, the trilinear commutation relations for one single Green parafermion field was first formalized by Green~\cite{green}. Later on, Greenberg and Messiah~\cite{greenberg} extended this theory to  several parafields. As a convention, we introduce  $\phi_\mu$ to denote $n$ Green parafermion fields, where $\mu \in \{ 1,2,\ldots,n\}$. In particular, when $n=1$, it is convenient to drop off the subscript $\mu$. 

According to Greenberg and Messiah~~\cite{greenberg},  for each pair of parafields, there are four distinct subsets:  B for the relative bilinear commutation relations between two different fields; F for the relative bilinear anticommutation relations between two different fields; PB for the relative trilinear anticommutation relations between two different fields;  PF for the relative trilinear commutation relations between two different fields. The last two are allowed only for parafields of equal order $p$, so they are not equivalent to the first two, respectively, except for ordinary fields when $p=1$. 

Our description below is adapted to a specific realization of Green parafermions in condensed matter, mainly for an atypical partition. To this end, we only focus on realizations in the PF subset, so different parafermion fields have the same order $p$. All field creation and annihilation operators satisfy the no-particle conditions and are described by the Green Ansatz, with the Green components obeying bilinear commutation or anticommutation relations. 

For our purpose, the key ingredient  is the defining trilinear commutation relations for one single Green parafermion field and the relative  trilinear commutation relations for any two different parafermion fields, if they fall into the PF subset.
However, some conceptual developments are not avoidable for specific realizations in condensed matter systems, which are embodied in two aspects. First, it appears necessary to allow a non-Hermitian realization of  the trilinear and relative trilinear commutation relations. Second, it is conceptually satisfying to introduce a notion - the rank $r$ for a specific realization of Green parafermions in the presence of the projection operator $\Pi$. As we shall show, the introduction of this notion makes it possible to formalize a modified form of the Pauli exclusion principle for  Green parafermions, due to the presence of the projection operator $\Pi$. Generically, the rank $r$ is different from the order $p$, and  a modified form of the Pauli exclusion principle is the Pauli exclusion principle for Green parafermions.

Introduce the creation and annihilation operators $a^*_{\mu, p,l}$ and $a_{\mu, p,l}$ for Green parafermions  in a state labeled by $\mu$ and $k$.  Here we use as the subscript $\mu$ ($\mu =1,2, \ldots,n$) to label $n$ Green parafermion fields, and $l$ is chosen to label different emergent unit cells in the real space, which will be changed to $k$ in the momentum space (cf.\;Appendix~\ref{realandmomentum}). Note that $p$ has been introduced to indicate the emergent unit cell size. Following Green~\cite{green}, the creation and annihilation operators $a^*_{\mu, p,l}$ and $a_{\mu, p,l}$ for the same field
satisfy the trilinear commutation relations (\ref{tcr2}), in addition to other relations that follow by swapping  $a^*_{\mu,p, l}$ and $a_{\mu, p,l}$ in both sides. Moreover, the trilinear commutation relations (\ref{jacobi}) for $[[a_{\mu, p,l}, a_{\mu, p,l'}]_{-}, a^*_{\mu,p, l''}]_-$ follows from
Jacobi's identity.

We now turn to the relative trilinear commutation relations between different fields. As Greenberg and Messiah stressed~\cite{greenberg}, there is considerable leeway in the choice of trilinear commutation relations between different fields. To narrow down this choice, they have stipulated three restrictive requirements, which lead to a set of relative trilinear commutation relations for different parafermion fields.

We focus on the relative trilinear commutation relations for any two parafermion fields $\phi_\mu$  and $\phi_{\mu'}$, with the total number of trilinear relations  involving $\phi_\mu$ twice and $\phi_{\mu'}$ once being 18. Similarly, there are also 18  trilinear relations  involving $\phi_{\mu'}$ twice and $\phi_\mu$ once. Following  Greenberg and Messiah~\cite{greenberg}, we shall adopt the three requirements as follows. (i) The left-hand side must have the form $[[O_A,O_B]_-,O_C]_-$, and the right-hand side must be linear. (ii) When the internal pair $[O_A,O_B]_-$ refers to the same field, it must commute with $O_C$ if $O_C$ refers to another field. (iii) These relations must be satisfied by ordinary fermion fields. Here $O_A,O_B$ and $O_C$ refer to the creation and annihilation operators $a^*_{\mu,p, l}$ and  $a_{\mu, p,l}$, respectively.

The relative trilinear commutation relations (\ref{tcr3}) involving $\phi_\mu$ twice and $\phi_{\mu'}$  once follow from the requirements (i) and  (ii).
From Jacobi's identity, together with the requirements (i) and (iii), we are led to the relative trilinear commutation relations (\ref{tcr4}).

For a Hermitian realization, we may perform Hermitian conjugation to the above equations to  yield
\begin{align}
	[[a_{\mu,p, l}, a^*_{\mu,p, l'}]_-, a^*_{\mu',p, l^{''}}]_- &= 0, \cr
	[[a^*_{\mu', p,l''}, a_{\mu,p, l}]_-, a^*_{\mu,p, l'}]_- &= 2\delta_{l l^{'}} a^*_{\mu', p,l^{''}},\cr
	[[a^*_{\mu, p,l'}, a^*_{\mu',p, l^{''}}]_-, a_{\mu,p, l}]_- &= - 2\delta_{l l^{'}} a^*_{\mu', p,l^{''}}.
	\label{tcr5}
\end{align}
We assume that the same relations hold for a non-Hermitian realization. In other words, they follow from swapping  $a^*_{\mu,p, l}$ and $a_{\mu, p,l}$ in both sides of the corresponding equations above.
We thus have 18 (relative) trilinear commutation relations in total for an Hermitian realization of Green parafermions when $*$  is identical to  $\dagger$. Further,  we can establish 18 relations involving $\phi_{\mu'}$ twice and $\phi_\mu$ once, if the subscripts $\mu$ and $\mu'$ are exchanged. However, for a non-Hermitian realization, $*$  is different from $\dagger$, so the total number of (relative) trilinear commutation relations should be doubled.

In addition, for an Hermitian realization, one may introduce a commutative set of Green parafermion number operators $n_{\mu,p, l}$ as follows
\begin{equation}
	n_{\mu,p, l} = \frac {1}{2} [a^*_{\mu, p,l}, a_{\mu,p, l}]_{-} {+} \frac{1}{2} p_\mu, \label{nomu}
\end{equation}
where $p_\mu$ represent the orders of Green parafermion fields $\phi_\mu$. The Green parafermion number operators $n_{\mu,p, l}$ satisfy 
\begin{align*}
	[n_{\mu,p, l}, a^*_{\mu,p, l'}]_-  &=  \delta_{l l'} a^*_{\mu, p,l'}, \cr 
    [n_{\mu, p,l}, a_{\mu,p, l'}]_-  &= - \delta_{l l'} a_{\mu, p,l'}.
\end{align*}
We remark that $n_{\mu,p, l}$ commute with each other: $[n_{\mu, p,l}, n_{\mu',p, l'}]_-=0$ for all $\mu, \mu', l$ and $l'$, when $p$ is fixed. However, for a non-Hermitian realization, we also need to introduce $n^\dagger_{\mu,p, l}$ for $a^\dagger_{\mu, p,l}$ and $(a^*_{\mu, p,l})^\dagger$. They satisfy
\begin{align*}
	[n^\dagger_{\mu,p, l}, (a^*_{\mu,p, l'})^\dagger]_-  &=  -\delta_{l l'} (a^*_{\mu, p,l'})^\dagger, \cr 
	[n^\dagger_{\mu, p,l}, a^\dagger_{\mu,p, l'}]_-  &= \delta_{l l'} a^\dagger_{\mu, p,l'}.
\end{align*}
As a result, for a non-Hermitian realization, we have two copies of creation and annihilation operators for each $p$, together with two copies of the Green parafermion number operators. Note that the two copies are Hermitian conjugated to each other.

Assume that for an Hermitian realization, there is a unique parafermion Fock vacuum $\vert \Omega_0 \rangle$: $a_{\mu, p,l} \vert \Omega_0 \rangle=0$ for all $\mu$ and $l$. We are thus led to the no-particle conditions 
\begin{align}
	a_{\mu,p, l} a^*_{\mu,p, l'} \vert \Omega_0 \rangle &= p_\mu \delta_{l l'} \vert \Omega_0 \rangle, \cr
	a_{\mu,p, l} a^*_{\mu',p,l'} \vert \Omega_0 \rangle &= 0, \cr
	a_{\mu',p, l} a^*_{\mu,p, l'} \vert \Omega_0 \rangle &= 0.    \label{noparticle2}
\end{align}
Note that the no-particle conditions (\ref{noparticle2}) can be derived from the (relative) trilinear commutation relations and from the uniqueness of the parafermion Fock vacuum $\vert \Omega_0 \rangle$~\cite{greenberg}. 
Therefore, for any $l$, we have
\begin{equation*}
	p_{\mu}=\langle \Omega_0 \vert a_{\mu,p,l} a_{\mu,p,l}^* \vert \Omega_0 \rangle.
\end{equation*}
As a result, $p_\mu$ may be interpreted as a norm, so it must be a positive integer,
as argued in Ref.~\cite{greenberg}. In fact, $p_\mu$ is independent of $\mu$. We thus have $p_\mu=p$.

For a non-Hermitian realization, we have to tackle both left and right eigenstates for a non-Hermitian operator. We assume that there is a unique parafermion Fock vacuum $\vert \Omega_0 \rangle$, with  $\langle \Omega_0 \vert$ being its Hermitian conjugation: $a_{\mu, p,l} \vert \Omega_0 \rangle=0$ and  $(a^*_{\mu, p,l})^\dagger \vert \Omega_0 \rangle=0$ for all $\mu$ and $l$, since both $a_{\mu, p,l}$ and  $(a^*_{\mu, p,l})^\dagger$ act as the annihilation operators in the two copies. Or equivalently, we have $\langle \Omega_0 \vert a^\dagger_{\mu, p,l}=0$ and $\langle \Omega_0 \vert a^*_{\mu, p,l}=0$ for all $\mu$ and $l$. We are thus led to the no-particle conditions for  the creation and annihilation operators $a^*_{\mu, p,l}$ and $a_{\mu, p,l}$, which are identical to Eq.\;(\ref{noparticle2}).  In addition, the no-particle conditions for  the creation and annihilation operators $a^\dagger_{\mu, p,l}$ and $(a^*_{\mu, p,l})^\dagger$ take the form
\begin{align}
	 (a^*_{\mu,p, l})^\dagger a^\dagger_{\mu,p, l'} \vert \Omega_0 \rangle &= p_\mu \delta_{l l'} \Omega_0 \rangle, \cr
     (a^*_{\mu,p, l})^\dagger a^\dagger_{\mu',p,l'}  \vert \Omega_0 \rangle&= 0, \cr
      (a^*_{\mu',p, l})^\dagger a^\dagger_{\mu,p, l'} \vert \Omega_0 \rangle &= 0.    \label{noparticle22}
\end{align}
One may thus construct the right eigenstates of the Green parafermiuon number operators $n_{\mu,p, l}$ from Eq.(\ref{noparticle2}) and the right eigenstates of the Green parafermiuon number operators $n^\dagger_{\mu,p, l}$ from  (\ref{noparticle22}). Note that the right eigenstates of the Green parafermiuon number operators $n^\dagger_{\mu,p, l}$ are nothing but the left eigenstates of the Green parafermiuon number operators $n_{\mu,p, l}$, as follows from the Hermitian conjugated counterparts for Eq.(\ref{noparticle22}). 
For any $l$, we have
\begin{equation*}
	p_{\mu}=\langle \Omega_0 \vert a_{\mu,p,l} a_{\mu,p,l}^* \vert \Omega_0 \rangle=\langle \Omega_0 \vert (a^*_{\mu,p,l})^\dagger a_{\mu,p,l}^\dagger \vert \Omega_0 \rangle.
\end{equation*}
As a result, $p_\mu$ must be real for any non-Hermitian realization.  In fact, as we shall show below from specific realizations in condensed matter systems under investigation, the order $p_\mu$ do not depend on $\mu$ and are identical to the emergent unit cell size, so the order $p$ is always a (positive) integer, regardless of being Hermitian or non-Hermitian.

Now we turn to the Green Ansatz (\ref{greenansatz}) that expresses the creation and annihilation operators $a^*_{\mu,p, l}$ and $a_{\mu,p, l}$ for Green parafermions in terms of the Green components  $b^*_{\mu,p, l,\alpha}$ and $b_{\mu,p,l,\alpha}$, introduced  for one single parafermion field~\cite{green} and for several parafermion fields~\cite{greenberg}.
Here  $b^*_{\mu,p,l,\alpha}$ and $b_{\mu,p,l,\alpha}$ satisfy the following anti-commutation or commutation relations:  for the same $\mu$ and $\alpha$, we have
\begin{align}
[b^*_{\mu,p,l,\alpha}, b_{\mu,p,{l'}, \alpha}]_+ &= \delta_{l l'}, \cr
[b_{\mu,p, l,\alpha}, b_{\mu,p, {l'}},\alpha]_+ &= 0 , \label{greencomp1}
\end{align}
and for the same $\mu$ but $\alpha \neq \alpha'$, we have
\begin{align}
	[b^*_{\mu,p, l,\alpha}, b_{\mu,p,{l'},{\alpha'}}]_- &=0, \cr
	[b_{\mu,p,l,\alpha}, b_{\mu,p,{l'},{\alpha'}}]_- &= 0, \label{greencomp2}
\end{align}
whereas for $\mu \neq \mu'$ but the same $\alpha$, we have 
\begin{align}
	[b^*_{\mu,p, l,\alpha}, b_{\mu',p,{l'},\alpha}]_+ &=0, \cr
	[b_{\mu,p, l,\alpha}, b_{\mu',p,{l'},\alpha}]_+ &= 0,  \label{greencomp3}
\end{align}
and for $\mu \neq \mu'$ and $\alpha \neq \alpha'$, we have
\begin{align}
	[b^*_{\mu,p,l,\alpha}, b_{\mu',p,{l'},{\alpha'}}]_- &=0, \cr
	[b_{\mu,p,l,\alpha}, b_{\mu',p,{l'},{\alpha'}}]_- &= 0. \label{greencomp4}
\end{align}
Here we emphasize that the Green Ansatz (\ref{greenansatz}) was originally introduced for an Hermitian realization~\cite{green,greenberg}, but it also works for a non-Hermitian realization.

Note that  the Green components  $b^*_{\mu,p, l,\alpha}$ and $b_{\mu,p,l,\alpha}$ are neither bosons nor fermions. As a result, it is impossible to realize the creation and annihilation operators for Green parafermions in terms of ordinary spin or fermion degrees of freedom. This opens up the possibilities for introducing auxiliary Majorana fermions to realize Green parafermions, together with  ordinary spin or fermion degrees of freedom. 
Here we stress that a realization of the Green components  $b^*_{\mu,p, l,\alpha}$ and $b_{\mu,p,l,\alpha}$ in terms of ordinary spin or fermion degrees of freedom makes it possible to introduce an internal symmetry group ${\rm U}(p)$, which are induced from a unitary transformation of combined objects of spin degrees of freedom and auxiliary Majorana fermions $\gamma _{l,\alpha}$ for quantum many-body spin systems or fermion degrees of freedom for strongly correlated itinerant electron systems, as discussed in Section~\ref{gscheme}. Note that 
the Green Ansatz for the creation and annihilation operators $a^*_{\mu,p, l}$ and $a_{\mu,p, l}$ does not introduce any composite structure, since it is linear, as argued by Greenberg and Messiah~\cite{greenberg}. Hence there are no internal degrees of freedom associated with the Green components 
$b^*_{\mu,p, l,\alpha}$ and $b_{\mu,p,l, \alpha}$. Indeed, the defining relations for the Green components  $b^*_{\mu,p, l,\alpha}$ and $b_{\mu,p,l,\alpha}$ are invariant if $b_{\mu,p,l,\alpha}$ are replaced by $\sum _{\alpha'=0} ^{p-1} U_{\alpha \alpha'} \gamma_{\alpha} \gamma_{\alpha'} b_{\mu,p,l,\alpha'}$ for any $p \times p$ unitary matrix $U$. The unitary transformations involving auxiliary Majorana fermions $\gamma_\alpha$ for Green parafermions may be regarded as an extension of the Bogoliubov transformations for ordinary fermions~\cite{pcoleman}, given that the defining anti-commutation relations for creation and annihilation operators for ordinary fermions are invariant under  unitary transformations, with their entries being only complex numbers.

As seen in Subsection~\ref{biquadratic-tasaki}, the ferromagnetic spin-1 biquadratic model involves only one single Green parafermion field, so  it is sufficient to restrict to the trilinear commutation relations between the  creation and annihilation operators for the same field. In other words,
the subscript $\mu$ may be dropped off.  It is readily seen that the creation and annihilation operators $a^*_{p, l}$ and $a_{p, l}$ for Green parafermions, with the Green components  $b^*_{p,l,\alpha}$ and $b_{p,l,\alpha}$ presented in Eq.\;(\ref{aobiquadratic}), satisfy the anti-commutation or commutation relations above, if we restrict ourselves to the constrained Hilbert space $V_1$. Here the order $p$ follows from $p=\langle \Omega_0 \vert a_{p, l} a_{p, l}^* \vert \Omega_0 \rangle$ for this specific realization, which is identical to the emergent unit cell size $p$.
In contrast, a
realization  appears in the $\rm {SU}(2)$ flat-band ferromagnetic Tasaki model, which involves both the trilinear commutation relations for the same field and the relative trilinear commutation relations for different parafermion fields. It thus falls into the PF subset. In other words, two Green parafermion fields are needed, with the subscript $\mu$ being identified as $\sigma$.  Indeed, the creation and annihilation operators $a^*_{\sigma,p, l}$ and $a_{\sigma,p, l}$ for Green parafermions, with the Green components   $b^*_{\sigma, p,l,\alpha}$ and $b_{\sigma,p, l,\alpha}$ presented in Eq.\;(\ref{aotasaki}), satisfy the anti-commutation or commutation relations above, in the original unconstrained Hilbert space $V_0$. 
Note that this realization is non-Hermitian, so we need to introduce the Hermitian conjugated forms $(a^*_{\sigma,p, l})^\dagger$ and $a^\dagger_{\sigma,p, l}$ of the creation and annihilation operators $a^*_{\sigma,p, l}$ and $a_{\sigma,p, l}$. However, the order $p_\sigma =p$ still follows from
$p_{\sigma}=\langle \Omega_0 \vert a_{\sigma, l} a_{\sigma, l}^* \vert \Omega_0 \rangle$, or equivalently, 
$p_\sigma =\langle \Omega_0 \vert (a^*_{\sigma,p,l})^\dagger a_{\sigma,p,l}^\dagger \vert \Omega_0 \rangle$, as 
follows from the consistency requirement. Note that $p_{\sigma}$ is independent of $\sigma$, since the two Green parafermion fields are time-reversed counterparts to each other.

In addition to the two typical examples under investigation,  Green parafermion states (up to a projection operation)  emerge as flat-band excitations in a broad class of condensed matter systems,  as long as they undergo SSB  with type-B GMs, with the ground state degeneracies being exponential in system size, irrespective of the type of boundary condition adopted. Further concrete examples may be found in Refs.~\cite{shi-so,shi-dtmodel,shiqq}.  Moreover,  it is convenient to interpret Green parafermion states (up to a projection operation) as flat-band excitations in the momentum space representation, as discussed in Appendix~\ref{realandmomentum}.

\subsection{The projected Green parafermion states for a non-Hermitian realization}~\label{ssblimit}

One of the key ingredients in our construction is to identify a state $\vert \eta \rangle$ with Green parafermion states $a^*_{\mu_1, p, k_1}a^*_{\mu_2,p, k_2} \ldots a^*_{\mu_m, p,k_m} \vert \Omega_0 \rangle$ or $a^\dagger_{\mu_1, p, k_1}a^\dagger_{\mu_2,p, k_2} \ldots a^\dagger_{\mu_m, p,k_m} \vert \Omega_0 \rangle$ in the momentum space representation or their counterparts in the real space representation. Note that these two states are identical for an Hermitian realization. However, they are different for a non-Hermitian realization.  This identification may be established in the scenario that  SSB is treated as a limit of explicit symmetry breaking, when an extra term $-h \;\Sigma_{c}$ is introduced by taking into account auxiliary Majorana fermions living in the constrained Hilbert space $V_1$ or the original unconstrained Hilbert space $V_0$, as described in Subsection~\ref{ssbscenario}. As already argued there, although this scenario
is well-known in the conventional description of SSB~\cite{zuber}, some novel features arise in the context of condensed matter systems undergoing SSB with type-B GMs, if the ground state degeneracies are exponential. Physically, this is due to the possibilities for introducing auxiliary Majoarana fermions on emergent unit cells, either physical or fictitious, where emergent unit cells themselves result from partial SSB of the translation symmetry under one lattice unit cell.
Here we shall encounter another feature that is relevant to the non-Hermitian nature of a specific realization of Green parafermions. This is particularly so if  auxiliary Majorana fermions are physical.

Recall that one may introduce an extra term $-h \;\Sigma_{c}$ into the model Hamiltonian  $\mathscr{H}$. For this purpose, we need to distinguish two cases. First,  if Green parafermions are realized in the constrained Hilbert space $V_1$, then $\Sigma_{c}$ is chosen to satisfy Eq.(\ref{sigma1}), which reduces to $\Pi_1 \Sigma_{c} \Pi_1 = \sum_\mu N_{\mu,p}$ for an Hermitian realization. Second, if Green parafermions are realized in the unconstrained Hilbert space $V_0$, then $\Sigma_{c}$ is  chosen to satisfy Eq.(\ref{sigma2}), which becomes $\Sigma_{c} = \sum_\mu N_{\mu,p}$ for an Hermitian realization. Here the Green parafermion number operators $n_{\mu, p, l}$ are Hermitian, namely $n^\dagger_{\mu, p, l}=n_{\mu, p, l}$, for an Hermitian realization, but not for a non-Hermitian realization. As a result, it is nontrivial to show that the projected Green parafermion states, namely 
$\Pi(a^*_{\mu_1, p, l_1}a^*_{\mu_2,p, l_2} \ldots a^*_{\mu_m, p,l_m} \vert \Omega_0 \rangle)$ and $\Pi(a^\dagger_{\mu_1, p, l_1}a^\dagger_{\mu_2,p, l_2} \ldots a^\dagger_{\mu_m, p,l_m} \vert \Omega_0 \rangle)$, are essentially identical for a non-Hermitian realization, up to a unitary transformation in each sector labeled by $m$.

Since the projection operator $\Pi$ commutes with  $\Pi \sum_\mu (N_{\mu,p}+N^\dagger_{\mu,p})\Pi/2$ and both of them are Hermitian, they must share simultaneous eigenstates. We remark that  $\Pi (a^*_{\mu_1, p, k_1}a^*_{\mu_2,p, k_2} \ldots a^*_{\mu_m, p,k_m}) \vert \Omega_0 \rangle$ and $\Pi (a^\dagger_{\mu_1, p, l_1}a^\dagger_{\mu_2,p, l_2} \ldots a^\dagger_{\mu_m, p,l_m}) \vert \Omega_0 \rangle$ are nontrivial eigenstates of $\Pi$, as long as $\Pi$ does not nullify   $a^*_{\mu_1, p, l_1}a^*_{\mu_2,p, l_2} \ldots a^*_{\mu_m, p,l_m} \vert \Omega_0 \rangle$ and $a^\dagger_{\mu_1, p, l_1}a^\dagger_{\mu_2,p, l_2} \ldots a^\dagger_{\mu_m, p,l_m} \vert \Omega_0 \rangle$. In addition,  note that $\Pi (a^*_{\mu_1, p, l_1}a^*_{\mu_2,p, l_2} \ldots a^*_{\mu_m, p,l_m}) \vert \Omega_0 \rangle$ or $\Pi (a^\dagger_{\mu_1, p, l_1}a^\dagger_{\mu_2,p, l_2} \ldots a^\dagger_{\mu_m, p,l_m}) \vert \Omega_0 \rangle$ for all possible atypical partitions and their counterparts for all possible atypical partitions produce a basis for the constrained Hilbert space $V$ specified by the projection operator $\Pi$, in the sense that both $\Pi (a^\dagger_{\mu_1, p, l_1}a^\dagger_{\mu_2,p, l_2} \ldots a^\dagger_{\mu_m, p,l_m}) \vert \Omega_0 \rangle$ and $\Pi (a^*_{\mu_1, p, l_1}a^*_{\mu_2,p, l_2} \ldots a^*_{\mu_m, p,l_m}) \vert \Omega_0 \rangle$
in each sector labeled by $m$ are linear combinations of  all the nontrivial basis states in each sector labeled by $m$. Generically,  these nontrivial basis states are not necessarily orthogonal to each other, but  they must be linear independent, as we shall see below from the $\rm {SU}(2)$ flat-band ferromagnetic Tasaki model.
As a result, $\Pi (a^\dagger_{\mu_1, p, l_1}a^\dagger_{\mu_2,p, l_2} \ldots a^\dagger_{\mu_m, p,l_m}) \vert \Omega_0 \rangle$ must be a linear combination of $\Pi (a^*_{\mu_1, p, l_1}a^*_{\mu_2,p, l_2} \ldots a^*_{\mu_m, p,l_m}) \vert \Omega_0 \rangle$
in each sector labeled by $m$, with the coefficients in this linear combination depending on auxiliary Majorana fermions. We are thus led to conclude that any eigenstate of  $\Pi (N_{\mu,p}+N^\dagger_{\mu,p})\Pi/2$, with the eigenvalue being $m$, must be a linear combination of   $\Pi (a^*_{\mu_1, p, l_1}a^*_{\mu_2,p, l_2} \ldots a^*_{\mu_m, p,l_m}) \vert \Omega_0 \rangle$ 
in each sector labeled by $m$.

For condensed matter systems, all known realizations of Green parafermions are Hermitian for quantum many-body spin models (cf.~Refs.~\cite{shi-so,shi-dtmodel} for other realizations, in addition to the realization specified by the Green components in Eq.(\ref{aobiquadratic}) for the ferromagnetic spin-1  biquadratic model). In contrast,  for  the $\rm {SU}(2)$ flat-band ferromagnetic Tasaki model, the realization, as specified by the Green components in Eq.(\ref{aotasaki}), is non-Hermitian.  Another model that entails a non- Hermitian realization of Green parafermions is its $\rm {SU}(n)$ variant~\cite{shiqq}. 
A tedious but straightforward calculation confirms that any eigenstate of  $\Pi (N_{\sigma,p}+N^\dagger_{\sigma,p})\Pi/2$, with the eigenvalue being $m$, must be a linear combination of   $\Pi (a^*_{\sigma_1, p, l_1}a^*_{\sigma_2,p, l_2} \ldots a^*_{\sigma_m, p,l_m}) \vert \Omega_0 \rangle$ 
in each sector labeled by $m$ for this non-Hermitian realization. Mathematically, this follows from the fact that  $\Pi (N_{\sigma,p}+N^\dagger_{\sigma,p})\Pi/2$ is identical to $\Pi {\hat {N}}_\sigma \Pi$, where $\Pi {\hat {N}}_\sigma \Pi$ are related with
the electron number operator $\hat{N} = {\hat {N}}_\uparrow + {\hat {N}}_\downarrow$ as the generator of the ${\rm U(1)}$ group in the charge sector and the Cartan generator  $\sum_j S^z = ({\hat {N}}_\uparrow - {\hat {N}}_\downarrow)/2$ of the ${\rm SU(2)}$ group in the spin sector (cf.~Section~\ref{ssbtype-b}). 
In fact, both of the expectation values of $\Pi (N_{\sigma,p}+N^\dagger_{\sigma,p})\Pi/2$ and $\Pi {\hat {N}}_\sigma \Pi$ in $\Pi (a^*_{\sigma_1, p, l_1}a^*_{\sigma_2,p, l_2} \ldots a^*_{\sigma_m, p,l_m}) \vert \Omega_0 \rangle$ are equal to $m$,  as long as $\Pi$ does not nullify   $a^*_{\mu_1, p, l_1}a^*_{\mu_2,p, l_2} \ldots a^*_{\mu_m, p,l_m} \vert \Omega_0 \rangle$ for an atypical partition. 
Note that this statement is valid for the Green parafermion states for any typical partition, with the only difference being that there are more than one species of Green parafermions, with their orders being $p_\nu$, each of which contains $n$ Green parafermion fields (cf.~Section~\ref{gscheme}).

If auxiliary Majorana fermions are physical, then what is realized in a specific condensed matter system is a flat-band excitation that represents a linear combination of the projected Green parafermion states  $\Pi (a^*_{\sigma_1, p, k_1}a^*_{\sigma_2,p, k_2} \ldots a^*_{\sigma_m, p,k_m}) \vert \Omega_0 \rangle$ 
in each sector labeled by $m$ for a non-Hermitian realization. Note that we have switched back to the momentum space representation here.
However, if auxiliary Majorana fermions are fictitious, then one may identify $\vert \eta \rangle$ with Green parafermion states $a^*_{\mu_1, p, l_1}a^*_{\mu_2,p, l_2} \ldots a^*_{\mu_m, p,l_m} \vert \Omega_0 \rangle$ or $a^\dagger_{\mu_1, p, l_1}a^\dagger_{\mu_2,p, l_2} \ldots a^\dagger_{\mu_m, p,l_m} \vert \Omega_0 \rangle$ for an atypical (periodic) partition in the real space representation or  $a^*_{\mu_1, p, k_1}a^*_{\mu_2,p, k_2} \ldots a^*_{\mu_m, p,k_m} \vert \Omega_0 \rangle$ or $a^\dagger_{\mu_1, p, k_1}a^\dagger_{\mu_2,p, k_2} \ldots a^\dagger_{\mu_m, p,k_m} \vert \Omega_0 \rangle$ for an atypical (periodic) partition in the momentum space representation, regardless of a realization being Hermitian or non-Hermitian.

\subsection{Subsystem gauge symmetries in the context of Green parafermions and the Elitzur theorem}~\label{elitzur1}
The Elitzur theorem states that no local gauge symmetry is allowed to be spontaneously broken~\cite{elitzur}. Here the meaning of local gauge symmetry is no doubt well-defined in gauge field theory~\cite{fradkin,weinberg}. As is well-known, in relativistic quantum field theory, external (spatial and temporal) symmetries are well separated from internal symmetries~\cite{weinberg} (see also Ref.~\cite{costa}). However, this is not true in the context of Green parafermions realized in condensed matter, as we shall see below. Here we note that it is possible to explain the origin of the division between internal and external symmetries in quantum field theory from an information-theoretic perspective~\cite{kober}.
Our discussion below is focused on auxiliary  (physical) Majorana fermions. However, it is valid for auxiliary (fictitious) Majorana fermions introduced on emergent unit cells for a specific partition, before they are removed from the Hilbert space.  

As we have demonstrated in Section~\ref{gscheme}, the introduction of auxiliary  (physical) Majorana fermions results in the dichotomy between internal and external degrees of freedom, with internal degrees of freedom being spin degrees of freedom plus auxiliary Majorana fermions $\gamma_{\mu,l,\alpha}$  or fermion degrees of freedom inside an emergent unit cell (cf. Eq.(\ref{aobiquadratic}) and Eq.(\ref{aotasaki}) for concrete examples) inside an emergent unit cell for any partition. In particular, for an atypical partition, an emergent unit cell consists of $p$ adjacent lattice unit cells. 
There are different ways to organize internal (local) spin or fermion degrees of freedom inside an emergent unit cell. This in turn leads to an internal symmetry group ${\rm U}(p)$.  In particular,  the symmetric group ${\rm S}_p$ consisting of all the permutations with respect to  $p$ lattice unit cells labeled by $\alpha$ inside an emergent unit cell labeled by $l$ forms a subgroup of ${\rm U}(p)$ for any $p$.  As a result, the ordering of $p$ adjacent lattice unit cells inside an emergent unit cell is irrelevant. Here we remark that this internal symmetry group ${\rm U}(p)$ act on local degrees of freedom located at different lattice unit cells, which are spatially well-separated.
Meanwhile, such an internal symmetry group ${\rm U}(p)$ may be interpreted as a subsystem gauge group ${\rm U}(p)$, since any ${\rm U}(p)$ unitary transformation may be independently carried out in each emergent unit cell. 
In other words, one way to organize internal (local) spin or fermion degrees of freedom inside an emergent unit cell corresponds to one choice of gauge for a Green parafermion system described by the Hamiltonian $-h \sum_\mu N_{\mu,p}$ and different ways are connected via a gauge transformation, under the condition that {\it only} observables invariant under subsystem gauge transformations are allowed. Otherwise, inaccessibility to internal degrees of freedom is violated.

If $p=1$, then we have a {\it local} gauge group ${\rm U}(1)$; if $p=L$, then we have a {\it global} gauge group ${\rm U}(L)$. Generically, for any $p$ between 1 and $L$, a subsystem gauge group interpolates a local gauge group ${\rm U}(1)$ and  a global gauge group ${\rm U}(L)$. It is readily seen that a subsystem gauge group ${\rm U}(p)$ is not allowed to be spontaneously broken for $p=1$, but it is allowed to be spontaneously broken for any $p$ ($2 \le p \le L$). In this sense, only a subsystem gauge group ${\rm U}(p)$ when $p=1$ is qualified to be interpreted as a local gauge symmetry, as required in the Elitzur theorem. Indeed, for $p \ge 2$, a subsystem gauge group ${\rm U}(p)$ always acts on spin or fermion degrees of freedom located at different lattice unit cells. Note that $p$ is allowed to be $L$ here,  with the caveat that an atypical partition with the period $p=L$  should be treated as non-periodic. Note that this is similar to what has been mentioned for atypical degenerate ground states in Section~\ref{ssbtype-b}.

We stress that a Green parafermion system described by the Hamiltonian $-h \sum_\mu N_{\mu,p}$ is revealed as a constituent of a hierarchical structure at a Hamiltonian level in the context of SSB with type-B GMs ( cf. Section~\ref{gscheme}). Many novel features arises in this scenario, which may be attributed to a novel type of interaction  representing communication between  spin or fermion degrees of freedom, with auxiliary (physical) Majorana fermions acting as a medium.
As a result, auxiliary (physical) Majorana fermions do not manifest themselves in the Hamiltonian explicitly. In addition, this novel type of interaction accounts for an apparent contradiction with a statement by Greenberg and Messiah~\cite{greenberg} that the Green Ansatz for the creation and annihilation operators does not introduce composite structure,  in the sense that it is linear.

Three remarks are in order. First, $-h \sum_\mu N_{\mu,p}$ is invariant under subsystem gauge transformations, which imposes constraints on any choice of the extra term  $-h \;\Sigma_{c}$, in addition to the requirement that $\;\Sigma_{c}$ is Hermitian. Second, the model Hamiltonian $\mathscr{H}$ is not invariant under any subsystem gauge transformation. A condensed matter system described by  the model Hamiltonian $\mathscr{H}$ is thus not a gauge theory in any sense. As such, we have built a bridge connecting a condensed matter system that is not invariant under subsystem gauge transformations with a Green parafermion system described by the Hamiltonian $-h \sum_\mu N_{\mu,p}$ in the scenario that SSB is regarded as a limit of explicit symmetry breaking.
A peculiar feature is that  all the sectors labeled by $m$ ($m=1,\ldots,M$), or equivalently, labeled by the eigenvalue(s) of the Cartan generator(s) of the symmetry group $G$ or by the electron number and the eigenvalue(s) of the Cartan generator(s) of the semi-simple subgroup of the symmetry group $G$, need to be taken into account simultaneously. 
Third, as already pointed out in Section~\ref{ssbtype-b}, condensed matter undergoing SSB with type-B GMs fall into two different categories, depending on the ground state degeneracies being exponential and polynomial with system size. A marked difference between them lies in the fact that it is possible to make a connection with a set of Green parafermion fields, subject to subsystem gauge symmetries, if the ground state degeneracies are exponential.

Our discussion above has been restricted to auxiliary (physical) Majorana fermions. If auxiliary Majorana fermions are fictitious, then they may be arranged on emergent unit cells for a specific partition. However, auxiliary Majorana fermions must be removed from the Hilbert space for one specific partition, before switching to another partition, in order to ensure that no auxiliary Majorana fermions are left in the Hilbert space eventually.
As a consequence, it still makes sense to introduce an internal symmetry group, which in turn may be interpreted as a subsystem gauge group for a Green parafermion system, with auxiliary (fictitious) Majorana fermions as a key ingredient. Note that such a Green parafermion system is well-defined for every (atypical or typical) partition, with all possible partitions being on the same footing. Indeed, for a specific partition, there are different ways to organize local spin or fermion degrees of freedom inside an emergent unit cell; one way corresponds to one choice of gauge for a Green parafermion system and different ways are connected via a gauge transformation, under the condition that {\it only} observables invariant under subsystem gauge transformations are allowed. Otherwise, inaccessibility to internal degrees of freedom is violated. Indeed, removing auxiliary Majorana fermions from the Hilbert space for one specific partition amounts to violating  inaccessibility to internal degrees of freedom, thus making it possible to switch to another set of auxiliary Majorana fermions on emergent unit cells for another partition.  This provides an alternative perspective for understanding why the absence of auxiliary (physical) Majorana fermions defined on an emergent unit cell of {\it fixed} size for a specific partition is equivalent to the presence of auxiliary (fictitious) Majorana fermions defined on any emergent unit cells of  different sizes for all possible partitions.

\subsection{The real and momentum space representations for Green parafermions}~\label{realandmomentum}

In order to interpret Green parafermion states (up to a projection operation)  emerge as flat-band excitations, one has to move to the momentum space representation.
Here we briefly discuss the real and momentum space representations for Green parafermions. In the real space representation,  a Green parafermion state is defined on an emergent unit cell consisting of $p$ adjacent lattice unit cells (cf.~Appendix~\ref{tcr}). In contrast, a Green parafermion state with fixed momentum  in the momentum space representation is defined on the entire lattice as a result of the uncertainty principle. Here we stress that PBCs have been adopted.

As already stressed in Section~\ref{gscheme}, an emergent unit cell, labeled by $l$, is not necessary to be identical to the lattice unit cell.  Here we restrict ourselves to one spatial dimension, though an extension to two and higher spatial dimensions is straightforward. Then the total number of emergent unit cells is $N$, with $N=L/p$, where we have assumed that $p$ divides $L$, so we have $l=1, 2,\ldots, N$. In this setting, we define the 
creation and annihilation operators $a^*_{\mu, p, l}$ and $a_{\mu, p, l}$, which satisfy the trilinear commutation relations (\ref{tcr2}) and (\ref{jacobi}) for the same field and the relative trilinear commutation relations (\ref{tcr3}) and (\ref{tcr4}) for different fields, when $p$ is fixed.
In the momentum space representation, we have to introduce the 
creation and annihilation operators $a^*_{\mu, p, k}$ and $a_{\mu, p, k}$ that create and annihilate a Green parafermion of order $p$ with momentum $k$, where $k= 2 \pi \delta /N$ ($\delta = 0,1,2,\ldots, N-1$).  

In fact, the creation and annihilation operators $a^*_{\mu, k, l}$ and $a_{\mu, k, l}$ for Green parafermions in the momentum space representation may be expanded into a sequence of the creation and annihilation operators $a^*_{\mu, p, l}$ and $a_{\mu, p, l}$ for Green parafermions, which are defined on an emergent unit cell in the real space representation. More precisely, we have
\begin{align}
	a^*_{\mu, p, k} &= \frac {1}{\sqrt N}\sum_{l=1}^N e^{-i2\pi k l} a^*_{\mu, p, l}, \cr
	a_{\mu, p, k} &= \frac {1}{\sqrt N} \sum_{l=1}^N e^{i2\pi k l} a_{\mu, p, l}.
\end{align}
Conversely, the creation and annihilation operators $a^*_{\mu, p, l}$ and $a_{\mu, p, l}$ for Green parafermions in the real space representation may be expressed in terms of the creation and annihilation operators $a^*_{\mu, k, l}$ and $a_{\mu, k, l}$ for Green parafermions in the momentum space representation. More precisely, we have 
\begin{align}
	a^*_{\mu, p, l} &= \frac {1}{\sqrt N}\sum_{k} e^{i2\pi k l} a^*_{\mu, p, k}, \cr
	a_{\mu, p, l} &= \frac {1}{\sqrt N} \sum_{k} e^{-i2\pi k l} a_{\mu, p, k}.
\end{align}
In other words, the creation and annihilation operators $a^*_{\mu, p, l}$ and $a_{\mu, p, l}$ in the real space representation are connected to their counterparts  $a^*_{\mu, p, k}$ and $a_{\mu, p, k}$  in the momentum space representation via the Fourier transformation.  

As discussed in~Appendix~\ref{tcr}, the creation and annihilation operators $a^*_{\mu,p, l}$ and $a_{\mu,p, l}$ for Green parafermions may be written in terms of the Green components  $b^*_{\mu,p, l,\alpha}$ and $b_{\mu,p,l, \alpha}$ in the real space representation. Similarly, one may also introduce 
the Green components  $b^*_{\mu,p, k,\alpha}$ and $b_{\mu,p,k, \alpha}$ to express the creation and annihilation operators $a^*_{\mu,p, k}$ and $a_{\mu,p, k}$ in the momentum space representation; they are connected to each other via a Fourier transformation. Mathematically, we have
\begin{align}
	a^*_{\mu,p, k} &= \sum_{\alpha =0}^{p-1} b^*_{\mu,p,k,\alpha}, \cr
	a_{\mu,p, k} &= \sum_{\alpha =0}^{p-1} b_{\mu,p,k,\alpha}, 
\end{align}
and
\begin{align}
	b^*_{\mu, p, l,\alpha} &= \frac {1}{\sqrt N}\sum_{k} e^{i2\pi k l} b^*_{\mu, p, k, \alpha}, \cr
	b_{\mu, p, l, \alpha} &= \frac {1}{\sqrt N} \sum_{k} e^{-i2\pi k l} b_{\mu, p, k, \alpha}.
\end{align}

We are thus led to conclude that the real and momentum space representations for Green parafermions are equivalent, because the trilinear commutation relations (\ref{tcr2}) and (\ref{jacobi}) for the same  parafermion field and the relative trilinear commutation relations (\ref{tcr3}) and (\ref{tcr4}) for different parafermion fields are invariant under any unitary transformation of the annihilation operators $a_{\mu,p,l}$. Mathematically,  the defining trilinear commutation relations for the same field and the relative trilinear commutation relations for any two different fields still hold for the creation and annihilation operators $a^*_{\mu, p, k}$ and $a_{\mu, p, k}$ for Green parafermions in the momentum space representation, as presented in Eqs.(\ref{tcr2}),~(\ref{jacobi}),~(\ref{tcr3}),~(\ref{tcr4}) (also cf. Eq.(\ref{tcr5}) in Appendix~\ref{tcr}), but $l$, $l'$ and $l''$ in these relations should be replaced by  $k$, $k'$ and $k''$, respectively.
Indeed, the Fourier transformation connecting  the 
creation and annihilation operators in the real and momentum space representations for Green parafermions is unitary. 
Here we have assumed that a unitary transformation of the creation operators $a^*_{\mu,p,l}$ is always conjugate to that of 
the annihilation operators $a_{\mu,p,l}$ for either an Hermitian or a non-Hermitian realization.  More precisely, the total Green parafermion number operator is invariant under the (unitary) Fourier transformation. That is, we have  $\sum_{\mu,k} n_{\mu,p, k} = \sum_{\mu,l} n_{\mu,p, l}$.
However, for a non-Hermitian realization, the Hermitian conjugated forms $(a^*_{\sigma,p, k})^\dagger$ and $a^\dagger_{\sigma,p, k}$ of the creation and annihilation operators $a^*_{\sigma,p, k}$ and $a_{\sigma,p, k}$ are also needed, together with the Hermitian conjugated form $n^\dagger_{\mu,p, k}$ of the Green paraferrmion number operators $n_{\mu,p, k}$.

For a specific condensed matter system, the order $p$ as an intrinsic character for Green parafermions does not depend on what representations are adopted, namely the real and momentum space representations.  Note that	the order $p$ follows from $p=p_\mu=\langle \Omega_0 \vert a_{\mu,p, k} a_{\mu, p,k}^* \vert \Omega_0 \rangle$ in the momentum space representation. For  one single Green parafermion field in the ferromagnetic spin-1 biquadratic model and two Green parafermion fields in the ferromagnetic $\rm {SU}(2)$ flat-band Tasaki model,  the order $p$ is equal to the emergent unit cell size in both the representations. In fact, one may introduce a symmetry operation $\tau_p$ that is defined to be the translation symmetry operation under $p$ lattice unit cells. That is, we have $\tau_p = \tau^p$, where $\tau$ denotes the translation symmetry operation under one lattice unit cell. One of the advantages for choosing to work in the momentum space representation lies in the fact that, after the projection operators $\Pi_1$ and $\Pi_2$ are implemented, the model Hamiltonian including an extra term becomes the total Green parafermion number operator $\sum_{\mu,l} n_{\mu,p, l}$ in the real space representation, which is in turn mapped to $\sum_{\mu,k} n_{\mu,p, k}$ in the momentum space representation via the Fourier transformation (cf. Section~\ref{gscheme}). One may thus identify $\vert \eta \rangle$ with the  Green parafermion states  $a^\dagger_{\mu, p, k_1}a^\dagger_{\sigma, p, k_2} \ldots a^\dagger_{\sigma, p,k_m} \vert \Omega_0 \rangle$ for Green parafermions of order $p$ in the momentum space representation, which are eigenvectors of the translation symmetry operation $\tau_p$ under $p$ lattice unit cells. As a result, Green parafermion states in the momentum space representation offers a natural description for the emergence of flat-band excitations in condensed matter systems, since $\tau_p$ commutes with a projection operator $\Pi$. Meanwhile, if we introduce permutation operations with respect to $\{k_1,k_2,\ldots,k_m\}$, then $\Pi$ also commutes with any permutation thus defined, in contrast to permutation operations with respect to $\{l_1,l_2,\ldots,l_m\}$. This is a useful feature when one examines a modified form of the Pauli exclusion principle for Green parafermions (cf.~Appendix~\ref{tcr}).

The discussion above provides a proper conceptual framework to identify Green parafermion states (up to a projection operation) in the real and momentum space representations as emergent flat-band excitations in condensed matter systems undergoing SSB with type-B GMs, subject to the exponential ground state degeneracies under both PBCs and OBCs. We remark that it only makes sense to investigate Green parafermion states in the real and momentum space representations for a realization of Green parafermions in condensed matter, in contrast to abstract Green parafermions~\cite{green,greenberg}.
Note that physical observables in the real space representation are {\it local}, defined on an emergent unit cell consisting of  $p$ adjacent lattice unit cells. In contrast, physical observables in the momentum space representation  are {\it global}, in the sense that they are defined on the entire lattice. Physically, this dichotomy stems from the wave-particle duality in quantum theory.

\subsection{ A connection between an emergent subsystem symmetry operation tailored to a specific degenerate ground state and the Elitzur theorem}~\label{elitzur}

As discussed in Section~\ref{gscheme}, if  Green parafermions live in the constrained Hilbert space $V_1$, then not all  generalized highest weight states are primary. Indeed, secondary generalized highest weight states may be generated from the action of emergent subsystem non-invertible subsystem symmetries on primary ones (cf.~Appendix~\ref{primary} for concrete examples). As it turns out, emergent subsystem non-invertible subsystem symmetries naturally follow from a restatement of the Elitzur theorem, as emergent subsystem invertible subsystem symmetries do~\cite{dimertrimer,jesse}.

Here we recall a notion - an emergent subsystem symmetry operation tailored to a specific degenerate ground state,  which has been introduced in Refs.~\onlinecite{dimertrimer,jesse} for  condensed matter systems undergoing SSB with type-B GMs. As already noticed in Ref.~\cite{jesse}, there is a hidden connection between an emergent subsystem symmetry operation tailored to a specific degenerate ground state and the Elitzur theorem~\cite{elitzur}. Note that the formulation there is restricted to an emergent subsystem invertible symmetry operation, so it may be regarded as the generator of a discrete group, such as ${\rm Z}_2$.  However, an emergent subsystem symmetry operation is not necessarily invertible, thus leading to an emergent subsystem non-invertible symmetry tailored to a specific degenerate ground state. Actually, emergent subsystem invertible and non-invertible symmetries are indispensable for a full understanding of  condensed matter systems undergoing SSB with type-B GMs, as long as the ground state degeneracies are exponential in system size under PBCs and OBCs.  Here we show that they follow from an equivalent restatement of the Elitzur theorem. 

For our purpose, we state a mathematical lemma as follows. If there exists a subsystem operation $g$ that does not commute with the model Hamiltonian $\mathscr{H}$, namely $[\mathscr{H},g]_- \neq 0$, but $C|\Psi_0\rangle=0$ is satisfied,  where $C$ is the commutator between the model Hamiltonian $\mathscr{H}$ and $g$, namely $C=[\mathscr{H},g]_-$.  By a subsystem operation we mean it is only defined on a specific subsystem instead of the system itself. Given $\mathscr{H}|\Psi_0\rangle=E_0|\Psi_0\rangle$ and $|\langle \Psi_0|g|\Psi_0\rangle|\neq 1$, we have $\mathscr{H}g|\Psi_0\rangle=E_0g|\Psi_0\rangle$, where $|\Psi_0\rangle$ is assumed to be a ground state that has been normalized, with $E_0$ being the ground state energy, if we do not assume that the Hamiltonian $\mathscr{H}$ is in a canonical form, i.e., the multiplicative constant is one and the additive constant is zero (cf. Appendix~\ref{lemma}). In other words, $g|\Psi_0\rangle$ is a degenerate ground state. We refer to $g$ as an emergent subsystem  symmetry operation tailored to a specific degenerate ground state $|\Psi_0\rangle$.
The converse is also true. That is, if there is a subsystem  operation $g$ such that $g|\Psi_0\rangle$, satisfying $|\langle \Psi_0|g|\Psi_0\rangle|\neq 1$, is a ground state degenerate with $|\Psi_0\rangle$, then
the commutator $C=[\mathscr{H},g]_-$ between the Hamiltonian $\mathscr{H}$ and $g$ nullifies $|\Psi_0\rangle$: $C|\Psi_0\rangle=0$. Hence $g$ is an emergent
subsystem  symmetry operation tailored to a specific degenerate ground state $|\Psi_0\rangle$. We stress that $g$ is not required to be a unitary operation, in contrast to the original definition in Refs.~\onlinecite{dimertrimer,jesse}.

Now we are ready to reveal a connection between an emergent subsystem  symmetry operation tailored to a specific degenerate ground state and the Elitzur theorem, as already noticed in Ref.~\cite{jesse}, though the discussion there is restricted to the situation when $g$ is unitary.
The Elitzur theorem states that no local gauge symmetry is spontaneously broken~\cite{elitzur}. It may be restated as follows. If a local operation acting on a specific ground state  generates a degenerate ground state, then this operation must either generate a (discrete) group but it is not a local gauge symmetry group or do not generate any group. In either case, we are thus led to an emergent subsystem symmetry operation tailored to a specific degenerate ground state, which has been defined above. In fact, two possibilities arise, as far as the nature of  an emergent subsystem symmetry operation is concerned: one possibility involves an emergent subsystem symmetry operation that generates a (discrete) group, but not a local gauge symmetry group, and the other possibility involves an emergent subsystem symmetry operation that is non-invertible, so it does not generate any (discrete) group. Indeed, an emergent subsystem symmetry operation may be interpreted as the generator of an emergent subsystem symmetry group for the first possibility and as an emergent subsystem non-invertible symmetry for the second possibility. As it turns out, both possibilities appear in the ferromagnetic spin-1 biquadratic model~\cite{jesse} and  the ferromagnetic $\rm {SU}(2)$ flat-band Tasaki model (also 
cf.~Appendix~\ref{primary} for the roles of emergent subsystem non-invertible symmetries in identifying primary generalized highest weight states).

We stress that emergent subsystem invertible and non-invertible symmetry operations tailored to  specific degenerate ground states
are directly relevant to highest and generalized highest weight states in condensed matter systems undergoing SSB with type-B GMs, thus leading to a mechanism for explaining the origin of the exponential ground state degeneracies in system size, but different under PBCs and OBCs. There are two subclasses of  condensed matter systems undergoing SSB with type-B GMs, subject to the exponential ground state degeneracies in system size, irrespective of the type of boundary condition adopted~\cite{jesse}. 
The first subclass arises if the symmetry group (modulo a discrete symmetry subgroup) is simple or semi-simple, with only one highest weight state. As a result, the number of degenerate ground states generated from the action of the generator(s) of the symmetry group $G$ on the unique highest weight state is polynomial. It follows that the number of  generalized highest weight states is exponential.
The second subclass arises if the symmetry group (modulo a discrete symmetry subgroup) is non-semi-simple, so there are exponentially many highest weight states, in addition to exponentially many generalized highest weight states. Moreover, although we have restricted ourselves to a discrete local gauge symmetry, the same argument also works for any continuous local gauge symmetry, when the Elitzur theorem is stated.

Here we mention that non-invertible symmetries~\cite{seiberg,shao0,shao1,nameki,oshikawa,sierra,zhou-ising,zhou-doubleinsing},  subsystem invertible symmetries~\cite{subsystemsymmetry,subsystemsymmetry1} and  subsystem non-invertible symmetries~\cite{noninvertiblesymmetry} are of current interest, which are under intensive investigations in different contexts. Here subsystem invertible symmetrie and  subsystem non-invertible symmetries appear to be in an emergent form, in a sense that they themselves do  not commute with the model Hamiltonian $\mathscr{H}$. 
In particular, the essential difference between topological and non-topological defects was clarified in Ref.~\cite{zhou-ising}, given that topological  defects, introduced in Refs.~\cite{fendley,seifnashri}, play a crucial role in constructing non-invertible Kramers-Wannier duality symmetries -- an important class of categorical symmetries~\cite{seiberg0,category1,category3,category4,category5}, as a result of the recent conceptual development from the original Kramers-Wannier duality transformation~\cite{kramers}. In particular, a lattice version of fusion rules has been constructed for the transverse-field Ising model, which flows to the Tambara-Yamagami ${\rm Z}_2$ fusion category~\cite{ty}.

Up to the present, we have mainly focused on emergent subsystem invertible or non-invertible symmetries. It is worthwhile to stress that, if a subsystem is the entire system itself, then symmetries become global. As follows from the equivalent restatement of the Elitzur theorem,  SSS is allowed for them, regardless of being invertible or non-invertible. In particular,  global invertible or non-invertible symmetries are not necessarily emergent, in the sense that they are allowed to commute with the model Hamiltonian. Indeed, there might be an inherent connection between SSB with type-B GMs, Green parafermions and non-invertible Kramers-Wannier duality symmetries~\cite{ue1}.

\subsection{Primary and secondary generalized highest weight states:  emergent subsystem non-invertible symmetries}~\label{primary}

The identification of a state $\vert \eta \rangle$ with imaginary Green parafermion states $a^*_{\mu_1, p, k_1}a^*_{\mu_2,p, k_2} \ldots a^*_{\mu_m, p,k_m} \vert \Omega_0 \rangle$ or $a^\dagger_{\mu_1, p, k_1}a^\dagger_{\mu_2,p, k_2} \ldots a^\dagger_{\mu_m, p,k_m} \vert \Omega_0 \rangle$ for an atypical partition and their counterparts for a typical partition in the momentum space representation enables to classify exponentially many degenerate ground states into different families, with only one family being primary  and all other families secondary, in the sense that the secondary families are derived from the primary family through the action of different types of symmetry operations. Note that our discussion also works for imaginary Green parafermion states $a^*_{\mu_1, p, l_1}a^*_{\mu_2,p, l_2} \ldots a^*_{\mu_m, p,l_m} \vert \Omega_0 \rangle$ or $a^\dagger_{\mu_1, p, l_1}a^\dagger_{\mu_2,p, l_2} \ldots a^\dagger_{\mu_m, p,l_m} \vert \Omega_0 \rangle$ for an atypical partition and their counterparts for a typical partition in the real space representation. In this regard, both the symmetry group and emergent subsystem (invertible and non-invertible) symmetries play crucial roles (for the definitions of  emergent subsystem (invertible and non-invertible) symmetries, cf.~Appendix~\ref{elitzur}). 

We define the primary family, as  a subset of exponentially many fully factorized degenerate ground states, to be consisting of all highest weight states and primary generalized highest weight states that appear as a subset of generalized highest weight states derivable from the identification of $\vert \eta \rangle$ with imaginary Green parafermion states for both atypical and typical partitions. In other words, any  highest weight state belongs to the primary family, but not all generalized highest weight states are the primary family members. We are thus led to a new notion - the primary generalized highest weight states. Once they are identified, all other (secondary) generalized highest weight states follow from the action of an emergent subsystem  non-invertible symmetry operation tailored to a specific primary generalized highest weight state. Afterwards, all exponentially many degenerate ground states follow from the action of the generators of the symmetry group $G$ on all highest and generalized highest weight states. Given the number of the primary family members is exponential, we are able to explain the origin of the exponential ground state degeneracies for condensed matter systems undergoing SSB with type-B GMs. Indeed, the exponential ground state degeneracies are different for a specific system under both PBCs and OBCs. We may thus expect that both the primary and secondary families and emergent subsystem  non-invertible symmetries depend on the types of boundary conditions adopted. Here we restrict ourselves to PBCs only for brevity.

In  addition to an emergent subsystem  non-invertible symmetry operation tailored to a specific primary generalized highest weight state, there exist emergent subsystem  non-invertible symmetries tailored to  some subset of the ground state subspace, which act as the ladder operators to generate all highest weight states and primary  generalized highest weight states from {\it one} chosen highest weight state.
For our purpose, we need to distinguish two situations: (1) Green paraferrmions live in the constrained Hilbert space $V_1$, with the ferromagnetic spin-1 biquadratic model as an example, and (2) Green paraferrmions live in the unconstrained Hilbert space $V_0$, with the ferromagnetic $\rm {SU}(2)$ flat-band Tasaki model as an example. 

\subsubsection{Emergent subsystem non-invertible symmetries: from the primary to secondary generalized highest weight states}

For the ferromagnetic spin-1 biquadratic model (\ref{hambq}), the identification of a state $\vert \eta \rangle$ with imaginary Green parafermion states $a^*_{p, k_1}a^*_{p, k_2} \ldots a^*_{p,k_m} \vert \Omega_0 \rangle$ or $a^\dagger_{p, k_1}a^\dagger_{p, k_2} \ldots a^\dagger_{p,k_m} \vert \Omega_0 \rangle$ for an atypical partition and their counterparts for a typical partition in the momentum space representation yields
a fully factorized  (degenerate) ground state $|\Psi_0\rangle$  that only consists of $\vert + \rangle_j$ and $\vert 0 \rangle_j$, with all the states containing the local states $\vert 00 \rangle_{j,j+1}$ on any two adjacent lattice sites being excluded, as a result of a modified form of the Pauli exclusion principle of Green parafermions in the presence of the projection operator $\Pi$. Note that such a fully factorized  (degenerate) ground state $|\Psi_0\rangle$ is a primary generalized highest weight state. For instance, if the total number of lattice sites in the local states $\vert 0 \rangle_j$ ($j=1,2,\ldots,L$) is two, then $|\Psi_0\rangle$ must take the form $|\Psi_0\rangle= \vert + \ldots + 0 + \ldots + 0 + \ldots + \rangle$, where the two lattice sites in the local states $\vert 0 \rangle_j$ are assumed to be located at $j=j_1$ and $j=j_2$ ($j_2 >j_1+1$), respectively.  If one chooses an emergent subsystem symmetry operation $g_{j_1,j_2}$ (tailored to this primary generalized highest weight state) to be 
\begin{equation}
	g_{j_1,j_2}= \left( S_B \right)^{j_2-j_1-1}, \label{g-operator}
\end{equation}
with
\begin{equation*}
S_B = \sum _{j=j_1+1}^{j_2-1} (-1)^j (S_j^-)^2,
\end{equation*}
then it is readily seen that $g_{j_1,j_2}$  does not commute with the Hamiltonian (\ref{hambq}). That is, we have $[\mathscr{H}, g_{j_1,j_2}]_- \neq 0$. However, the commutator $C_{j_1,j_2}=[\mathscr{H},g_{j_1,j_2}]_-$ nullifies $|\Psi_0\rangle$, namely $C_{j_1,j_2}|\Psi_0\rangle=0$.  Obviously, $g_{j_1,j_2}$ is not invertible, so $g_{j_1,j_2}$ is an emergent subsystem non-invertible symmetry (tailored to this specific degenerate ground state $|\Psi_0\rangle$), in sharp contrast to emergent subsystem invertible symmetries tailored to degenerate ground states discussed in Ref.~\cite{jesse} (also cf.~Appendix~\ref{elitzur}). If the total number of lattice sites in the local states $\vert 0 \rangle_j$ is three,  labeled as $j_1$, $j_2$ and $j_3$ ($j_1<j_2<j_3$), then one may define emergent subsystem symmetries $g_{j_1,j_2}$, $g_{j_2,j_3}$ and $g_{j_3,j_1}$, all of which take the same form as (\ref{g-operator}), subject to the condition that, when $g_{j_3,j_1}$ is defined, $j$ should be identified as $j-L$ if $j$ exceeds $L$. That is, $j$ should be understood as $j \; ({\rm mod }\; L)$.  It is readily seen that $g_{j_1,j_2} *g_{j_2,j_3} = g_{j_2,j_3}*g_{j_1,j_2}$,  $g_{j_1,j_2} *g_{j_2,j_3}=K*g_{j_3,j_1}$ and 
$g_{j_1,j_2} *g_{j_2,j_3}*g_{j_3,j_1}=K$, where $*$ denotes the combined operation and $K$ represents the time-reversal symmetry operation that flips spins from $\vert + \rangle_j$ to $\vert - \rangle_j$ and vice versus, but leaves $\vert 0 \rangle_j$ intact when it acts on all spins in the entire lattice.
This construction may be exten)ded to any primary generalized highest weight state $|\Psi_0\rangle$ that appears as a fully factorized (degenerate) ground state defined above. Indeed, it is possible to have a certain number of lattice sites in the local states $\vert 0 \rangle_j$, as long as they are not adjacent to each other. As a result, there is at least one lattice site in the local state  $\vert + \rangle_j$ between any two lattice sites in the local states $\vert 0 \rangle_j$. This implies that the total number of lattice sites in the local states $\vert 0 \rangle_j$ must not be greater than $L/2$ for even $L$ and  $(L-1)/2$ for odd $L$ (cf. Appendix~\ref{pipi}). 

Meanwhile, $g^\dagger_{j_1,j_2}$ undoes what $g_{j_1,j_2}$ does, in the sense that it flips all spins between any two closest lattice sites in the local states $\vert 0 \rangle_{j_1}$ and $\vert 0 \rangle_{j_2}$ from  $\vert - \rangle_{j}$ to  $\vert + \rangle_{j}$ ($j_1<j<j_2$). Here we stress that  $g_{j_1,j_2}$ acts on a  primary generalized highest weight state $|\Psi_0\rangle$, whereas $g^\dagger_{j_1,j_2}$ acts on the image of $|\Psi_0\rangle$ under the time-reversal symmetry $K|\Psi_0\rangle$.  Mathematically, we have $K*g_{j_3,j_1}= g^\dagger_{j_3,j_1}*K$.

We are now allowed to flip all spins between any two lattice sites in the local states $\vert 0 \rangle_{j_1}$ and $\vert 0 \rangle_{j_2}$ from  $\vert + \rangle_{j}$ to  $\vert - \rangle_{j}$ and vice versus, but leave $\vert 0 \rangle_{j}$ ($j_1<j<j_2$) intact (if any), by introducing the combined operation of emergent subsystem non-invertible symmetries, which represents the time-reversal symmetry operation that {\it only} acts on
spin states in a block consisting of  adjacent lattice unit cells, namely
all spins between any two lattice sites in the local states $\vert 0 \rangle_j$.
With this fact in mind, one may regard an emergent subsystem non-invertible symmetry  $g_{j_1,j_2}$ as a {\it fragmented}  time-reversal symmetry operation, in the sense that they produce the same result. In addition, as seen from Eq.\;(\ref{g-operator}), the symmetry generator  $\sum _{j=1}^{L} (-1)^j (S_j^-)^2$ is also fragmented, although it only commutes with the Hamiltonian (\ref{hambq}) when $L$ is even. 
We speculate that the so-called Hilbert space fragmentation, as revealed in Ref.~\cite{moudgalya} for the ferromagnetic spin-1 biquadratic model, is relevant to this time-reversal symmetry fragmentation.

As a consequence, the primary family consists of primary generalized highest weight states defined to be fully factorized (degenerate) ground states $|\Psi_0\rangle$,  which only contain $\vert + \rangle_j$ and $\vert 0 \rangle_j$, as long as the local states $\vert 00 \rangle_{j,j+1}$ on the two adjacent lattice sites are excluded, in addition to the unique highest weight state, namely the fully polarized state
$\ket{\psi_0} \equiv  \otimes_{j=1}^L  \vert+\rangle_j$. All secondary generalized highest weight states follow from primary  generalized highest weight states by acting emergent subsystem non-invertible symmetries. As a consequence, all generalized highest weight states, which appear as fully factorized  degenerate ground states~\cite{goldensu3,dimertrimer,jesse}, are derived.

For the ferromagnetic $\rm {SU}(2)$ flat-band Tasaki model (\ref{hamtasaki}),  the identification of a state $\vert \eta \rangle$ with imaginary Green parafermion states $a^*_{\sigma_1, p, k_1}a^*_{\sigma_2,p, k_2} \ldots a^*_{\sigma_m, p,k_m} \vert \Omega_0 \rangle$ or $a^\dagger_{\sigma_1, p, k_1}a^\dagger_{\sigma_2,p, k_2} \ldots a^\dagger_{\sigma_m, p,k_m} \vert \Omega_0 \rangle$ for an atypical partition and their counterparts for a typical partition in the momentum space representation 
yields a sequence of (fully factorized) degenerate ground states
$\hat{a}_{q_1,\sigma_1}^\dagger  \hat{a}_{q_2,\sigma_2}^\dagger \ldots \hat{a}_{q_m,\sigma_m}^\dagger |\otimes_{y \in \Lambda} 0_y\rangle$ in a sector labeled by the eigenvalues $m$ and $\sigma/2$ ($\sigma =\sum_\sigma \sigma_\alpha $) of the electron number operator ${\hat N}= {\hat N}_{\uparrow} +{\hat N}_{\downarrow}$ and the $z$-projection of the total spin $S^z=({\hat N}_{\uparrow} - {\hat N}_{\downarrow})/2$, where $(q_1,\sigma_1), \ldots, (q_m,\sigma_m)$ are subject to the condition that no local states with both spin up and spin down are allowed on the same lattice site and on the two adjacent lattice sites in the external sublattice,  as a result of a modified form of the Pauli exclusion principle for Green parafermions, due to the presence of the projection operator $\Pi$~(cf.~Subsection~\ref{hierarchical}). Note that this identification also works for imaginary Green parafermion states $a^*_{\sigma_1, p, l_1}a^*_{\sigma_2,p, l_2} \ldots a^*_{\sigma_m, p,l_m} \vert \Omega_0 \rangle$ or $a^\dagger_{\sigma_1, p, l_1}a^\dagger_{\sigma_2,p, l_2} \ldots a^\dagger_{\sigma_m, p,l_m} \vert \Omega_0 \rangle$ for an atypical partition and their counterparts for a typical partition in the real space representation. 

We remark that all the highest and generalized highest weight states, together with all the lowest and generalized lowest weight states, may be derived directly from this identification. Note that this is quite different from what one encounters in the ferromagnetic spin-1 biquadratic model, as seen above. However, we may only focus on the primary family, with its members denoted as $|\Psi_0\rangle$, as long as a hierarchical structure among all fully factorized degenerate ground states is concerned.  Intriguingly, $|\Psi_0\rangle$ take the form $|\Psi_0\rangle= \hat{a}_{q_1,\uparrow}^\dagger  \hat{a}_{q_2,\uparrow}^\dagger \ldots \hat{a}_{q_m,\uparrow}^\dagger |\otimes_{y \in \Lambda} 0_y\rangle$ in a sector labeled by the eigenvalues $m$ and $m/2$ ($\sigma =\sum_\sigma \sigma_\alpha $) of the total number of electrons ${\hat N}= {\hat N}_{\uparrow} +{\hat N}_{\downarrow}$ and the $z$-projection of the total spin $S^z=({\hat N}_{\uparrow} - {\hat N}_{\downarrow})/2$. As a result, the number of highest weight states in this sector is $C_L^m$. The highest weight state is thus unique in the sector when $m=L$ and $m=0$, namely at zero and quarter fillings. However,  there are more than one highest weight state at other fillings, when $m \neq 0$ and $L$. 

As it turns out, all the highest weight states at non-zero fillings may be constructed from the unique highest weight state at zero filling, namely the fermionic Fock vacuum $\vert \otimes_{x \in \Lambda} 0_x\rangle$. In fact, for any subset $A_\uparrow$ of $\mathscr{E}$, one may define an emergent subsystem symmetry operation $g_{A_\uparrow}$ tailored to the fermionic Fock vacuum $\vert \otimes_{x \in \Lambda} 0_x\rangle$ as follows 
\begin{equation}
	g_{A_\uparrow}=  \prod _{q \in A_\uparrow} \hat{a}_{q,\uparrow}^\dagger. \label{g-operator-0} 
\end{equation}
In fact, they may be generated successively from the fermionic Fock vacuum $\vert \otimes_{x \in \Lambda} 0_x\rangle$, if one interprets $\hat{a}_{q,\uparrow}^\dagger$ as the raising operator, as explained below. Mathematically, we have $g_{A_\uparrow^1} * g_{A_\uparrow^2} = g_{A_\uparrow^1 \cup A_\uparrow^2}$, if $A_\uparrow^1 \cap A_\uparrow^2 = \phi$, where $A_\uparrow^1$ and $A_\uparrow^2$ are two subsets of $\mathscr{E}$, without any intersection.

We now turn to the construction of generalized highest weight states from a specific highest weight state via an emergent subsystem symmetry operation. If one considers a highest weight state $|\Psi_0\rangle$ when $m < L-2$, then there are at least two lattice unit cells, on which no creation operators $\hat{a}_{q,\uparrow}^\dagger$ act. In other words, there are at least two lattice sites in the external sublattice  $\mathscr{E}$ that are in the local states   $\vert 0 \rangle_q$ ($q \in \mathscr{E}$). For instance,  if the total number of lattice sites in the local states $\vert 0 \rangle_q$ is two, which are labeled by $q_1$ and $q_2$ ($q_1 < q_2$), then one may choose an emergent subsystem symmetry operation $g_{q_1,q_2}$ to be 
\begin{equation}
	g_{q_1,q_2}=  (S_B^-)^{q_2-q_1-1}, \label{g-operator-1}
\end{equation}
with
\begin{equation*}
	S_B^-= \sum _{q=q_1+1}^{q_2-1} S_q^-+\sum _{u=q_1+1/2}^{q_2-1/2} S_u^-. 
\end{equation*}
Note that $g_{q_1,q_2}$ does not commute with the Hamiltonian  (\ref{hamtasaki}), namely  $[\mathscr{H}, g_{q_1,q_2} \neq 0$. However, the commutator $C_{q_1,q_2}=[\mathscr{H},g_{q_1,q_2}]_-$ nullifies $|\Psi_0\rangle$, namely $C_{q_1,q_2}|\Psi_0\rangle=0$.  In addition, $g_{q_1,q_2}$ is non-invertible. Hence $g_{{\tilde q}_1,{\tilde q}_2}$  is an emergent subsystem non-invertible symmetry (tailored to this specific degenerate ground state $|\Psi_0\rangle$), in sharp contrast to emergent subsystem invertible symmetries discussed in Ref.~\cite{jesse} (also cf.~Appendix~\ref{elitzur}).
If the total number of lattice sites in the local states $\vert 0 \rangle_q$ is three,  labeled as $q_1$, $q_2$ and $q_3$ ($q_1<q_2<q_3$), then one may define emergent subsystem symmetries $g_{q_1,q_2}$, $g_{q_2,q_3}$ and $g_{q_3,q_1}$, all of which take the same form as (\ref{g-operator-1}), subject to the condition that, when $g_{q_3,q_1}$ is defined, $q$ should be identified as $q-L$ if $q$ exceeds $L$.  It is readily seen that $g_{q_1,q_2} *g_{q_2,q_3} = g_{q_2,q_3}*g_{q_1,q_2}$,  $g_{q_1,q_2} *g_{q_2,q_3}=K*g_{q_3,q_1}$ and 
$g_{q_1,q_2} *g_{q_2,q_3}*g_{q_3,q_1}=K$, where $*$ denotes the combined operation and $K$ represents the time-reversal symmetry operation that flips spins from $\vert \uparrow \rangle_q$ to $\vert \downarrow \rangle_q$ and vice versus, when it acts on all spin configurations in the entire lattice $\Lambda$. We remark that there is a drastic difference between the ferromagnetic spin-1 biquadratic model and  the ferromagnetic $\rm {SU}(2)$ flat-band Tasaki model. For the former,  all the states containing the local states $\vert 00 \rangle_{j,j+1}$ on any two adjacent lattice sites are projected out, whereas for the latter an arbitrary number of the local states $\vert 0 \rangle_q$ are allowed to be present on any region consisting of adjacent lattice unit cells. As a result, $g_{q_1,q_2}$ might be trivial when $q_2 = q_1+1$, since this choice is not excluded.
This construction may be extended to any highest weight state $|\Psi_0\rangle$ defined above.

Meanwhile, $g^\dagger_{q_1,q_2}$ undoes what $g_{q_1,q_2}$ does, in the sense that it flips all spins between any two closest lattice sites in the local states $\vert 0 \rangle_{q_1}$ and $\vert 0 \rangle_{q_2}$ from  $\vert \downarrow \rangle_{q}$ to  $\vert \uparrow \rangle_{q}$. Here we stress that  $g_{q_1,q_2}$ acts on a  primary highest weight state $|\Psi_0\rangle$, whereas $g^\dagger_{q_1,q_2}$ acts on the image of  $|\Psi_0\rangle$ under the time-reversal symmetry $K|\Psi_0\rangle$.  Mathematically, we have $K*g_{q_3,q_1}= g^\dagger_{q_3,q_1}*K$.

We are now allowed to flip all spin configurations between any two lattice sites in the local states $\vert 0 \rangle_{q_1}$ and $\vert 0 \rangle_{q_2}$ from  $\vert \uparrow \rangle_{q}$ to  $\vert \downarrow \rangle_{q}$ and vice versus, but leave $\vert 0 \rangle_{q}$ ($q_1<q<q_2$) intact (if any), by introducing the combined operation of emergent subsystem non-invertible symmetries, which represents the time-reversal symmetry operation that {\it only} acts on fermion states in a block consisting of  adjacent lattice unit cells, namely
all spin configurations between any two lattice sites in the local states $\vert 0 \rangle_q$.
Hence emergent subsystem non-invertible symmetries may be regarded as a {\it fragmented}  time-reversal symmetry operation, in the sense that they produce the same result. In addition, as seen from Eq.\;(\ref{g-operator-1}), the symmetry generator  $\sum _{x \in \Lambda} S_x^-$ is also fragmented. As a consequence, one may expect that, the so-called Hilbert space fragmentation, as revealed in Ref.~\cite{moudgalya} for the ferromagnetic spin-1 biquadratic model, must be present in the ferromagnetic $\rm {SU}(2)$ flat-band Tasaki model, due to the relevance to this time-reversal symmetry fragmentation. Note that the translation-invariant Hamiltonian densities are replaced by the non-translation-invariant ones, with disorder being incorporated in the coupling constants~\cite{moudgalya}, as already mentioned in Appendix~\ref{frustration-free}.

We are thus led to conclude that the primary family consists of all the highest weight states for the ferromagnetic $\rm {SU}(2)$ flat-band Tasaki model. The secondary family contains generalized highest weight states and lowest weight states by acting emergent subsystem non-invertible symmetries on the primary family members.

\vspace{10mm}

We have revealed emergent subsystem non-invertible symmetries tailored to a specific primary family member as a degenerate ground state for both of the two illustrative models, which allows to derive secondary generalized highest weight states from the primary family. In fact, emergent subsystem non-invertible symmetries tailored to a specific primary family member form a semigroup, if the combined operation $*$ is regarded as the defining multiplication operation. This reflects a common hierarchical structure underlying highest and generalized highest weight states for condensed matter systems undergoing SSB with type-B GMs, if the ground state degeneracies are exponential. 

\subsubsection{The  ladder operators as emergent subsystem  non-invertible symmetries}

In addition to emergent subsystem  non-invertible symmetries  tailored to a specific primary family member as a degenerate ground state, there are also  other emergent subsystem  non-invertible symmetries tailored to some subset of the entire ground state subspace. They may be interpreted as the ladder operators, including the raising and lowering operators.
In other words, one may identify emergent subsystem  non-invertible symmetries that act as the ladder operators to generate all the highest and generalized highest weight states successively, starting from one chosen highest weight state. This construction leads to  all primary family members for the ferromagnetic spin-1 biquadratic model, when Green parafermionns live in the constrained Hilbert space $V_1$, and to all highest and generalized  weight states for the ferromagnetic $\rm {SU}(2)$ flat-band Tasaki model, when  Green parafermionns live in the original unconstrained Hilbert space $V_0$. In this sense, the ladder operator approach developed here may be viewed as an alternative way to construct exponentially many (fully factorized) degenerate ground states for condensed matter systems undergoing SSB with type-B GMs.

For the ferromagnetic spin-1 biquadratic model (\ref{hambq}),  one may choose emergent subsystem symmetry operations $g_j$ to be 
$g_j = \sigma _j^-$, where  $\sigma _j^-$ represents $\Pi_1 S_j^- \Pi_1$, then it is readily seen that $g_j$  do not commute with the Hamiltonian (\ref{hambq}). That is, we have $[\mathscr{H}, g_j]_- \neq 0$. However, the commutator $C_j=[\mathscr{H}, g_j]_-$ nullifies $|\psi_0\rangle$, namely $C_j|\psi_0\rangle=0$, if $|\psi_0\rangle$ is chosen to be the unique highest weight state $\vert \otimes +_j \rangle$.  Since $g_j$ are not invertible, $g_j$ are  emergent subsystem non-invertible symmetries. This argument may be extended to  emergent subsystem symmetry operations $g_{j_1,j_2,\ldots,j_m} = \sigma _{j_1}^-\sigma _{j_2}^-\ldots \sigma _{j_m}^-$. Note that there are two different interpretations for $g_{j_1,j_2,\ldots,j_m} = \sigma _{j_1}^-\sigma _{j_2}^-\ldots \sigma _{j_m}^-$ to be an emergent subsystem symmetry operation. One is to interpret it as an emergent subsystem non-invertible symmetry operation tailored to the highest weight state $|\psi_0\rangle$. The other is to interpret $\sigma _{j_m}^-$ as
an emergent subsystem non-invertible symmetry operation tailored to $\sigma _{j_1}^-\sigma _{j_2}^-\ldots \sigma _{j_{m-1}}^-|\psi_0\rangle$
successively, as $m$ increases. We are thus able to reproduce all fully factorized degenerate ground states $S _{j_1}^- S_{j_2}^-\ldots S_{j_m}^- |\Psi_0\rangle$, as long as all the local states $\vert 00 \rangle_{j,j+1}$ on the two adjacent lattice sites are excluded,
as follows from a modified form of the Pauli exclusion principle for Green parafermions, due to the presence of the projection operator $\Pi$~(cf.~Subsection~\ref{hierarchical}).
Indeed, they are primary generalized highest weight states. 
Actually, $g_j$ act as the raising operators, in the sense that, if they act on one highest or generalized highest weight state, then other ones are generated successively. We remark that the lowering operators $g_j^\dagger = \sigma _j^+$, where  $\sigma _j^+$ represents $\Pi_1 S_j^+ \Pi_1$, are Hermitian conjugated to the raising operators $g_j = \sigma _j^-$. This reflects the fact that the realization of Green parafermions in the ferromagnetic spin-1 biquadratic model  is Hermitian. 

We turn to the ferromagnetic $\rm {SU}(2)$ flat-band Tasaki model (\ref{hamtasaki}). If one chooses emergent subsystem symmetry operations defined by the operators $g_{q,\sigma}$ to be 
$g_{q,\sigma} = \hat{a}_{q,\sigma}^\dagger$ ($\sigma = \uparrow$ and $\downarrow$), then it is readily seen that $g_{q,\sigma}$  do not commute with the Hamiltonian (\ref{hamtasaki}). That is, we have $[\mathscr{H}, g_{q,\sigma}]_- \neq 0$. However, the commutator $C_{q,\sigma}=[\mathscr{H}, g_{q,\sigma}]_-$ nullifies $|\Psi_0\rangle$, namely $C_{q,\sigma}|\Psi_0\rangle=0$, if $|\Psi_0\rangle$ is chosen to be the fermionic Fock vacuum $\vert \otimes_{x \in \Lambda}  0_x \rangle$.  Obviously $g_{q,\sigma}$ are not invertible, so $g_{q,\sigma}$ are  emergent subsystem non-invertible symmetries. This argument may be extended to  emergent subsystem symmetry operations defined by the operators $g_{\{ q\},\{\sigma\}} = \hat{a}_{q_1,\sigma_1}^\dagger  \hat{a}_{q_2,\sigma_2}^\dagger \ldots \hat{a}_{q_m,\sigma_m}^\dagger$.
Again, there are two different interpretations for $g_{\{ q\},\{\sigma\}} = \hat{a}_{q_1,\sigma_1}^\dagger  \hat{a}_{q_2,\sigma_2}^\dagger \ldots \hat{a}_{q_m,\sigma_m}^\dagger$ to be an emergent subsystem symmetry operation. One is to interpret it as an emergent subsystem non-invertible symmetry operation tailored to the fermionic Fock vacuum $\vert \otimes_{x \in \Lambda}  0_x \rangle$. The other is to interpret $\hat{a}_{q_m,\sigma_m}^\dagger$ as an emergent subsystem non-invertible symmetry operation tailored to $\hat{a}_{q_1,\sigma_1}^\dagger  \hat{a}_{q_2,\sigma_2}^\dagger \ldots \hat{a}_{q_{m-1},\sigma_{m-1}}^\dagger \vert \otimes_{x \in \Lambda}  0_x \rangle$
successively, as $m$ increases.  We are thus able to  reproduce all degenerate ground states $\hat{a}_{q_1,\sigma_1}^\dagger  \hat{a}_{q_2,\sigma_2}^\dagger \ldots \hat{a}_{q_m,\sigma_m}^\dagger \vert \otimes_{x \in \Lambda}  0_x\rangle$, which act as the highest weight states in the sector, if all $m$ spins are up, as the lowest weight states in the sector, if all $m$ spins are down, and as generalized highest weight states in the sector, if $m_{\uparrow}$ spins are up and the other $m_{\downarrow}$ spins are down ($m= m_{\uparrow}+ m_{\uparrow}$), where $(q_1,\sigma_1), \ldots, (q_m,\sigma_m)$ are subject to the condition that no local states with both spin up and spin down are allowed on the same lattice site and on the two adjacent lattice sites in the external sublattice, as follows from a modified form of the Pauli exclusion principle for Green parafermions, due to the presence of the projection operator $\Pi$~(cf.~Subsection~\ref{hierarchical}). 

Actually, $g_{q,\sigma} = \hat{a}_{q,\sigma}^\dagger$ act as the raising operators, in the sense that, if they act on one highest or generalized highest weight state, then other ones are generated. However, the lowering operators are not their Hermitian conjugated operators. This reflects the fact that the ferromagnetic $\rm {SU}(2)$ flat-band Tasaki model entails a non-Hermitian realization of Green parafermions. In particular, the Green parafermion number operators are non-Hermitian.
Instead, the lowering operators  $g^*_{q,\sigma} = \hat{c}_{q,\sigma}$ ($\sigma = \uparrow$ and $\downarrow$) act as  emergent subsystem  non-invertible symmetries. It is readily seen that $g^*_{q,\sigma}$ are non-invertible and do not commute with the Hamiltonian (\ref{hamtasaki}). That is, we have $[\mathscr{H},g^*_{q,\sigma}]_- \neq 0$. However, the commutator $C_{q,\sigma}=[\mathscr{H},g^*_{q,\sigma}]_-$ nullifies the unique highest weight state $\hat{a}_{q_1,\uparrow}^\dagger  \hat{a}_{q_2,\uparrow}^\dagger \ldots \hat{a}_{q_L,\uparrow}^\dagger \vert \otimes_{x \in \Lambda}~0_x~\rangle$  or
the unique lowest weight state $\hat{a}_{q_1,\downarrow}^\dagger  \hat{a}_{q_2,\downarrow}^\dagger \ldots \hat{a}_{q_L,\downarrow}^\dagger \vert \otimes_{x \in \Lambda}~0_x~\rangle$
at quarter filling (modulo those generated from the action of the lowering or raising operator of the symmetry group $\rm{SU(2)}$ in the spin sector on the highest or lowest weight state), namely $C_{q,\sigma} (\hat{a}_{q_1,\uparrow}^\dagger  \hat{a}_{q_2,\uparrow}^\dagger \ldots \hat{a}_{q_L,\uparrow}^\dagger) \vert \otimes_{x \in \Lambda}~0_x~\rangle=0$ or $C_{q,\sigma} (\hat{a}_{q_1,\downarrow}^\dagger  \hat{a}_{q_2,\downarrow}^\dagger \ldots \hat{a}_{q_L,\downarrow}^\dagger) \vert \otimes_{x \in \Lambda}  0_x \rangle=0$.  This argument may be extended to  emergent subsystem symmetry operations defined by the operators $g^*_{\{ q\},\{\sigma\}} = \hat{c}_{q_1,\sigma_1}  \hat{c}_{q_2,\sigma_2} \ldots \hat{c}_{q_m,\sigma_m}$. We are thus able to  reproduce all highest weight states $\hat{a}_{q_1,\uparrow}^\dagger  \hat{a}_{q_2,\uparrow}^\dagger \ldots \hat{a}_{q_m,\uparrow}^\dagger \vert \otimes_{x \in \Lambda}  0_x\rangle$ or all lowest weight states $\hat{a}_{q_1,\downarrow}^\dagger  \hat{a}_{q_2,\downarrow}^\dagger \ldots \hat{a}_{q_m,\downarrow}^\dagger \vert \otimes_{x \in \Lambda}  0_x\rangle$ in the sector labeled by the eigenvalues $m$ of the electron number operator ${\hat N}= {\hat N}_{\uparrow} +{\hat N}_{\downarrow}$.

\vspace{10mm}
We note that the ladder operators allow to generate all primary family members for a condensed matter system undergoing SSB with type-B GMs, if Green parafermions live in the constrained Hilbert space $V_1$, as happens to the spin-1 ferromagnetic biquadratic model. One thus still needs to resort to 
emergent subsystem non-invertible symmetries tailored to a specific primary family member to yield all generalized highest weight states. In contrast, 
both emergent subsystem non-invertible symmetries tailored to a specific primary family member and the ladder operators provide an alternative means to yield  all generalized highest weight states for a condensed matter system undergoing SSB with type-B GMs, if Green parafermions live in the unconstrained Hilbert space $V_0$, as happens to the ferromagnetic $\rm {SU}(2)$ flat-band Tasaki model.

\subsection{Real and imaginary Green parafermions in the first quantization formalism}~\label{symmetricgroup}

As already mentioned in Section~\ref{intro},  Stolt and Taylor~\cite{stolt} showed that  for Green paraparticles of finite order,  the first and second quantization formalisms yield consistent results. However, this equivalence is yet to be established for Green paraparticles of infinite order. 
Recall that the order of Green parafermions realized in a condensed matter system is identical to the emergent unit cell size. Hence we are led to conclude that a specific realization of real Green parafermions in a condensed matter system, as presented in the second quantization formalism, may be equivalently formalized in the first quantization formalism, as long as the emergent unit cell size $p$ is finite. As a consequence,  real Green parafermions on the (periodic) emergent unit cells for an atypical partition follow from a high-dimensional representation of the symmetric group $S_N$ ($N=L/p$).

However,  for a specific realization of imaginary Green parafermions in a condensed matter system, the key ingredient is auxiliary (fictitious) Majorana fermions. As argued in Section~\ref{gscheme}, auxiliary (fictitious) Majorana fermions are not included in the Hilbert space, in contrast to auxiliary (physical) Majorana fermions. In particular, they are defined on any emergent unit cells for any possible (atypical and typical) partitions, and must be removed eventually from the Hilbert space. Hence imaginary Green parafermions are not treated as an entity that is regarded as composite particles, in contrast to real Green parafermions. Consequently, the approach developed by Stolt and Taylor~\cite{stolt} is not suitable for imaginary Green parafermions realized in a condensed matter system.

Instead, an alternative approach to imaginary Green parafermions needs to be developed in the first quantization formalism. 
Consider a wave function $\Psi(\{ j_\alpha \}_{\alpha =1}^m)$ for the ferromagnetic spin-1  biquadratic model (\ref{hambq}) or  $\Psi(\{ q_\alpha \sigma_\alpha\}_{\alpha =1}^m)$ for the ferromagnetic $\rm {SU}(2)$ flat-band Tasaki model (\ref{hamtasaki}), where $m=0,1,\ldots,L$. 
We remark that $\Psi(\{ j_\alpha \}_{\alpha =1}^m)$ and $\Psi(\{ q_\alpha \sigma_\alpha\}_{\alpha =1}^m)$ correspond to  $|\psi_m^{j_1,j_2,\ldots,j_m}\rangle \equiv S_{j_1}^-S_{j_2}^- \ldots S_{j_m}^- \ket{\psi_0}$ and $\hat{a}_{q_1,\sigma_1}^\dagger  \hat{a}_{q_2,\sigma_2}^\dagger \ldots \hat{a}_{q_m,\sigma_m}^\dagger \vert \otimes_{x \in \Lambda}  0_x\rangle$ in the second quantization formalism, respectively 
(cf. Section~\ref{fictitious}). Note that they follow from the Green parafermion states for all possible partitions. Here no constraints are imposed on $\{ j_1,j_2,\ldots,j_m\}$ or $\{q_1,q_2,\ldots,q_m\}$ and $\{ \sigma_1,\sigma_2,\ldots,\sigma_m\}$. Accordingly, all these states span the constraint Hilbert space $V_1$,  where the dimension of $V_1$ is $2^L$ for the ferromagnetic spin-1  biquadratic model and $4^L$ for the ferromagnetic $\rm {SU}(2)$ flat-band Tasaki model.  Actually, for the former, the constraint Hilbert space $V_1$ specified by the projection operator $\Pi_1$ is decomposed into $L+1$ sectors $V_1^m$ labeled by $m$, with the dimension being the binomial coefficients $C_L^m$. This is also valid for the latter, if all the spins are up or down. Note that $\sum_m C_L^m=2^L$. 

For the symmetric group  ${\rm S}_L$ acting on $j$ or $q$, we only need to consider $L-1$ adjacent exchanges $P_{j j+1}$ or $P_{q q+1}$ ($j$ or $q=1,2,\ldots,L-1$). This is due to the fact that they act as the generators of the symmetric group  ${\rm S}_L$, in the sense that any non-adjacent exchange may be decomposed into a sequence of adjacent exchanges. Suppose the action of $P_{j j+1}$ on $\Psi(\{ j_\alpha \}_{\alpha =1}^m)$ yields
\begin{equation*}
P_{j j+1} \Psi(\{ j_\alpha \}_{\alpha =1}^m) = \sum _{\{ j'_\alpha \}_{\alpha =1}^m} (R_{jj+1})_{j_1,j_2,\ldots,j_m}^{j'_1,j'_2,\ldots,j'_m}  \Psi(\{ j'_\alpha \}_{\alpha =1}^m),
\end{equation*}
then the matrices $R_{jj+1}$ satisfy the constraints~\cite{hazzard}: (i) $R_{jj+1}^2$ must be the identity; (ii) $R_{j-1 j}R_{jj+1}R_{j-1 j}= R_{jj+1}R_{j-1 j}R_{jj+1}$;
(iii) $R_{i i+1}R_{jj+1}= R_{jj+1}R_{i i+1}$ for any $i$ and $j$ if $|i-j| \ge 2$. Similarly, the action of $P_{q q+1}$ on $\Psi(q_1 \sigma_1,q_2\sigma_2,\ldots,q_m\sigma_m)$ yields the matrices $R_{qq+1}$ satisfying the above three constraints, with $j$ replaced by $q$.
These constraints imply that  the matrices $R_{jj+1}$ or $R_{qq+1}$ constitute a representation of the symmetric group ${\rm S}_L$.

Now we turn to the explicit expressions of $\Psi(\{ j_\alpha \}_{\alpha =1}^m)$ for distinct but fixed $m$ ($m=1,2,\dots,L$). As already argued in Ref.~\cite{exactmps}, $\Psi(\{ j_\alpha \}_{\alpha =1}^m)$ are unentangled states, so they are subject to an exact matrix product state representation with the bond dimension being one. For $m=0$, there is only one state,  namely the highest weight state $\vert \psi_0 = \otimes _j \vert + \rangle_j$, so this sector constitutes a one-dimensional representation of the symmetric group $S_L$, with the matrices $R_{jj+1}$ being 1 for all $j$. For $m=1$, we have $L$ states in total, which take the form $\otimes _{i=1}^{j-1} \vert + \rangle_i
\vert 0 \rangle_j \otimes _{i=j+1}^{L}\vert + \rangle_i$. They constitute an $L$-dimensional representation of the symmetric group  ${\rm S}_L$, with the matrices $R_{jj+1}$ being $L\times L$ matrices. Generically, for arbitrary $m$, we have $C_L^m$ states, which constitute a $C_L^m$-dimensional representation of the symmetric group ${\rm S}_L$, with the matrices $R_{jj+1}$ being $C_L^m\times C_L^m$ matrices. This process continues until $m$ hits $L$.
For fixed $m$, it is readily seen that the corresponding matrices $R_{jj+1}$ satisfy the three constraints (i), (ii) and (iii) above.
A similar conclusion is valid for $\Psi(\{ q_\alpha \sigma_\alpha\}_{\alpha =1}^m)$, with a caveat that they are generically entangled (at least locally), due to the fact that the regions on which $\hat{a}_{q,\sigma}^\dagger$ are incommensurate with the lattice unit cells. This amounts to stating that they are subject to an exact graded matrix product state representation with the bond dimension being greater than one~\cite{TypeBtasaki}.

However, what has been realized in the two models is the projected Green parafermion states. As stressed in Section~\ref{ssbscenario}, the presence of the projection operator $\Pi$ affects parastatistics, leading to a modified form of the Pauli exclusion principle. Hence we have to deal with the projected Green parafermion states. As shown in Section~\ref{fictitious}, for the ferromagnetic spin-1  biquadratic model,  $\{ j_1,j_2,\ldots,j_m\}$ are subject to the constraints that $j_1$ is not less than 1, $j_{\beta+1}$ not less than $j_{\beta}+2$ ($\beta =1,2,\ldots,m-2$), and $j_m$  not less than $j_{m-1}+2$, but less than $L$. For the ferromagnetic $\rm {SU}(2)$ flat-band Tasaki model,  $(q_1,\sigma_1), \ldots, (q_m,\sigma_m)$ are subject to the condition that no local states with both spin up and spin down are allowed on the same lattice site and on the two adjacent lattice sites in the external sublattice.
Accordingly, we have to deal with a subset of the set consisting of all the states
$\Psi(\{ j_\alpha \}_{\alpha =1}^m)$ or  $\Psi(\{ q_\alpha \sigma_\alpha\}_{\alpha =1}^m)$ for fixed $m$,  in the sense that all the states containing the local states $\vert 00 \rangle_{j,j+1}$ on the two adjacent lattice sites must be excluded.  As a result, they do not constitute any representation of the symmetric group ${\rm S}_L$, since not all permutation operations result in a (degenerate) ground state when they are acted on a specific (degenerate) ground state. Instead, for fixed $m$,  a subgroup of  the symmetric group ${\rm S}_L$ arises, which consists of all the permutations that yield a (degenerate) ground state when they are acted on each of (degenerate) ground states. In this sense, a modified form of the Pauli exclusion principle is associated with a subgroup of  the symmetric group ${\rm S}_L$, as a result of the presence of the projection operator $\Pi$.


\begin{thebibliography}{10}
	
\bibitem{messiahbook} A. Messiah,  {\it Quantum Mechanics} (Dover Publications, 1999).

\bibitem{Haag} K. Dr\"uhl, R. Haag, and J.E. Roberts, Commun. Math. Phys. \textbf{18}, 204 (1970).

\bibitem{leinaas} J. M. Leinaas and J. Myrheim, Nuovo Cim. B  \textbf{37}, 1 (1977).

\bibitem{wilczek1} F. Wilczek, Phys. Rev. Lett. \textbf{48}, 1144 (1982); Phys. Rev. Lett. \textbf{49}, 957 (1982).

\bibitem{wilczek2} F. Wilczek,  {\it Fractional Statistics and Anyon Superconductivity} (World Scientific, Singapore, 1990).

\bibitem{nayak} C. Nayak, S.H. Simon, A. Stern, M. Freedman, and S. Das Sarma, Rev. Mod. Phys.  \textbf{80}, 1083 (2008).

\bibitem{green} H. S. Green, Phys. Rev. \textbf{90}, 270 (1953).

\bibitem{kamefuchi0} S. Kamefuchi and J. Strathdee,  Nucl. Phys. \textbf{42}, 166 (1963).

\bibitem{messiah} A. M. L. Messiah and O. W. Greenberg,  Phys. Rev.  \textbf{136}, B248 (1964).

\bibitem{greenberg} O. W. Greenberg and A. M. L. Messiah, Phys. Rev. \textbf{138}, B1155 (1965).

\bibitem{hartle} J. B. Hartle and J. R. Taylor, Phys. Rev. \textbf{178}, 2043 (1969).

\bibitem{stolt} R. H. Stolt and J. R. Taylor, Nucl. Phys. B\textbf{19}, 1 (1970).

\bibitem{bialynicki} I. Bialynicki-Birula, Nucl. Phys. \textbf{49}, 605 (1963).

\bibitem{Yamada} M. Yamada,  Nucl. Phys. B\textbf{6}, 596 (1968).

\bibitem{landshoff} P. V. Landshoff and H. P. Stapp,  Ann. Phys. B\textbf{45}, 72 (1967).

\bibitem{kamefuchi}  Y. Ohnuki and S. Kamefuchi,  Ann. Phys. B\textbf{51}, 337 (1969).

\bibitem{tony} A. J. Bracken and H. S. Green, Nuovo Cimento A \textbf{9}, 349 (1972).

\bibitem{mark} M. D. Gould and J. Paldus, Phys. Rev. A \textbf{34}, 804 (1986).

\bibitem{mark1}  M. D. Gould and J. Paldus, J. Math. Phys. \textbf{28}, 2304 (1987). 


\bibitem{cyabe1} V. G. Turaev, Invent. Math.  \textbf{92}, 527 (1988).

\bibitem{cyabe2} S. Majid, Int. J. Mod. Phys. A  \textbf{05}, 1 (1990).

\bibitem{cyabe3} P. Etingof, T. Scheller, and A. Soloviev,  V.G. Turaev, Duke Math. J.  \textbf{100}, 169 (1999).

\bibitem{hazzard} Z. Wang and K. A. Hazzard, Nature,  \textbf{637}, 314 (2025).

\bibitem{hazzard1} B. Sunder, B. Gadway, and  K. A. Hazzard, Sci. Rep.  \textbf{8}, 3422 (2018).

\bibitem{hazzard2} B. Sunder, M. Thibodeau, Z. Wang, B. Gadway, and K. A. Hazzard, Phys. Rev. A  \textbf{99}, 013624 (2019).

\bibitem{jesse} H.-Q. Zhou, J. J. Osborne,  Q.-Q. Shi, and I. P. McCulloch, arXiv: 2502.14605 (2025).

\bibitem{goldensu3}
H.-Q. Zhou, Q.-Q. Shi, I. P. McCulloch, and M. T. Batchelor,  J. Phys. A: Math. Theor.  \textbf{58}, 39LT01 (2025).


\bibitem{TypeBtasaki}  H.-Q. Zhou, Q.-Q Shi, I. P. McCulloch, and J. O. Fj{\ae}restad,  arXiv: 2412.15724 (2024).

\bibitem{spinorbitalsu4} Q.-Q. Shi, H.-Q. Zhou, I. P. McCulloch, and  M. T. Batchelor,  arXiv: 2309.04973 (2023).

\bibitem{finitesize} H.-Q. Zhou, Q.-Q. Shi, I. P. McCulloch, and M. T. Batchelor,  arXiv: 2304.11339 (2023).

\bibitem{dimertrimer}  H.-Q. Zhou, Q.-Q. Shi, I. P. McCulloch, and M. T. Batchelor, arXiv:2403.06825.

\bibitem{wen} X.-G. Wen, {\it Quantum Field Theory of Many-Body Systems: From the Origin of Sound to an Origin of Light and Electrons} (Oxford,  2004). 

\bibitem{barber}  M. N. Barber and M. T. Batchelor,  Phys. Rev. B \textbf{40}, 4621 (1989).

\bibitem{barber1} M. T. Batchelor and M. N. Barber, J. Phys. A: Math. Gen. \textbf{23}, L15 (1990).

\bibitem{barber2} A. Kl\"{u}mper, Europhys. Lett. \textbf{9}, 815 (1989).



\bibitem{saleur} N. Read and H. Saleur, Nucl. Phys. B, \textbf{777}, 263-315 (2007).

\bibitem{aufgebauer}  B. Aufgebauer and A. Kl\"{u}mper,  J. Stat. Mech. P05018 (2010).

\bibitem{moudgalya} S. Moudgalya and O. I. Motrunich, Phys. Rev. X  \textbf{12}, 011050 (2022).

\bibitem{tasaki} H. Tasaki, Phys. Rev. Lett. \textbf{69}, 1608 (1992).	

\bibitem{dtmodel} Y.-T. Oh, H. Katsura, H.-Y. Lee, and J. H. Han, Phys. Rev. B, \textbf{96}, 165126 (2017).

\bibitem{So4} K. I. Kugel and D. I. Khomskii, Sov. Phys. Usp. \textbf{25}, 2319 (1982).



\bibitem{tasakibook}  H. Tasaki, {\it Physics and Mathematics of Quantum Many-Body Systems} (Springer, 2020).

\bibitem{FMGM}  Q.-Q. Shi, Y.-W. Dai, H.-Q. Zhou, and I. P. McCulloch,  J. Phys. A \textbf{58}, 05LT01 (2025).

\bibitem{hqzhou} H.-Q. Zhou, Q.-Q. Shi, J. O. Fj{\ae}restad, and I. P. McCulloch,  Phys. Rev. A \textbf{111}, 022425 (2025).

\bibitem{2dtypeb}  H.-Q. Zhou, Q.-Q Shi, I. P. McCulloch, and M. T. Batchelor, arXiv: 2412.06396 (2024).


\bibitem{baxterbook} R. J. Baxter, {\it Exactly Solved Models in Statistical Mechanics} (Academic Press, London, 1982).

\bibitem{tla} H. N. V. Temperley and E. H. Lieb,  Proc. R. Soc. Lond. \textbf{A322}, 251 (1971).

\bibitem{martin} P. Martin, {\it Potts Models and Related Problems in Statistical Mechanics} (World Scientific, Singapore, 1991).

\bibitem{faddeev} L. A. Takhtadzhan and L. D. Faddeev,  Russ. Math. Surv. \textbf{34}, 11 (1979).



\bibitem{goldstone} J. Goldstone, Nuovo Cimento \textbf{19}, 154 (1961).

\bibitem{goldstone1}	J. Goldstone, A. Salam, and S. Weinberg, Phys. Rev. \textbf{127}, 965 (1962).

\bibitem{goldstone2}	Y. Nambu and G. Jona-Lasinio, Phys. Rev. \textbf{122}, 345 (1961).
		
	
\bibitem{watanabe} H. Watanabe and H. Murayama,  Phys. Rev. Lett. \textbf{108}, 251602 (2012).

\bibitem{watanabe1} H. Watanabe and H. Murayama, Phys. Rev. X \textbf{4}, 031057 (2014).
	
\bibitem{NG} Y. Hidaka, Phys. Rev. Lett. \textbf{110}, 091601 (2013).

\bibitem{NG1}	T. Hayata and Y. Hidaka,  Phys. Rev. D \textbf{91}, 056006 (2015).

\bibitem{NG2}	D. A. Takahashi and M. Nitta,   Ann. Phys. \textbf{354}, 101 (2015).
	
	
\bibitem{nambu} Y. Nambu,  J. Stat. Phys. \textbf{115}, 7 (2004).
	
\bibitem{nielsen} H. B. Nielsen and S. Chadha, Nucl. Phys. B \textbf{105}, 445 (1976).
	
\bibitem{schafer}  T. Schafer, D. T. Son, M. A. Stephanov, D. Toublan, and J. J. M. Verbaarschot, Phys. Lett. B \textbf{522}, 67 (2001).
	
\bibitem{miransky} V. A. Miransky and I. A. Shovkovy,  Phys. Rev. Lett. \textbf{88}, 111601 (2002).
	
\bibitem{nicolis}	 A. Nicolis and F. Piazza,  Phys. Rev. Lett. \textbf{110}, 011602 (2013).
	
	
\bibitem{brauner-watanabe} H. Watanabe, T. Brauner, and H. Murayama, Phys. Rev. Lett. \textbf{111}, 021601 (2013).
	
\bibitem{brauner-watanabe1}	 H. Watanabe and T. Brauner, Phys. Rev. D \textbf{84}, 125013 (2011).
	

\bibitem{mwc} N. D. Mermin and H. Wagner,  Phys. Rev. Lett. \textbf{17}, 1133 (1966).

\bibitem{mwc1} S. R. Coleman, Commun. Math. Phys. \textbf{31}, 259 (1973).


\bibitem{afflecksun} I. Affleck, J. Phys.: Condens. Matter \textbf{2}, 405 (1990).


\bibitem{mielke} A. Mielke, J. Phys. A \textbf{24}, 3311 (1991).

\bibitem{popkov} V. Popkov and M. Salerno,  Phys. Rev. A \textbf{71}, 012301 (2005).

\bibitem{popkov1} V. Popkov, M. Salerno, and G. Sch\"{u}tz,  Phys. Rev. A \textbf{72}, 032327 (2005).

\bibitem{doyon} O. A. Castro-Alvaredo and B. Doyon, Phys. Rev. Lett. \textbf{108}, 120401 (2012).

\bibitem{doyon1} O. A. Castro-Alvaredo and B. Doyon,  J. Stat. Mech.  P02016 (2013).
  
\bibitem{shi-so} H.-Q. Zhou { \it et al.}, A realization of Green parafermions in the ferromagnetic spin-orbital model and beyond,
  in preparation.
  
\bibitem{shi-dtmodel} H.-Q. Zhou { \it et al.}, A realization of Green parafermions in the quantum spin-1 system with competing dimer and trimer interactions,
  in preparation.
  
 \bibitem{shiqq} H.-Q. Zhou { \it et al.},  A realization of Green parafermions in the ferromagnetic $\rm {SU}(n)$ flat-band Tasaki model,
  in preparation.

\bibitem{pcoleman} P. Coleman, {\it Introduction to Many-body Physics} (Cambridge University Press, Cambridge, 2015).

\bibitem{franz} S. R. Elliott and M. Franz, Rev. Mod. Phys. \textbf{87}, 137 (2015).

\bibitem{zuber} C. Itzykson and J.-B. Zuber, {\it Quantum Field Theory}  (McGraw-Hill, New York, 1985).

\bibitem{elitzur} S. Elitzur, Phys. Rev. D \textbf{12}, 3978 (1975).
      
\bibitem{tasakidegeneracy} O. Derzhko, A. Honecker, and J. Richter, Phys. Rev. B \textbf{76}, 220402 (R) (2007).
   
\bibitem{tasakidegeneracy1}  V. Baliha, J. Richter and O. Derzhko, Acta Physica Polonica A,  \textbf{132}, 1256-1260 (2017). 


\bibitem{wzhang} R. Liu, W. Nie, and W. Zhang, Sci. Bull. \textbf{64}, 1490 (2019).

\bibitem{tamura} K. Tamura and H. Katsura, Phys. Rev. B \textbf{100}, 214423 (2019).

\bibitem{tamura1} K. Tamura and H. Katsura, J. Stat. Phys.  \textbf{182}, 16 (2021).

\bibitem{ganesh} A. J. Raja and R. Ganesh, the Kitaev-AKLT model, arXiv:2510.12880.

\bibitem{hqz} H.-Q. Zhou, Projected Green parafermion states: a realization in the Kitaev-AKLT model, preprint.

\bibitem{hall} B. C. Hall, {\it Lie Groups, Lie Algebras, and Representations: An Elementary Introduction} (Springer, 2015).

\bibitem{knapp} A. W. Knapp, {\it Representation Theory of Semisimple Groups}  (Princeton University Press, 1986).
  

\bibitem{exactmps} H.-Q. Zhou, Q.-Q. Shi, and I. P. McCulloch, arXiv: 2403.09458 (2024).

\bibitem{svd} R. A. Horn and C. R. Johnson, {\it Matrix Analysis} (Cambridge University Press, Cambridge, 1985).

\bibitem{sd-orth}  G. H. Golub and C. F. Van Loan,   {\it Matrix Computations (3rd ed.)} (Johns Hopkins, 1996).

\bibitem{fradkin} E. Fradkin, {\it Field Theory of Condensed Matter Physics (2nd ed.)}, Cambridge University Press (Cambridge, 1995).

\bibitem{weinberg} S. Weinberg, {\it The Quantum theory of fields} Vol. 1: Foundations, Cambridge University Press (Cambridge, 1995).


\bibitem{costa} G. Costa and G. Fogli, {\it Symmetries and Group Theory in Particle Physics: An Introduction to Space-Time and Internal Symmetries}, Lecture Notes in Physics 823 (2012).

\bibitem{kober} M. Kober,  About the Origin of the Division between Internal and External Symmetries in Quantum Field Theory, arXiv:0910.3303 (2009). 



\bibitem{seiberg} N. Seiberg, S. Seifnashri, and S.-H. Shao, SciPost Phys.  \textbf{16}, 154 (2024).

\bibitem{shao0}  N. Seiberg and S.-H. Shao, SciPost Phys. 16, 064  (2024).

\bibitem{shao1} S.-H. Shao, arxiv:2308.00747.

\bibitem{nameki} S. Sch\"afer-Nameki, Physics Reports \textbf{1063}, 
1 (2024). 

\bibitem{oshikawa} L. Li, M. Oshikawa and Y. Zheng, Phys. Rev. B \textbf{108}, 214429 (2023).

\bibitem{sierra} H.-C. Zhang and G. Sierra,  JHEP \textbf{2025}, 5 (2025).

\bibitem{zhou-ising} H.-Q. Zhou and Q.-Q. Shi, Non-invertible symmetries and boundary conditions for the transverse-field Ising model, preprint.

\bibitem{zhou-doubleinsing} H.-Q. Zhou and Q.-Q. Shi, {\it Non-invertible Kramers-Wannier symmetries for the double transverse-field Ising model and beyond}, preprint. 

\bibitem{subsystemsymmetry} J. Iaconis, S. Vijay, and R. Nandkishore,  Phys. Rev. B \textbf{100}, 214301 (2019).


\bibitem{subsystemsymmetry1} A. Khudorozhkov, A. Tiwari, C. Chamon, and T. Neupert,  SciPost Phys.  \textbf{13}, 098 (2022).

\bibitem{noninvertiblesymmetry} W. Cao, L. Li, M. Yamazaki, and Y. Zheng,  SciPost Phys.  \textbf{15}, 155 (2023).


\bibitem{fendley} D. Aasen, R. S. K. Mong, and P. Fendley, J. Phys. A \textbf{49}, 354001 (2016).


\bibitem{seifnashri} S. Seifnashri,	SciPost Phys. \textbf{16}, 098 (2024).
 

\bibitem{seiberg0} D. Gaiotto, A. Kapustin, N. Seiberg, and B. Willett, JHEP  \textbf{02}, 172 (2015).

\bibitem{category1} P. Etingof, S. Gelaki, D. Nikshych, and V. Ostrik, Tensor Categories. Mathematical Surveys and Monographs. American Mathematical Society, 2016.

\bibitem{category3} L. Bhardwaj and Y. Tachikawa, JHEP 03 (2018) 189.

\bibitem{category4} R. Thorngren and Y. Wang, arXiv:1912.02817.

\bibitem{category5} G. W. Moore and N. Seiberg, Commun. Math. Phys. 123 (1989) 177.

\bibitem{kramers} H. A. Kramers and G. H. Wannier, Phys. Rev. \textbf{60}, 252 (1941).

\bibitem{ty} D. Tambara and S. Yamagami, J. Algebra \textbf{209}, 692 (1998).

\bibitem{ue1} H.-Q. Zhou, {\it Symmetry group factorization and unitary equivalence among Temperley-Lieb integrable models}, preprint.


\end{thebibliography}
\end{document}